\documentclass[a4paper,fleqn,usenatbib]{mnras}

\usepackage[T1]{fontenc}
\usepackage{ae,aecompl}

\usepackage{graphicx}	
\usepackage{amsmath}	
\usepackage{amssymb}	
\usepackage{bm}        

\newcommand{\beqn}{\begin{equation}}
\newcommand{\eeqn}{\end{equation}}

\title[Thermal irradiation induced wind]{Thermal irradiation induced wind outflow in a geometrically thin accretion disk: A hydrodynamic study}

\author[Kumar and Mukhopadhyay]{Nagendra Kumar$^{}$\thanks{nagendra.bhu@gmail.com}
  and Banibrata Mukhopadhyay$^{}$\thanks{bm@iisc.ac.in}
  \\
 Department of Physics, Indian Institute of Science, Bangalore 560012, India}

\begin{document}

\date{}
\pagerange{\pageref{firstpage}--\pageref{lastpage}} 

\maketitle

\label{firstpage}

  \begin{abstract}
    Many astrophysical sources, e.g., cataclysmic variables, X-ray binaries, active galactic nuclei, 
exhibit a wind outflow, when they reveal a multicolor blackbody spectrum, hence harboring 
a geometrically thin Keplerian accretion disk. Unlike an advective disk, in the thin disk, 
the physical environment, like, emission line, external heating, is expected to play a key role to drive the wind outflow.
We show the wind outflow in a thin disk
attributing a disk irradiation effect, probably from the inner to outer disks. 
We solve the set of steady, axisymmetric disk model equations in cylindrical coordinates 
along the vertical direction for a given launching radius $(r)$ from the midplane, introducing 
irradiation as a parameter.  
We obtain an acceleration solution, for a finite irradiation in the presence of a fixed but tiny initial
vertical velocity (hence thin disk properties practically do not alter) at the midplane, upto 
a maximum height  ($z^{max}$). 
We find that wind outflow mainly occurs from the outer region of the disk 
and its density decreases with increasing launching radius, and for a 
given  launching radius with increasing ejection height. Wind power decreases with increasing ejection height. 
For $z^{max} < 2r$, wind outflow is ejected tangentially 
(or parallel to the disk midplane) in all directions with the fluid speed same as the azimuthal speed.
This confirms mainly, for low mass X-ray binaries, (a) wind outflow should be 
preferentially observed in high-inclination sources, 
(b) the expectation of red and blue shifted absorption lines.

\end{abstract}  


\section{Introduction}
Jets and outflows are ubiquitous in astrophysics. Astrophysical jets are generally exhibited in the low-hard (LH) state
of an accreting system, particularly around black holes. However, matter is also evident to be emanating with speed 
much lower than that of a jet with much less collimation compared to jets, from an accreting system, called wind outflow. 
This outflow is sometimes exhibited from the 
high-soft (HS) state of low mass X-ray binaries (LMXBs) with speed 0.001--0.04$c$, where $c$ is the 
speed of light %
\cite[e.g.,][]{Remillard-McClintock2006, Done-etal2007, Yuan-Narayan2014, Trigo-Boirin2016}. 
Although the exact origin of jet is still under dispute, there are many theories
and models explaining successfully important features of jets and underlying accretion processes. 
As jets are mostly 
seen in the LH state of an accretion flow, they seem to be producing when the accretion flow
deviates from its Keplerian disk structure, in the presence of advection of matter in the 
geometrically thick flow. 
The underlying physics associated with advection along with positive Bernoulli's number and
magnetic fields, often tied up with underlying general relativistic effects, are argued to be 
the basic building block of unbounded matter and jet.
On the other hand, wind outflows are to be originated from the Keplerian disk which is
geometrically thin without advection. Also necessarily following Kepler's law to exhibit
soft photons, there is no chance to have even moderate magnetic fields in the underlying
accretion flow. Hence, the question arises, how matter emanates from such a colder disk?
With this question in mind, we study the wind outflow in a  Keplerian accretion disk \citep[][]{Shakura-Sunyaev1973} 
attributing a source of external heating or irradiation.

Wind outflows are observed in many astrophysical systems, e.g., 
protoplanetary discs, cataclysmic variables (CVs), X-ray binaries (XRBs)
ultra-luminous X-ray sources (ULXs), and active galactic nuclei (AGNs) \cite[][]{Knigge-etal1995, Alexander-etal2006,  Miller-etal2006, King-etal2013, Tombesi-etal2015, Pinto-etal2016}.  
In LMXBs, wind is usually %
inferred from the presence of blueshifted
absorption lines of ions %
in the high resolution X-ray spectra, primarily observed with {\it Chandra, Suzaku} and {\it XMM-Newton}. %
Mainly, Fe {\sc xxv}, Fe {\sc xxvi} ions are detected
\cite[][]{Lee-etal2002, Neilsen-2013, Trigo-Boirin2016}. 
In some sources, jet and wind are also observed simultaneously  \cite[][]{Romanova-etal2009, Tombesi-etal2014, Homan-etal2016}. Winds actually show more diversity and
variability. \cite{Miller-etal2015} reported a doublet absorption line profile of Fe {\sc xxvi} in GRO J1655-40. %
\cite{trigo-etal2014} found that the winds are not present consistently in the HS state,
i.e., for some times it disappears \cite[see also][]{Gatuzz-etal2019}.
In addition, winds are more likely to be present in high-inclination LMXB sources, though in a few
low-inclination sources it is also observed \cite[][]{Ponti-etal2012, Degenaar-etal2016, Trigo-Boirin2016}.

Apart from a jet outflowing model, many authors investigated also the inflow-outflow solutions for advective typed accretion disk
with the motivation that an outflow is possible when the cooling factor $f$, which is the fraction of heat contained with respect to viscous heating, tends to unity \cite[e.g.,][]{Narayan-Yi1995, Yuan-Narayan2014}.
In their approach, they simplify 
the set of governing equations in such a way that it becomes a set of ordinary differential equations (ODEs), either by
%
assuming a self-similar approach or by parameterization or other ways \cite[][]{Misra-Taam2001, Ghosh-Mukhopadhyay2009,  Bhattacharya-etal2010, Jiao-Wu2011, Kumar-Gu2018, Mondal-Mukhopadhyay2018}. 
However, in a geometrically thin disk, %
one has to inspect the physical environment around the disk,
e.g., a radiation force (mediated primarily by spectral lines) term has been added in the governing equations by previous authors \cite[e.g.,][]{Pereyra-etal1997, Proga-etal1998} for a wind outflow in CVs (or underlying thin disk). This line driven wind is unlikely for LMXBs due to the presence of highly ionized gas by X-ray irradiation \cite[][]{Proga-Kallman2002, Trigo-Boirin2016}.

In LMXBs, the wind outflow from a thin disk can be driven via thermal, radiative or
magnetic accelerations.
In thermally driven wind outflow, when the disk temperature (due to the irradiation) rises
enough that the corresponding thermal velocity exceeds the escape velocity, then
the wind outflow will be arisen at the midplane of the disk and the corresponding radius terms as
 Compton radius $R_{IC}$. Thermal-wind outflow can also be started from a small radius, like 0.1 $R_{IC}$, at
 some height, where thermal velocity is comparable to the Keplerian velocity (\citealt{Begelman-etal1983, Woods-etal1996}; see also, \citealt{Done-etal2018}).
However, \cite{Miller-etal2006} showed that
observed wind outflows in GRO J1655-40 cannot be thermally driven due to a dense
outflow close to the black hole \cite[see also,][]{Reynolds2012, Neilsen-2013}.
Recently by considering a frequency
dependent attenuation of irradiated spectral energy distribution (SED), \cite{Higginbottom-etal2018}  \cite[see also,][]{Dyda-etal2017}
showed that the thermal wind can be a viable mechanisms for wind in LMXBs \cite[see also,][]{Gatuzz-etal2019} with $R_{IC} \sim 10^6 R_g$.
On the other hand, in a Compton thick wind or radiation driven wind, pre-existed wind property
(like, ionization parameter) gets changed due to X-ray irradiation
(from the inner disk) via multiple Compton scattering, and results in a
strongly blueshifted iron absorption line \cite[][and references therein]{Tatum-etal2012}. However, \cite{Reynolds2012} argued that Compton thick wind
is not a viable mechanism for sub-Eddington black hole XRBs and AGNs.
Finally, the magneto-centrifugal driven wind outflows are widely studied in literature \cite[e.g.,][]{Miller-etal2006, Reynolds2012, Yuan-etal2015, Chakravorty-etal2016}.   
Apart from the physical/theoretical consistency, the merit/demerit of above described 
model, in principle, will be decided based on observations, especially by, wind density, absorption lines profile, preferential occurrence in high-inclination
accretion disks.

In this work, we study a thermally driven wind, mainly from the outer region
of the thin accretion disk by advocating an external heating or an irradiation
 effect. 
This irradiation is possible from the inner disk to the outer region. As a result, the disk flow deviates from a pure hydrostatic equilibrium, but with a very small extent. 
We also consider a finite initial vertical speed (which is very small compared to the sound speed of the medium) to start with, which practically does not
alter Keplerian velocity profile. We obtain a wind solutions in outer region of the disk ($\gtrsim 800 R_g$, here $R_g =\frac{GM}{c^2}$ is a gravitational radius, $G$ is the gravitational constant, and $M$ is the mass of the compact object) and constrain the model free parameters from the observationally inferred wind characteristics, like, wind speed, wind density.
In the next section, we discuss the model and  the solution procedure. In section \S 3, we describe the general properties of  the model results. Finally in section \S 4 we present the wind solutions and comparison with the observationally derived wind parameters, followed by conclusions in section \S 5.   


\section{Model}
To explore the origin of wind outflow in details, we consider a 2.5-dimensional accretion disk 
formalism in cylindrical coordinates
($r,\phi,z$). 
We assume a steady
\big($\frac{\partial}{\partial t}$ $\equiv$ 0\big) and axisymmetric 
\big($\frac{\partial}{\partial \phi}$ $\equiv$ 0\big) flow. 
The equation of continuity (\ref{eq:conti}), the momentum balance equations (\ref{eq:radmom}-\ref{eq:varmom}) and the
energy equation (\ref{eq:energ}) are %
given as follows \cite[e.g.,][]{Bisnovatyi-Lovelace2001, Mondal-Mukhopadhyay2018}:

\beqn\label{eq:conti}
\frac{1}{r}\frac{\partial (r\rho v_r)}{\partial r} + \frac{\partial (\rho v_z)}{\partial z} = 0,
\eeqn
\beqn \label{eq:radmom}
v_r \frac{\partial v_r}{\partial r} + v_z \frac{\partial v_r}{\partial z} - \frac{\lambda^2}{r^3} + \frac{1}{\rho}\frac{\partial p}{\partial r} + F_r  =\frac{1}{\rho}\frac{\partial W_{rz}}{\partial z},
\eeqn
\beqn \label{eq:azimom}
v_r \frac{\partial \lambda }{\partial r} + v_z \frac{\partial \lambda }{\partial z} = \frac{r}{\rho}\left[\frac{1}{r^2}\frac{\partial (r^2 W_{\phi r})}{\partial r} + \frac{\partial W_{\phi z} }{\partial z}\right],
\eeqn
\beqn \label{eq:varmom}
v_r \frac{\partial v_z }{\partial r} + v_z \frac{\partial v_z}{\partial z} + \frac{1}{\rho} \frac{\partial p}{\partial z} + F_z = \frac{1}{r\rho}\frac{\partial rW_{zr}}{\partial r},
\eeqn
\beqn \label{eq:energ}
\frac{v_r}{\Gamma_3 - 1} \left[\frac{\partial p}{\partial r}- \Gamma_1\frac{p}{\rho}\frac{\partial \rho}{\partial r}\right] + \frac{v_z}{\Gamma_3 -1}\left[\frac{\partial p}{\partial z} - \Gamma_1 \frac{p}{\rho} \frac{\partial \rho}{\partial z}\right] = 0. 
\eeqn
Here, the flow variables are radial velocity $v_r$, specific angular
momentum $\lambda$ (=$r v_\phi$, where $v_\phi$ is the azimuthal velocity), vertical velocity $v_z$,  mass density $\rho$, fluid
pressure $p$. $F_r$ and $F_z$ are magnitudes of the radial and vertical
components of Newtonian gravitational force by the compact object
respectively. $\Gamma_1$ and $\Gamma_3$ are adiabatic exponents. We consider a gas pressure dominated regime such that $p \gg p_{rad}$, where $p_{rad}$ is the radiation pressure. The equation of state is $p$ = $k\rho T/\mu m_p$, where $k$ is the Boltzmann constant, $m_p$ is the mass of proton, $\mu$ is the mean molecular weight, $T$ is the temperature.
The sound speed of the medium is $c_s\sim \sqrt{p/\rho}$. W$_{ij}$ is the viscous shearing stress, where first subscript identifies the direction of the stress, and the second represents the outward normal to the surface on which it acts. In $\alpha$-prescriptions, proposed by \citet[][]{Shakura-Sunyaev1973}, the tangential shear stress W$_{\phi r}$ is expressed as $W_{\phi r}$ $\left(=\eta r \frac{\partial \Omega}{\partial r}\right)$  = $\alpha p$, 
where $\eta =\alpha c_s h \ \rho$ is the dynamical viscosity, $\alpha$ is the Shakura-Sunyaev viscosity parameter \cite[][]{Shakura-Sunyaev1973},
$\Omega$ is the Keplerian angular velocity, $h$ is the scale height of the Keplerian disk at radius $r$. The another tangential shear stress $W_{\phi z}$ (=$\eta r\frac{\partial \Omega}{\partial z}$) can be approximated in terms of $W_{\phi r}$, given by $ W_{\phi z} \approx \frac{z}{r} W_{r\phi}$, based on 
$\frac{W_{\phi z}}{W_{\phi r}}$ $\approx$ $\frac{\partial \Omega}{\partial z}\left / \frac{\partial \Omega}{\partial r} \right.$  $\approx$ $\frac{z}{r}$ \cite[see also,][]{Ghosh-Mukhopadhyay2009}. %
Other shearing stress can be generated in $r-$ and $z-$ directions, 
by varying $v_r$ and $v_z$ respectively. 
Since, $v_z$, $v_r$ $\ll$ $v_\phi$, these shear stresses are negligible compare to the $W_{\phi r}$. In the first approximation, we assume that $W_{rz}$ = $W_{zr}$ $\approx$ 0, or  
\beqn \label{eq:rzstress}
\frac{\partial v_r}{\partial z} + \frac{\partial v_z}{\partial r} = 0
\eeqn

In the right hand side of the energy equation (\ref{eq:energ}), we assume, like the Keplerian disk, that the rate of heat generation per unit volume  $q^+$ by viscous heating is immediately radiated out, i.e., $q^+ = q^-$, here $q^-$ is a rate of radiated energy density. The optical depth $\tau$ %
within the disk scale height is very greater than unity, i.e., $\tau \gg 1$, and disk cools vertically by blackbody radiation within the scale height.
The viscous heating rate per unit volume due to tangential shearing stresses is $q_{\phi r}^+$ = $W_{\phi r} r \frac{\partial \Omega}{\partial r}$ and $q_{\phi z}^+$ = $  W_{\phi r} z \frac{\partial \Omega}{\partial z}$. Clearly $q_{\phi r}^+$ $\gg$ $q_{\phi z}^+$ or $q^+$  $\approx q^+_{\phi r}$ for $z/r \ll$ 1, it will hold always within the disk scale height. 
Hence, like the Keplerian disk, the viscous heat dominantly generates at  midplane of the disk and disk immediately cools locally by black body emission.
Another cooling process, like bremsstrahlung cooling
 is negligible above the scale height. The bremsstrahlung cooling rate is proportional to the square of the number density (of ion/electron) and above the disk scale height the density falls rapidly \cite[][]{Rajesh-Mukhopadhyay2010}.  
Next, we assume that the fluids are slightly deviated from the vertical hydrostatic equilibrium, which is expressed as    
\beqn \label{eq:hydro}
\frac{1}{\rho}\frac{\partial p}{\partial z}= -(1-x)F_z
\eeqn
Here, $x$ ($\ll$ 1) is a number, and for $x$ = 0 the medium is purely in vertical
hydrostatic equilibrium. To interpret the physical meaning of above equation (\ref{eq:hydro}), we reexpress it at a given height $z$ as,   
$x = 1 + \left.\left(\frac{1}{\rho F_z}\frac{\Delta p}{\Delta z}\right)\right|_{z}$, here $\Delta p$ = [$p(z+\Delta z) - p(z)$] and $\Delta z$ is a small increment at height $z$. For a given pressure profile in the vertical direction, 
by varying $\Delta z$ the quantity $x$ will not change, i.e, $p$ vs $z$ profile will be different for different $x$. %
Particularly, for a given height $z$, the pressure will increase with increasing $x$.
This situation can be arisen by external heating, i.e., the raised in temperature due to external heating leads to an enhancement in pressure.
Below, we show that the internal energy of fluid increases with $x$ (see $c_s/c$ curve of Figure 2; also \S3.3), and  we have estimated the enhancement in rate of internal energy per unit volume by equation (\ref{eq:int_enhance}). Thus by
an introducing $(1 - x)$ factor in hydrostatic equilibrium equation we properly account for the external heating effect. We do not introduce an extra heating in the energy equation (\ref{eq:energ}). However, \cite{Begelman-etal1983} accounted for the external heating effect in energy equation.
The plausible source of external heating in the disk is a compact central X-ray source around inner region of the disk. Since a thin accretion disk has a concave shape, this will permit the irradiation of outer region of the disk by inner region. 
In principle, the irradiation by the inner disk can introduce a radiation pressure $p_{rad}^{irr}$, which would be appeared in the radial and momentum balance equations \cite[e.g.,][]{Proga-Kallman2002, Dannen-etal2020}. 
However, we find that at outer region, $p_{rad}^{irr}$ is very small in comparison with the gas pressure (see equation (\ref{eq:irr_en})). 
Hence, in some sense, $x$ is an index for the outer disk irradiation. In this calculation, for simplicity, we assume that the disk irradiation starts from the midplane of the disk, and throughout the disk height (at a given radius) $x$ remains constant.     

Combining all the above equations (\ref{eq:conti})-(\ref{eq:hydro}) we obtain
\begin{eqnarray}
\label{eq:master}
\nonumber
&&\frac{\partial v_z}{\partial z}\left[\frac{v_z^2-v_r^2}{v_r}\frac{(-\alpha r)\Gamma_1c_s^2}{v_r^2-\Gamma_1c_s^2}\right]=\frac{3 W_{r\phi}}{\rho}+ \alpha z\frac{1}{\rho} \frac{\partial p}{\partial z}\\ \nonumber &+& \alpha r \left[v_z\frac{xF_z}{v_r} +f^r_{bal} -\frac{v_r^2}{r}- 
\frac{1}{\rho} \frac{\partial p}{\partial z}\frac{v_zv_r}{\Gamma_1 c_s^2} \right]\frac{\Gamma_1 c_s^2}{ v_r^2-\Gamma_1 c_s^2 }\\ 
&-&v_r \frac{\partial \lambda}{\partial r}-v_z \frac{\partial \lambda}{\partial z},   
\end{eqnarray}

\noindent where $f^r_{bal}$ = $-\frac{\lambda^2}{r^3}+F_r$.
Above, $\frac{\partial v_z}{\partial z} $ is expressed in terms of $\frac{1}{\rho}\frac{\partial \rho}{\partial r} $, $\frac{\partial \lambda}{\partial r} $, and $\frac{\partial \lambda}{\partial z} $; all these quantities have to
be computed in advance.
First, we compute the specific angular momentum as a function of height at
a given radius.
  Since $F_r$ decreases with height, and in the present case, the pressure increases with $x$ for a given height, it is possible that after some height, the radial gradient of pressure can be comparable to $F_r$ for an appropriate $x$.
With this, we  take an account for the radial component of pressure gradient for supporting the rotations other than the gravity, and it expresses as follows

 %
\beqn \label{eq:lamb}
\frac{\lambda^2}{r^3} = F_r + \frac{1}{\rho} \frac{\partial p}{\partial r},
\eeqn
assuming $v_r\frac{\partial v_r}{\partial r} + v_z\frac{\partial v_r}{\partial z}$ $\approx$ 0 (see equation \ref{eq:radmom}), and we evaluate the derivatives
of $\lambda(z)$ \big(i.e.,$\frac{\partial \lambda}{\partial r}, \frac{\partial \lambda}{\partial z}$\big)  by neglecting the higher order derivatives.
Here we like to mention that in above 
equation (\ref{eq:lamb}), if we consider $\frac{1}{\rho} \frac{\partial p}{\partial r}$ the term associated with $(1\pm y)\frac{1}{\rho} \frac{\partial p}{\partial r}$ with $y < 10^{-5}$, then we still attain an acceleration solution. 
Finally, 
  we assume that the term $\frac{1}{\rho}\frac{\partial \rho}{\partial r}$ does not vary with height, i.e.,$\frac{1}{\rho}\frac{\partial \rho}{\partial r} (r,z) = \frac{1}{\rho}\frac{\partial \rho}{\partial r}(r) $. %

\subsection{Solution procedure} \label{sub:sol}

We aim at studying outflow for a given launching radius.
We solve the governing equations along the $z$-axis, compute the flow variables and their derivatives as functions of height. 
At a height $z$, the fluid moves with speed  $\sqrt{v_r^2 + v_\phi^2 + v_z^2}$ 
dominated by a circular path (see Figure \ref{fig:wind_geo} for a detailed geometry).
We consider a finite but tiny initial vertical speed $v_z$ at the launching radius on the midplane, 
whose magnitude is very less than the sound speed of the medium ($v_z$ $\ll c_s$). 
However, we parameterize the magnitude of the initial vertical speed in terms of the radial velocity (as, $v_r$ $\ll c_s$ also), which is given as 
\beqn \label{eq:vz}
v_z = f_v |v_r|,
\eeqn
here, $f_v$ is a number. For this choice of  $v_z$, %
we find that 
$v_z\frac{\partial v_r}{\partial z}$, $v_r\frac{\partial v_r}{\partial r}$ $\ll$ $\frac{1}{\rho}\frac{\partial p}{\partial r}$ (while, $\frac{1}{\rho}\frac{\partial p}{\partial r}$  $\ll$ $F_r$ already);
and $v_z\frac{\partial v_z}{\partial z}$, $v_r\frac{\partial v_z}{\partial r}$ $\ll$ $\frac{1}{\rho}\frac{\partial p}{\partial z}$ near the midplane. Thus the governing equations (\ref{eq:conti})-(\ref{eq:energ}) of the disk  become equivalent to the Keplerian disk, at least near to the midplane, assuring observed HS spectral state. We use this as an initial condition for solving the equations.
Therefore, we take the respective Keplerian values of flow variables,  
$v_r, \lambda$ and $c_s$ at the launching radius $r$ on midplane, 
e.g., $v_r(r,z=0)=v_r(r) $ according to \citet[][]{Shakura-Sunyaev1973}, and so on. The initial values of these variables would be a function of $\dot{M}$, $M_c$ and $\alpha$, here  $\dot{M}$ is the mass accretion rate, $M_c$ is mass of the compact object. In short, we begin to solve the governing equations for wind outflow
from the midplane of the disk. However, in a similar exploration, \cite{Woods-etal1996} assumed the base of the wind is above the disk midplane.   


The main focus here is to explore the wind outflow as a consequence of 
an external heating, mainly by the  inner disk irradiation.
We essentially initialize the flow variables with the solution set prescribed for the outer-region solutions of the Keplerian disks \cite[][]{Shakura-Sunyaev1973}, that is, the opacity $\kappa$ comes mainly from the free free absorption $\sigma_{ff}$  
which is the Rosseland mean opacity. 
The minimum radius for outer region of the Keplerian disk $r_{bc}$ is given by $r_{bc}$ 
$\gtrsim$ 2.5 $\times 10^7 \dot{m}_{16}^{2/3} M_{co}^{1/3} \left(1-\frac{\lambda_{in}}{\lambda}\right)^{8/3}$ cm \cite[e.g.,][]{Shakura-Sunyaev1973, Novikov-Thorne1973, Frank-etal2002}, where $\dot{M}_{16} =\frac{\dot{M}}{10^{16} g/s}$, and $M_{co} = \frac{M_c}{M_{\odot}}$ with $M_{\odot}$ the  solar mass. The minimum radius $r_{bc}$ is $\approx$ 150, 750 $R_g$ for $\dot{M}_{16}$ = 10, 100 respectively for $M_{co}$ = 10.

We solve simultaneously $\frac{\partial v_z}{\partial z}$, $\frac{\partial c_s}{\partial z}$ and $\frac{\partial v_r}{\partial z}$ treating them as partial  differentials.
  That is, the solution technique implicitly carries the information of $r$-derivative of the 
flow variables 
(e.g., $\frac{\partial p}{\partial r}$) as functions of height. 
We check the consistency of results obtained at a fixed $r$ based on the proposed numerical
analysis. We take two adjacent grid points in the $r$-direction (like, $r-\Delta r$  and $r+\Delta r$ with $\frac{\Delta r}{r} \ll 1$) as the launching radii, along with $r$, and compare results. We find that although computations are
carried out for a fixed radial coordinate, effectively the solutions capture
the variation of variables in the radial directions while propagating in the
vertical direction. Hence, the results are consistent within the approximations, see the appendix. The present solutions give a complete approximate pictures of flow variations in the
$z$-direction.
We  adopt 
the convention that the radially inflow velocity $v_r$ is negative, and 
vertical outflow velocity $v_z$ is positive. 
In this sign convention, to  ensure the angular momentum conservation, 
prescribed by \citet{Bisnovatyi-Lovelace2001}, we take a negative $\alpha$. 
We  illustrate a few points below to understand the solutions.

\indent  \textbf{(a) Critical point of  $\frac{\partial v_z}{\partial z}$:}
The equation (\ref{eq:master}) has a singular point at a height $z$ where  $v_z(z) \ = v_r(z)$.
To have a smooth velocity field at that $z$, the RHS  of equation (\ref{eq:master}) must be zero, which is written as  
%
\begin{align}
\frac{3 W_{r\phi}}{\rho}+ \alpha z\frac{1}{\rho} \frac{\partial p}{\partial z}&-v_r \frac{\partial \lambda}{\partial r}-v_z \frac{\partial \lambda}{\partial z} = -\alpha r \left[v_z\frac{xF_z}{v_r} +f^r_{bal} \right. \nonumber \\
& \qquad \left.   -\frac{v_r^2}{r}-  \frac{1}{\rho} \frac{\partial p}{\partial z}\frac{v_zv_r}{\Gamma_1 c_s^2} \right]\frac{\Gamma_1 c_s^2}{ v_r^2-\Gamma_1 c_s^2 }
\end{align}
\noindent For $v_r$ $\ll$ $\Gamma_1 c_s^2$ the above condition is 
always satisfied due to equation (\ref{eq:azimom}). Hence, $\frac{\partial v_z}{\partial z} $ exists at that height, where $v_z(z) \ = v_r(z)$ and $v_r(z)^2 \ll \Gamma_1 c_s^2$. 

\indent \textbf{(b) Sign flip of $\frac{\partial p}{\partial r}$:} In the Keplerian disk, $\frac{1}{\rho}\frac{\partial p}{\partial r}$ is negative, acting in radially outward direction, and $|\frac{1}{\rho}\frac{\partial p}{\partial r}|$ $\ll$ $F_r$ ($\equiv$ $\frac{\lambda^2}{r^3})$.
The quantity $\frac{\partial p}{\partial r}$ flips  sign at around 0.92 $h$, if one computes $\frac{1}{\rho}\frac{\partial p}{\partial r}$ as a function of height,  considering a constant $\lambda$ over the height, using the relation 
$\frac{1}{\rho}\frac{\partial p}{\partial r}$  $= -F_r +\left.\frac{\lambda^2}{r^3}\right|_{z=0} + \left.\frac{1}{\rho}\frac{\partial p}{\partial r}\right|_{z=0}$.
We compute 
$\frac{1}{\rho}\frac{\partial p}{\partial r}$ in the vertical direction 
for two values of $f_v$ = 0.1, 1.02 for $x$ =0.
We find, the sign flip occurs at around $0.89h$ and $0.85h$ for $f_v$ = 0.1 and 1.02 respectively, 
which is consistent with the result of Keplerian disk. 
For any $x$, by using equation (\ref{eq:azimom}), the condition for the sign flip of $\frac{\partial p}{\partial r}$ $\left(\text{or}\ \frac{\partial p}{\partial r} = 0\right)$ at height $z$ = $z_f$ is 
 \beqn \label{eq:sign}
3\alpha c_s^2  = - \left(\alpha z_f \frac{1}{\rho} \frac{\partial p}{\partial z}-v_r \frac{\partial \lambda}{\partial r}-v_z \frac{\partial \lambda}{\partial z}\right). 
 \eeqn

\noindent \textbf{(c) Acceleration and deacceleration in the vertical direction: 
 }
In equation (\ref{eq:master}), the coefficient of $\frac{\partial v_z}{\partial z}$ can be positive or negative depending on the relative magnitude of $v_z$ and $v_r$, 
near to the midplane, where $v_r^2 < \Gamma_1 c_s^2$. 
The dominated  RHS terms 
are $\frac{3W_{r\phi}}{\rho}$, $\left(\alpha z \frac{1}{\rho} \frac{\partial p}{\partial z}-v_r \frac{\partial \lambda}{\partial r}-v_z \frac{\partial \lambda}{\partial z}\right)$ and $-\alpha r \frac{1}{\rho} \frac{\partial p}{\partial r} \frac{\Gamma_1 c_s^2}{ v_r^2-\Gamma_1 c_s^2 }$, 
in which we find numerically that the first term is negative, second term is positive and last term
can be either positive ($\frac{\partial p}{\partial r} < 0$) or negative ($\frac{\partial p}{\partial r} > 0$), also the second term increases with height. 

To understand the acceleration/deacceleration  behavior of $v_z$ for a
given $x$, we consider a case where $v_r > v_z$ throughout the disk height, i.e., the coefficient of $\frac{\partial v_z}{\partial z}$ is negative.
As mentioned, the radial pressure gradient flips the sign from negative to positive 
above the height $z_f$. Within the height $z_f$, the third term 
mentioned above is positive
and the sum of second and third terms is less than the first term, hence
we have an accelerating solution.
Above the height $z_f$, third term becomes negative, and due to first law of
thermodynamics 
(or equation \ref{eq:energ}), first term will decrease, as  $v_r$ or $v_z$  increases with height. 
Eventually, at some large height, $v_z$ or $v_r$ becomes comparable to the sound speed. 

As mentioned earlier, at a given $z$ the pressure increases with increasing $x$, so also the first term. 
For sufficiently  large $x$, above $z_f$, the first term gets blown up (instead of decreasing) and again satisfies the equation (\ref{eq:sign}), which makes $\frac{\partial p}{\partial r}$ to flip the sign from positive to negative, as a result third term becomes positive.
Above this height, the deacceleration of $v_z$ or $v_r$  starts and finally $v_z$ or $v_r$ gets decreased to zero.
Thus for a given $f_v$, we have an acceleration solution for a range of $x$ ($x^{min}$ to $x^{max}$). 
e.g., for $M_{c}$ = 10$M_\odot$, $r$ = 300 $R_g$, we find that the acceleration solution exists for 
 0 $< x < 4.7\times 10^{-8}$ at $f_v \sim$ 1. 
Here, we like to stress that $f_v$ =1 is associated with a critical point. 
Above description is valid for $f_v < 1$ as well as $f_v>1$, as indeed 
we notice that for $f_v > 1$, $v_r$ becomes larger than $v_z$ above the
mid-plane, where still $v_r^2 \ll \Gamma_1 c_s^2$ (e.g., see the upper left panel of  Figure \ref{fig:x_fin}). 

\indent  \textbf{(d) Solution behavior at height where $v_r^2$ $\rightarrow$ $\Gamma_1 c_s^2$:} For $v_r^2$ tends to $\Gamma_1 c_s^2$, the equation (\ref{eq:master}) is reduced to 
 \begin{equation} \label{eq:max}
  \frac{\partial v_z}{\partial z}\left[\frac{v_z^2-v_r^2}{v_r}\right] =  -v_z\frac{xF_z}{v_r} -f^r_{bal}  +\frac{v_r^2}{r} + \frac{1}{\rho} \frac{\partial p}{\partial z}\frac{v_zv_r}{\Gamma_1 c_s^2}.
\end{equation} 
The above equation (\ref{eq:max}) has a singular point for $v_z = v_r$. For a smooth velocity field at singular point, the RHS of equation  (\ref{eq:max}) must be zero, which is written as   
\[ f^r_{bal} \approx -v_z\frac{xF_z}{v_r}  +\frac{v_r^2}{r} + \frac{1}{\rho} \frac{\partial p}{\partial z} \]
 \begin{equation}\label{eq:cond1}
      \text{or} \ \ \frac{1}{\rho} \frac{\partial p}{\partial r} + \frac{1}{\rho} \frac{\partial p}{\partial z} \approx -\frac{v_r^2}{r} \quad  \text{or} \quad  \left|\frac{1}{\rho} \frac{\partial p}{\partial r}\right| \approx \left|\frac{1}{\rho} \frac{\partial p}{\partial z}\right| = |F_z|. 
\end{equation} 
Here, $f^r_{bal}$ = $-\frac{1}{\rho} \frac{\partial p}{\partial r}$, $x$ $\ll$ 1, and $\frac{v_r^2}{r}$ $<$ $\frac{1}{\rho} \frac{\partial p}{\partial z} \approx F_z$.
The radial component of pressure gradient is expressed  by using equations (\ref{eq:conti}), (\ref{eq:radmom}) and (\ref{eq:energ}) as
\begin{equation}\label{eq:cond2}
   \frac{1}{\rho} \frac{\partial p}{\partial r}\left(\frac{v_r^2}{\Gamma_1 c_s^2}-1\right)=v_z \frac{\partial v_r}{\partial z}+f^r_{bal}-\frac{v_r^2}{r}-v_r \frac{\partial v_z}{\partial z}-\frac{1}{\rho} \frac{\partial p}{\partial z} \frac{v_r v_z}{\Gamma_1 c_s^2}.
   \end{equation}
\noindent Using equations (\ref{eq:cond1}) and (\ref{eq:cond2}), we find $v_z \frac{\partial v_r}{\partial z} \approx v_r \frac{\partial v_z}{\partial z} $. With
 this result, we obtain the relations $\left|\frac{1}{\rho} \frac{\partial p}{\partial z}\right| \approx \left|v_z \frac{\partial v_z}{\partial z}\right|$ and  $\left|\frac{1}{\rho} \frac{\partial p}{\partial r}\right| \approx \left|v_r \frac{\partial v_r}{\partial r}\right|$ by analyzing equations (\ref{eq:varmom}) and (\ref{eq:radmom}) magnitudewise respectively.
     In summary, at a height where $v_r$ or $v_z$ is comparable to the sound speed, we obtain mainly two results (i)  $\left|\frac{1}{\rho} \frac{\partial p}{\partial z}\right| \approx \left|v_z \frac{\partial v_z}{\partial z}\right|$ and  $\left|\frac{1}{\rho} \frac{\partial p}{\partial r}\right| \approx \left|v_r \frac{\partial v_r}{\partial r}\right|$, (ii) $\frac{1}{\rho} \frac{\partial p}{\partial r} + \frac{1}{\rho} \frac{\partial p}{\partial z} \approx -\frac{v_r^2}{r}$.
%

%

 Here the driver for acceleration is the pressure gradient,
also the pressure is gas dominated. As mentioned earlier, the flow speed 
increases on the expense of the internal energy (or kinetic energy of the molecular motion) following the energy conservation equation (\ref{eq:energ}). For a given external heating in the dynamical time-scale of wind outflow ($t_w$), the internal energy increases by a fixed extent. When
the fluid speed approaches to the sonic speed, its acceleration halts because at this point
its kinetic energy becomes comparable to the internal energy (which is 
reflected by the condition (i)) or fluid reaches to the equipartition of energy states. In other words, above the sonic point there is no 
acceleration or no pressure gradient, and an isobaric regime arises.
This is not the case for 
the sonic point in radial direction, where the acceleration towards compact object is mainly because of the gravity (acting as a driver). Also in the latter case,
  $\left|\frac{1}{\rho} \frac{\partial p}{\partial r} \right| < \left|v_r \frac{\partial v_r}{\partial r}\right|$ at the sonic point \cite[e.g.,][]{Chakrabarti-Titarchuk1995, Narayan-Yi1995, Rajesh-Mukhopadhyay2010, Mondal-Mukhopadhyay2018, Mondal-Mukhopadhyay2020}.
However in the solutions, we can not show directly that the pressure
gradient components become zero 
at or just above the sonic point 
as a consequence of arriving at the isobaric regime, since we
have expressed the vertical pressure gradient in terms of vertical gravitational
force by equation (\ref{eq:hydro}). Moreover, for  
$\frac{1}{\rho} \frac{\partial p}{\partial r}$ $\rightarrow$ 0
or $f^r_{bal}$  $\approx$ 0 (for $F_r$ $\gg$
$F_z$), by using the results (ii), 
we can show that at the singular point the
magnitude of $\frac{1}{\rho} \frac{\partial p}{\partial z}$ decreases sharply to $\frac{v_r^2}{r}$ from $F_z$. 
In short, at a height where $v_r$ approaches to $\Gamma_1 c_s^2$ and comparable to $v_z$, we find an isobaric regime arrived (due to the condition (i)), i.e. no pressure gradient or further no acceleration. We term this height as the maximum possible height for an acceleration and denote by $z^{max}$. Above the height $z^{max}$, there is a no point of interest.
Without loss of generality and results, we perform all acceleration calculations upto the height near to $z^{max}$, to
avoid the numerical uncertainty due to a singular point at $z$ =$z^{max}$.

\section{General Results}
Mainly two parameters, the initial vertical speed (parameterized by $f_v$) and the index of external heating $x$  characterize the acceleration/deacceleration solution of $v_z$. 
In this section, we explore the general behavior of solutions for $x$ and $f_v$, and also the relation between $x$ and $f_v$. 
However, first we intend to compare the vertical structure of the model disk %
with that of the Keplerian disk. For this, we take $x = 0$, since it refers a vertically hydrostatic equilibrium, and also in this limit our governing equations of disk are similar to the Keplerian disk.
Without loss of generality, we explore it 
for fixed  launching radius $r$ = 300$R_g$, accretion rate $\dot{m}$ = $10^{17} g/s$, compact object mass $M_c = 10M_{\odot}$ and coefficient of viscosity $\alpha$ = 0.1. 

\subsection{Vertical disk structure for $x$ = 0}\label{ver:x0}
  
In the Keplerian disk, the central disk  temperature $T_c$ is
computed by assuming that the radiative transfer is a dominant process for energy transport.
Since within the scale height, the optical depth is very large, $\tau \gg$ 1,  the temperature at the disk surface (or at $h$) can be approximated to be $T_c$, i.e., an isothermal disk. 
In the isothermal Keplerian disk at a given radius $r$, the density (or pressure) varies with height as \cite[see,][]{Pringle1981}, $\rho(z,r) /\rho_c(r) = p(z,r)/p_c(r) = \exp \left(\frac{-z^2}{2h^2}\right)$,
here $p_c (r)$, $\rho_c (r)$ are the pressure and density on the midplane respectively. The pressure and density scale height both are same as $h$  
(by definition, here, the scale height is a height at which pressure or density falls by a factor $e^{0.5}$
with respect to its respective  midplane values). 

\begin{figure*}
\centering
\begin{tabular}{lcr}\hspace{-0.9cm}
  \includegraphics[width=0.36\textwidth]{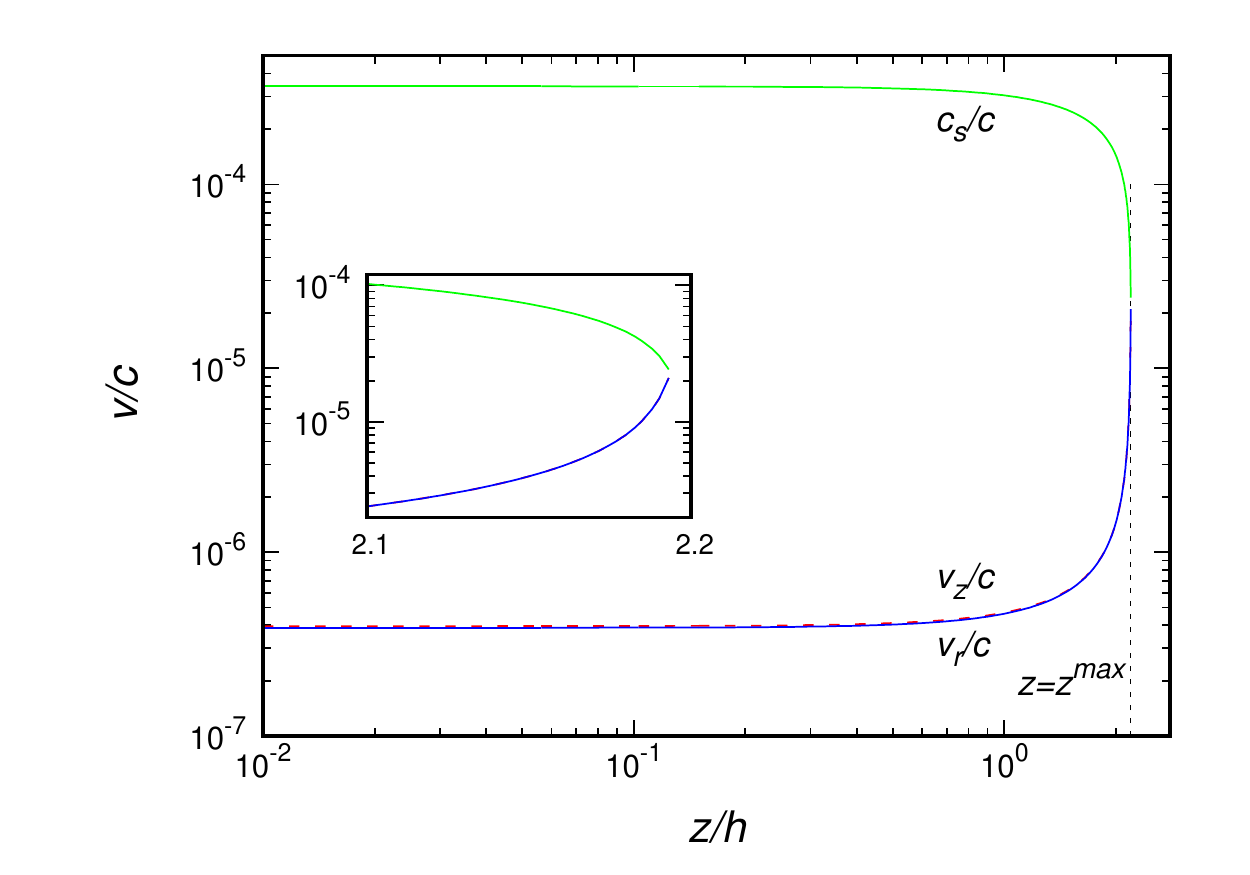}&\hspace{-0.90cm}
  \includegraphics[width=0.36\textwidth]{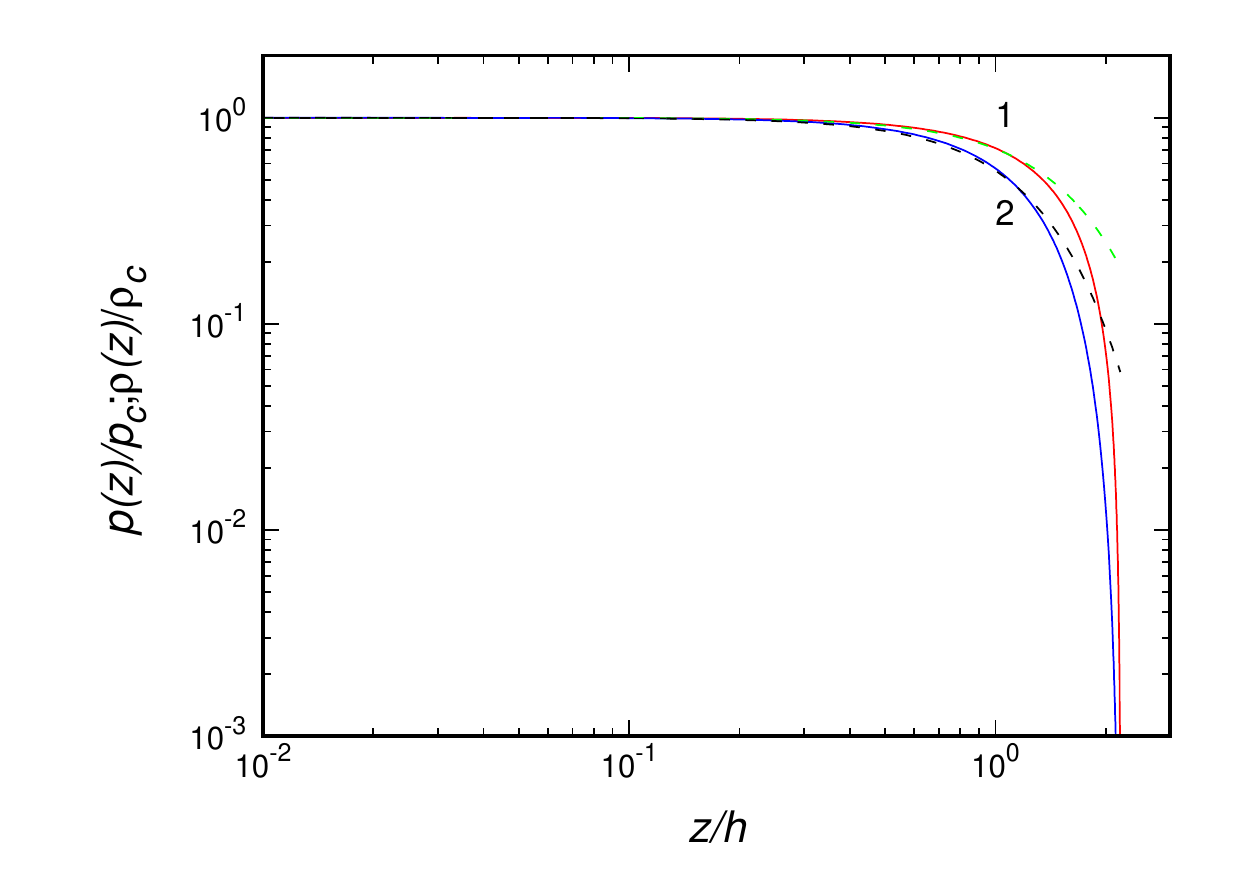}&\hspace{-0.90cm}
   \includegraphics[width=0.36\textwidth]{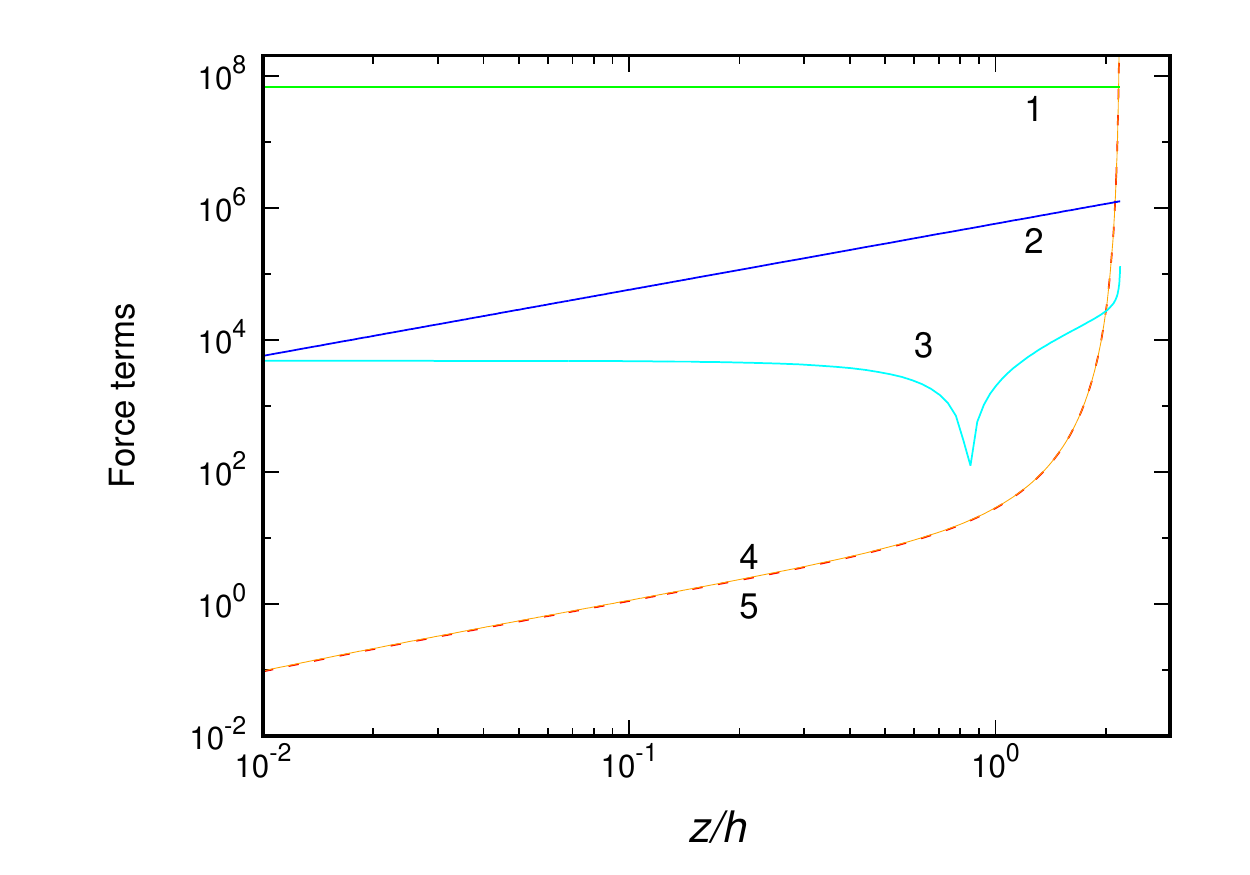}\\ 
\end{tabular}\vspace{-0.3cm}
\caption{The solutions of our model equations for $x$ = 0, $r$ = 300$R_g$.
The left panel is for three different velocities ($v_z$, $|v_r|$, $c_s$) as
 functions of $z$ (measured in units of the Keplerian scale height $h$, here $r/h$ $\sim$118). The middle panel is for pressure $p/p_c$ and density $\rho/ \rho_c$, which  are shown by solid curves 2 and 1 respectively. The dashed curves 2 and 1 are corresponding model curves  $\exp{\left(\frac{-z^2}{2(0.92h)^2}\right)}$ and $\exp{\left(\frac{-z^2}{2(1.2h)^2}\right)}$ respectively.  The right panel shows the comparison between $v_r \frac{\partial v_r}{\partial r}$, $v_z \frac{\partial v_z}{\partial z}$ and force terms  $\frac{1}{\rho} \frac{\partial p}{\partial r}$, $F_z$, and $F_r$, which  are shown by the curves 5, 4, 3, 2 and 1 respectively. In left panel, we have marked the $z^{max}$ (shown by vertical line) where $v_r$ and $v_z$ become comparable to the sound speed. }
\label{fig:x0}
\end{figure*}

We compare the above vertical structure of the Keplerian disk in the present model 
having $f_v$ $\sim$ 1 and $x$ = 0. This choice  ensures that
the considered disk is also an isothermal disk within the disk scale height, like a Keplerian disk. 
However, in reality we expect a small decrement in temperature (or $c_s$ also, which is
shown by $c_s$-curve  in the left panel of Figure \ref{fig:x0}) within the
scale height $h$ due to the energy conservation (as $v_r$ and $v_z$ are
increasing). As a consequence, we find a different isothermal pressure and density  profiles
which are shown in
the middle panel of Figure \ref{fig:x0}. 
The pressure and density  profiles behave as %

\begin{footnotesize}
  \beqn p(r,z) = p_c (r) \exp{\left(\frac{-z^2}{2 (0.92h)^2}\right)};\ \ \  \rho(r,z) = \rho_c (r) \exp{\left(\frac{-z^2}{2 (1.2h)^2}\right)}.\eeqn
\end{footnotesize}  

These model profiles are over plotted on respective
numerical results in Figure \ref{fig:x0}. 
Here, the pressure and density scale  heights of the disk are different and these are  $\sim$0.92$h$ and 1.2$h$ respectively. Above the scale height, both fall rapidly. In the previous section, we have found that the radial pressure gradient flips
  the sign at height $z_f$, and for $f_v \sim$  1, $z_f$
= 0.85$h$.  We have observed that pressure scale height $h_p$ and $z_f$ both are related each other as, $z_f = h_p^2 / h$.  

 Next, we  examine the validation of assumption for initializing the variables  to their respective Keplerian values at the launching radius $r$.  
 In the right panel of  Figure \ref{fig:x0}, we  show the variations of $v_z\frac{\partial v_z}{\partial z}$, $v_r\frac{\partial v_r}{\partial r}$, $\frac{1}{\rho}\frac{\partial p}{\partial r}$, $F_r$ and $F_z$ as functions of height $z$.  We notice $v_r\frac{\partial v_r}{\partial r}$, $v_z\frac{\partial v_r}{\partial z}$ $\ll$ $\frac{1}{\rho}\frac{\partial p}{\partial r}$ $\ll$ $F_r$; also   $v_z\frac{\partial v_r}{\partial z}$ $\ll$ $F_z$. Hence, the Keplerian limits are valid, atleast within the pressure scale height, for $f_v$ $\sim$ 1. 
In general, we find that it is valid even at greater value of $f_v$ $\sim$ 10.

 In the left panel of Figure \ref{fig:x0}, the profiles of velocities $v_r$, $v_z$ and $c_s$ are shown as functions of height $z$. 
We find that $v_z$ and $v_r$ become comparable to the sound speed at height
$z$ = 2.2$h$, hence, the maximum attainable height for acceleration $z^{max}$
is 2.2$h$. At $z^{max}$, $v_z$ and $v_r$ are accelerated to the maximum value,
just about 15 times less than the sound speed at midplane; in another way,
the sound speed is $\sim$ 15 times smaller than own midplane value.
 In addition,  for $z > z_f$, 
 $\frac{1}{\rho}\frac{\partial p}{\partial r}$ is positive, therefore it acts radially inward direction or opposes 
 the rotational effect.
 As discussed in  points (d) of \S \ref{sub:sol},  at $z^{max}$
     the equation (\ref{eq:master}) has a singular point (due to $v_r \sim v_z$ and
     $v_r^2 \approx \Gamma_1c_s^2$) but it has a smooth solution when the conditions  $\left|\frac{1}{\rho}\frac{\partial p}{\partial z}\right| \approx \left|v_z \frac{\partial v_z}{\partial z}\right| $; $\left|\frac{1}{\rho}\frac{\partial p}{\partial r}\right| \approx \left|v_r \frac{\partial v_r}{\partial r}\right| $ are
satisfied. We obtain this condition 
around $z^{max}$ as shown in right panel of Figure \ref{fig:x0}, thus we have a smooth solution, also an isobaric regime above $z^{max}$. In addition, we truncate the calculation just before the $z^{max}$ due to reaching an isobaric regime, also to avoid the numerical uncertainty, 
as mentioned in same subsection. 
Above $z^{max}$, there is no any pressure gradient 
 and only $F_z$ and $F_r$ act 
  on particles. 
  Since at $z^{max}$, $\frac{1}{\rho}\frac{\partial p}{\partial r}$ $\ll$ $F_r$, $F_r$
  is able to balance the rotation just above $z^{max}$ and the disk material is rotationally bound.    
In short, the pressure (or density) scale height will change if one  considers
a small initial vertical motion ($v_z \ll c_s$, and $f_v <$ 10) in the irradiated
Keplerian disk and the disk can maximally extend upto height 2.2$h$ for $f_v \sim$  1 at any radius.
For $x$ = 0, the model disk is consistent with the Keplerian disk, as the pressure and density  profiles follow isothermal profile 
and the sign flip of $\frac{\partial p}{\partial r}$ occurs
around the pressure scale height. 

\subsection{Vertical disk structure for fixed $x$ and $f_v$}\label{ver:xmax} 

\begin{figure*}
\centering
\begin{tabular}{lr}
 \includegraphics[width=0.46\textwidth]{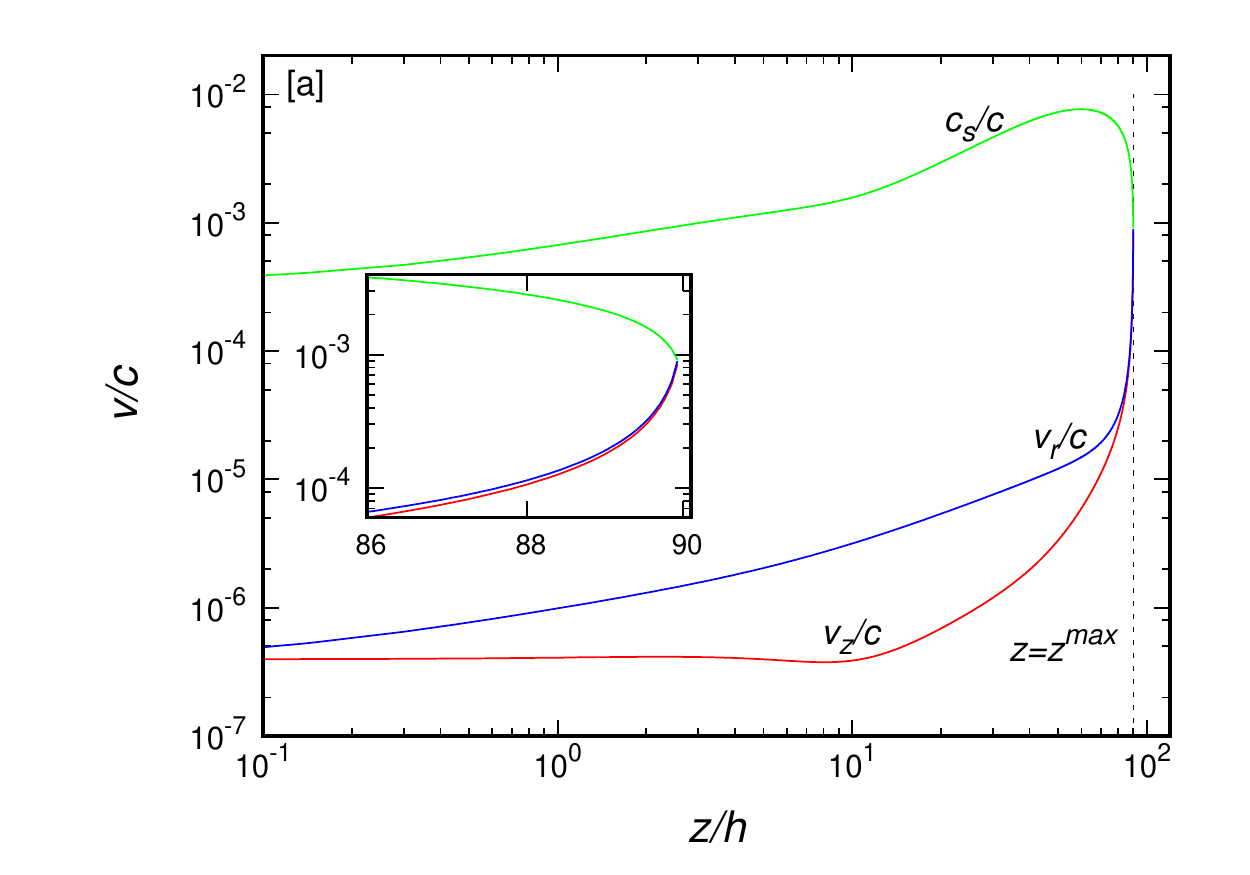}&\hspace{-0.90cm}
  \includegraphics[width=0.46\textwidth]{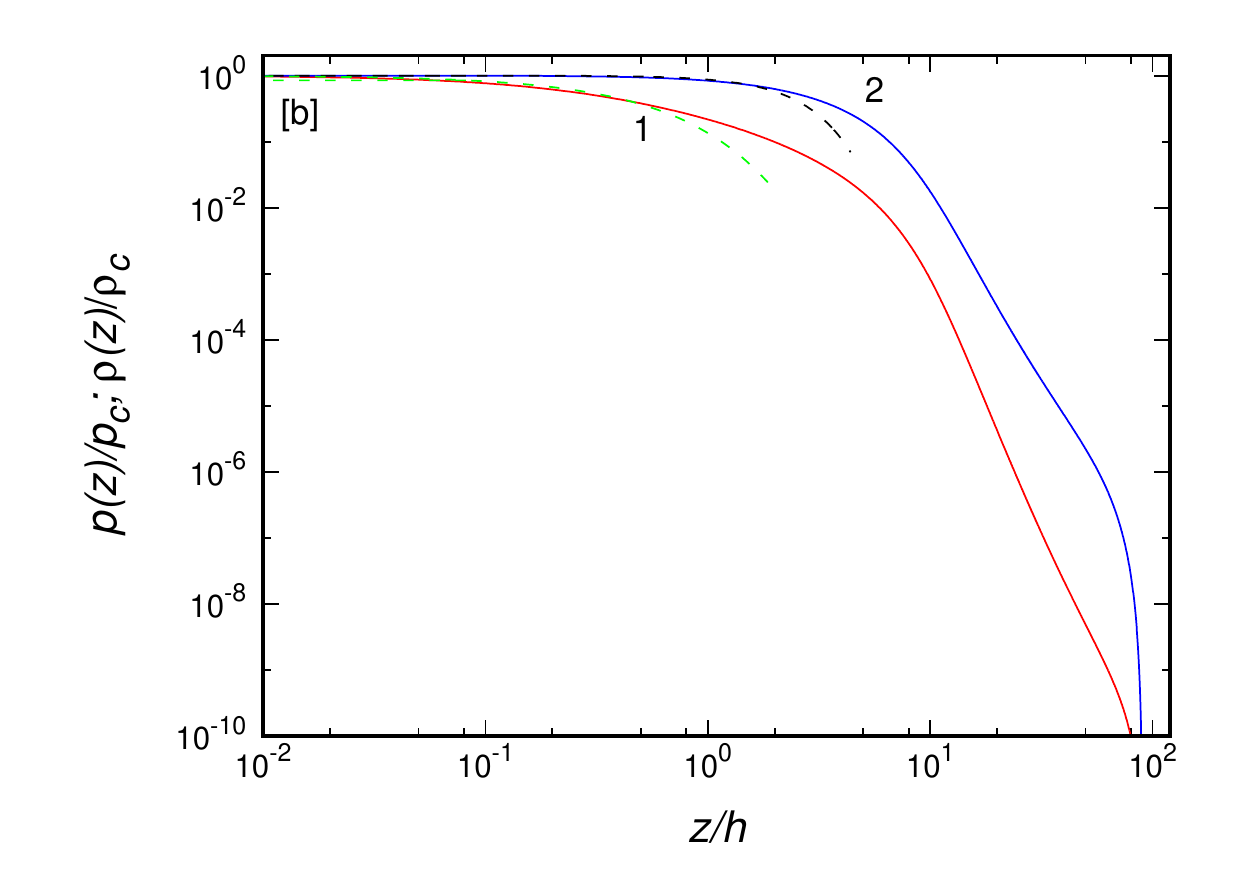}\\
  \includegraphics[width=0.46\textwidth]{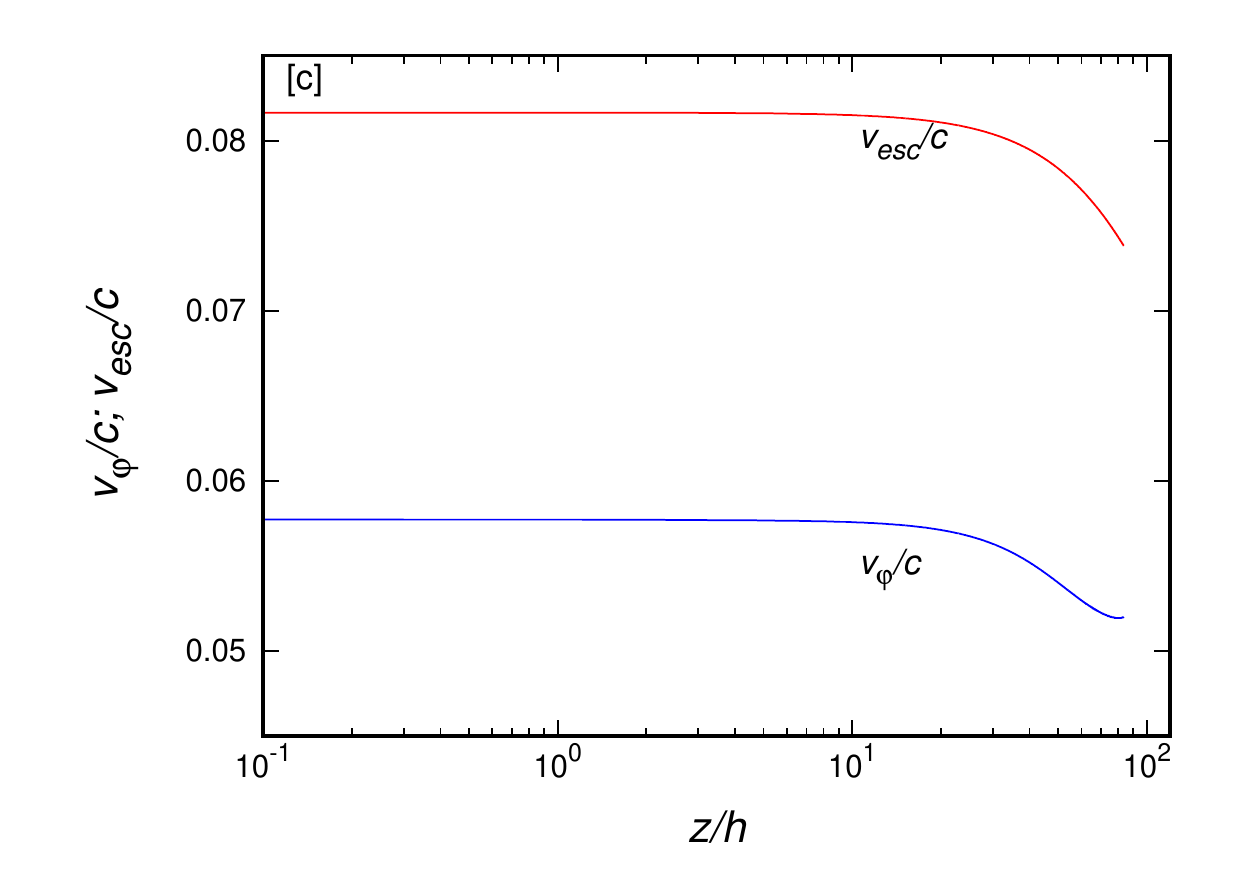}&\hspace{-0.90cm} 
  \includegraphics[width=0.46\textwidth]{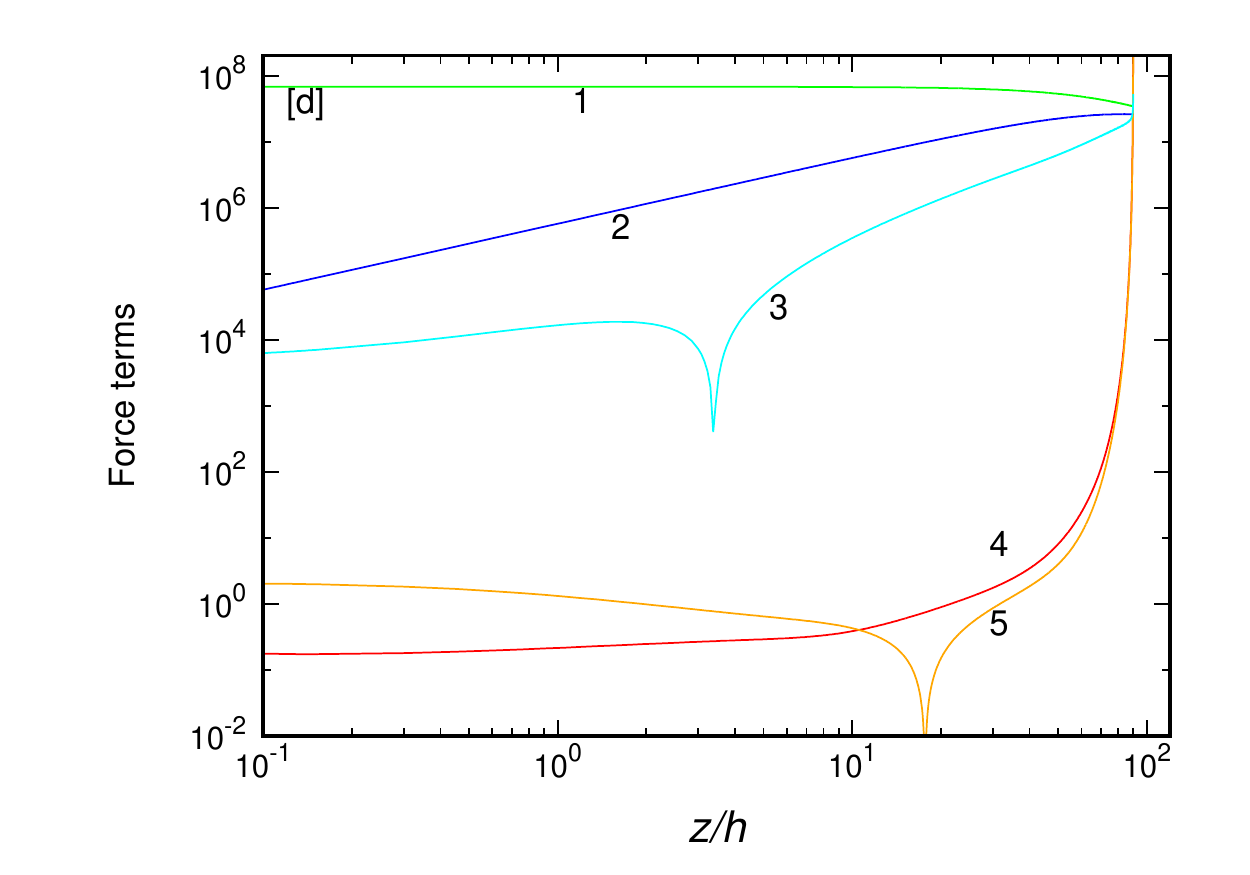}\\
\end{tabular}\vspace{-0.3cm}
\caption{The solutions of our model equations for $x$ = 4.65706 $\times$ 10$^{-7}$ (or $z^{max} \sim$ 92$h$) , $r$ = 300 $R_g$. The panels [a], [b] and [d] are same as the left, middle and right panels of Figure \ref{fig:x0}. The dashed curves 2 and 1 of panel [b] are model curves with $=\exp{\left(\frac{-z^2}{2h^23.8}\right)}$ and $\exp{\left(\frac{-z}{0.5h}\right)}$ respectively.
The panel [c] shows the variations of $v_\phi$ and $v_{esc}$= $\sqrt{\frac{2GM}{\sqrt{r^2+z^2}}}$ with height.   
}
\label{fig:x_fin}
\end{figure*}

We present here the above similar exercise for higher $x$ and  $f_v$ $\sim$ 1. 
For the considered set of parameters, we have an acceleration solution for
the range of $x$ $\equiv [0, 4.65707\times 10^{-7}]$.
We take $x =4.65706 \times$ 10$^{-7}$ ($\approx x^{max}$) for the presentation 
purpose. The results are shown in Figure \ref{fig:x_fin}, in which $z^{max}$ is around 92$h$.    
The sound speed increases slowly, almost by 1.5 times when height increases to $z$ = 2$h$ from the midplane (which is shown in Figure \ref{fig:x_fin}a). 
The interpretation of external heating for $x$ is justifiable as pressure and temperature increase with $x$ for a given $z$ and 
as a consequence the density falls with $z$. 
We find that the density and pressure profiles follow an isobaric and isothermal
profiles and their functional  forms are $\rho_c(r)\exp{\left(-\frac{z}{0.5h}\right)}$ and $p_c(r)\exp{\left(-\frac{z^2}{2h^2\ 3.6}\right)}$ respectively.
The model curves are overplotted on their numerical results, shown in Figure
\ref{fig:x_fin}b. For this maximal $x$, the disk scale height is $\sim$0.5$h$ for density and $\sim\sqrt{3.6} h$ $(= h_p)$  for pressure.
In the pressure profile,
the factor 3.6 is related to that height where the sign of $\frac{\partial p}{\partial r}$ changes, $z_f$ = 3.6$h$, as shown in  Figure \ref{fig:x_fin}d.
In another way, it also holds the previous expression $z_f = h_p^2/h$.
We also check the assumption for initializing the flow variables to
theirs respective Keplerian values in Figure \ref{fig:x_fin}d,
and notice $v_r\frac{\partial v_r}{\partial r}$, $v_z\frac{\partial v_r}{\partial z}$ $\ll$ $\frac{1}{\rho}\frac{\partial p}{\partial r}$ $\ll$ $F_r$ within the scale height. We find that, like $x$ = 0 case, for maximal $x$ the Keplerian approximation is still valid within the pressure scale height of the disk.

In Figure \ref{fig:x_fin}c, the variation of $v_\phi$ and escape velocity $v_{esc} = \sqrt{\frac{2GM_c}{r^*}}$ are shown, here $r^*$ = $\sqrt{r^2+z^2}$ is
the distance from the compact object. At $z$ = $z^{max}$, $v_\phi$ is much larger (almost 60 times) than $v_r$ and $v_z$. However, here $v_r$ and $v_z$ have been accelerated more, and the magnitudes of $v_r$ and $v_z$ are almost 2.5 times larger than the sound speed at midplane of the disk, i.e., $v_r(r, z=z^{max}) = 2.5c_s(r, z=0)$. 
Moreover $v_z$ is always less than the $v_r$ near to $z^{max}$, which is shown in the inset of
Figure \ref{fig:x_fin}a.  Like previous section, we have a smooth solution around $z^{max}$ as we obtain the requisite condition, e.g.,  $|\frac{1}{\rho}\frac{\partial p}{\partial z}| \approx |v_z\frac{\partial v_z}{\partial z}|$ as shown in Figure \ref{fig:x_fin}d (which also assures an isobaric regime, as mentioned in points (d) of \S\ref{sub:sol}). 
For $z > 3.6 h$,  $\frac{1}{\rho}\frac{\partial p}{\partial r}$ is positive 
and at $z^{max}$ it becomes order of $F_r$ (as shown in  Figure \ref{fig:x_fin}d). %
  Hence, near to the $z^{max}$ the radial pressure gradient is balancing 
the rotations substantially along with $F_r$. However, just above $z^{max}$, as there is no pressure gradient, 
the radial gravitational force can not support the rotations alone and the matter would be blown off with speed $v_{wind}= \sqrt{v_r^2+v_\phi^2+v_z^2}$. Here $v_{wind} \sim v_\phi < v_{esc}$, hence the fluid can not be escaped the system. 
In general, at $z$ = $z^{max}$ if $\frac{1}{\rho}\frac{\partial p}{\partial r}$ $\ll$ $F_r$, then the system is rotationally bound (like $x$ = 0 case), otherwise it is rotationally unbound.

\subsection{Vertical disk structure for $f_v$}\label{ver:fv} 

\begin{figure}
\centering
\begin{tabular}{l}
  \includegraphics[width=0.46\textwidth]{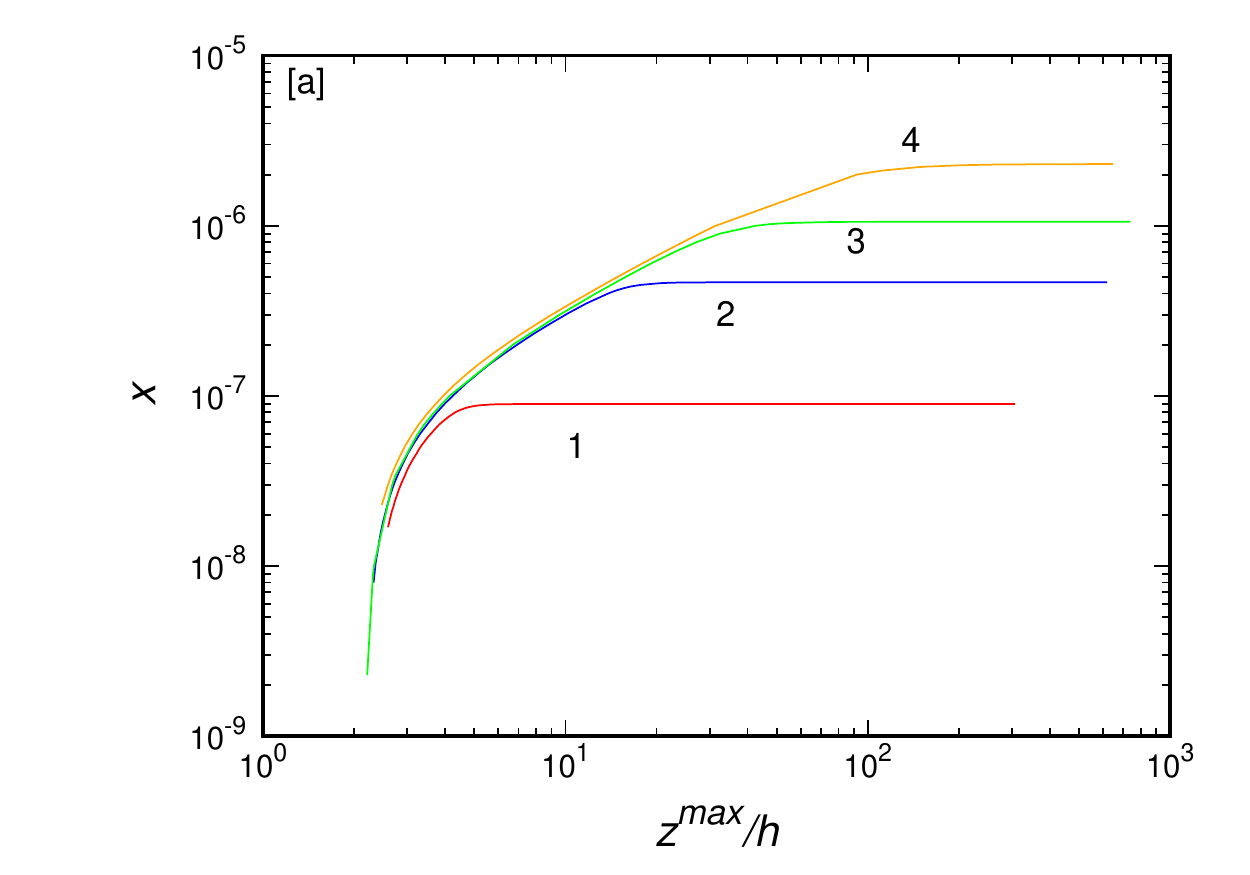}
\end{tabular}\vspace{-0.3cm}
\caption{The possible range of $x$ for acceleration solutions of equation (\ref{eq:master}) and corresponding $z^{max}$ for four different $f_v$ at $r$ = 300 $R_g$. Here the curves 1, 2, 3 and 4 are for $f_v$ = 0.1, 1, 3 and 10 respectively.  } 
\label{fig:fv_x}
\end{figure}

Next, we explore the connection between $f_v$ and possible range of $x$ for acceleration.
In  Figure \ref{fig:fv_x}, we show the variation of $z^{max}$ with $x$ 
for four different values of $f_v$. 
We notice,  $z^{max}$ increases with $x$ for a given $f_v$. In addition, for a given $z^{max}$, $x$ increases with $f_v$, which  signifies that both are attributed from same external heating. Loosely, the external heating (parameterized by $x$)  unrests the hydrostatic equilibrium which leads to a movement in the vertical direction (i.e., seeding the initial vertical speed). 
The maximum limit of $x$, $x^{max}$, for acceleration is 8.9646$\times 10^{-8}$, 4.65708$\times 10^{-7}$, 1.05778$\times 10^{-6}$ and 2.30485 $\times 10^{-6}$ for $f_v$ = 0.1, $\sim$1, 3 and 10 respectively.
We find that after some higher $z^{max}$, the small increment in $x$ leads to a large deviation in $z^{max}$; it occurs when $x$ tends to $x^{max}$. To identify the saturation of $x$ against $z^{max}$,  
we define a minimum $z^{max}$, termed as $z^{max}_t$, at which $x$  starts to tend $x^{max}$. %
Here, $z^{max}_t$ is around 5, 20, 50 and 100$h$ for $f_v$ = 0.1, 1, 3 and 10 respectively. 
In the saturation limit of $x$, $z^{max}$  varies significantly even by decimal increment in $x$, e.g., for $x$ = (0.46, 0.465, 0.4657 and 0.465707) $\times 10^{-7}$ the corresponding $z^{max}$ are $\sim$ 20, 28, 71 and 110$h$ respectively at $f_v$ $\sim$ 1 (shown in curve 2). Here, we like to stress that if we increase the above values of $x$ very little, i.e., $x$ = (0.47, 0.466, 0.4658, 0.46571) $\times 10^{-7}$, we have a deaccelerated solutions. 
In general, for a given $z^{max}$, $x$ increases with increasing $f_v$ for any launching radius, which confirms that $x$ and initial vertical speed are intimately related with external heating.   %

\begin{figure}
\centering
\begin{tabular}{l}
  \includegraphics[width=0.46\textwidth]{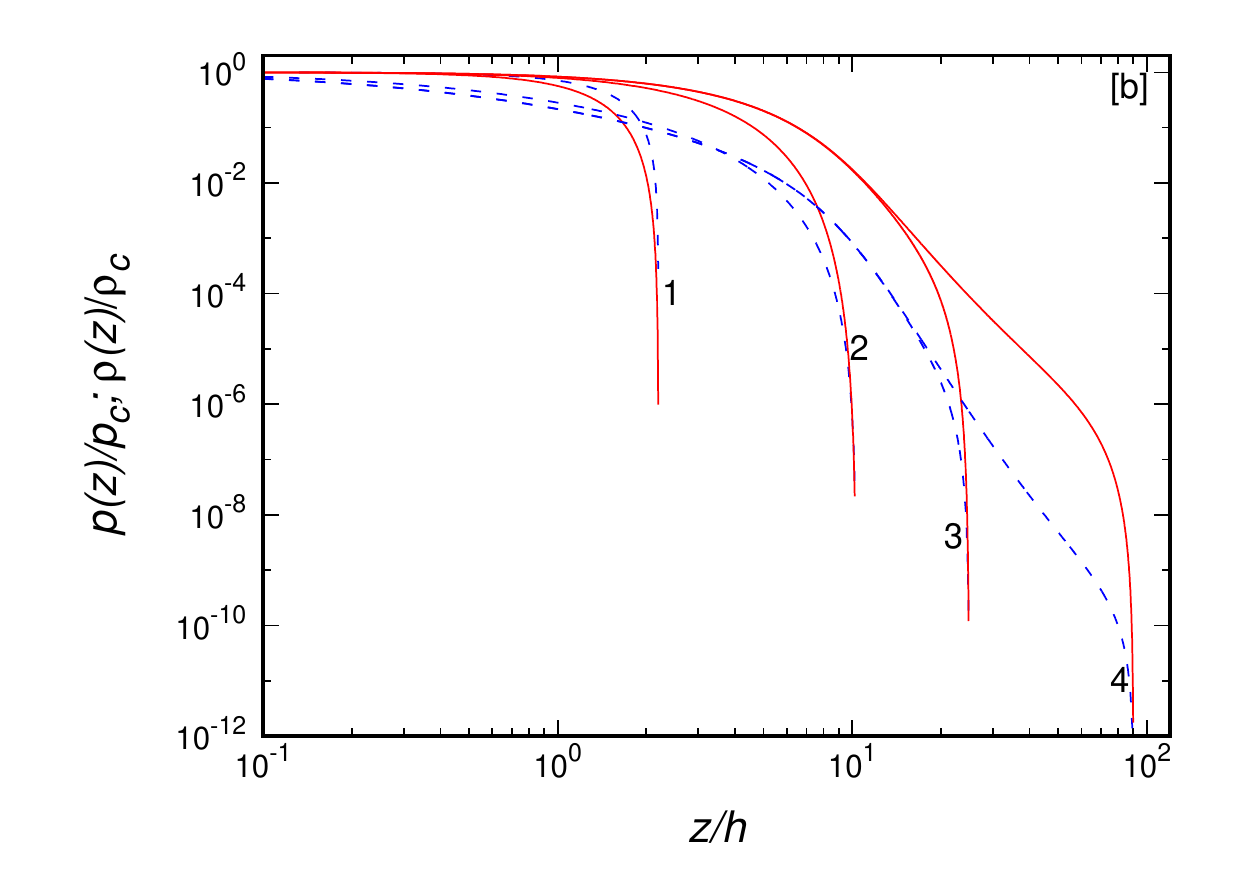}
\end{tabular}\vspace{-0.3cm}
\caption{ The pressure (solid curve) and density (dashed curve) profiles in vertical direction for different $x$ (or $z^{max}$) for $f_v$ $\sim$ 1. Here the curves 1, 2, 3 and 4 are for $x$ ($z^{max}$) = 0 (2.2$h$), 3.05414 $\times 10^{-7}$ (10$h$), 4.64 $\times 10^{-7}$ (25$h$) and 4.65706$\times 10^{-7}$ (92$h$) respectively. 
 } 
\label{fig:p_x}
\end{figure}

In Figure \ref{fig:p_x}, we show the variations of pressure (solid curve) and density (dashed curve) when $x$ changes from 0 to $x^{max}$ for $f_v\sim$1 (by considering four different  values of $x$). The curve 1 is for $x$ = 0 or $z^{max}$ = 2.2$h$ and curves 2, 3 and 4 are for $z^{max}$ = 10, 25 and 100$h$ respectively. Here, the pressure is increasing with $x$ which is consistent with the interpretation of equation (\ref{eq:hydro}).
$z^{max}_t$ for $f_v = 1$ is around 20$h$ (shown by curve 2 in Figure
\ref{fig:fv_x}a).
The pressure scale height for curves 1, 2 and 3 is $\sqrt{0.96}, \sqrt{2.2}$
  and $\sqrt{3.8}h$ respectively, and $\frac{\partial p}{\partial r}$ flips the sign around 0.96, 2.2 and 3.8$h$ respectively. In appendix, we elaborate the sign flip
behaviour of $\frac{\partial p}{\partial r}$ for curve 2 by obtaining the solutions for two adjacent $r (=300R_g)$:  $r-\Delta r$ and $r+\Delta r$, with $\Delta r = 0.1R_g$.

We find that the density or pressure profile changes only
for $z^{max} < 20h$ while for $z^{max} > 20h$, they settle to the profile  corresponding to $x$ = $x^{max}$.
The pressure is dropped by 10$\%$ from its midplane value at a height $z^p_{10}\sim4.5h$ and $\sim6.5h$ for $z^{max}$ = $10h$ and $25h$ respectively.
  For a given $r$, $W_{\phi r} \propto \rho c_s h'$ (here, $h'$ is the maximum turbulent eddy's size). Relatively, the averaged value of the quantity $\rho c_s h'$
  in the region $h< h' < z^p_{10}$ (or windy region) is small but not negligible in comparison to the magnitude corresponding to the disk region $0 < h' < h$. 
  $z^p_{10}$ is, in general, quite larger than $h$.
    Note that accretion flows are turbulent and viscosity $\alpha$ is the 
turbulent viscosity. The same $\alpha$ is also appearing in the model equations explaining flow
in, e.g., $h<z<z^p_{10}$, when the viscosity therein is also of turbulence 
origin as molecular viscosity is negligible. 
In the previous section, we have noticed that the
density profile is isothermal for $x$ = 0 and isobaric  for $x^{max}$,
while pressure profile is always isothermal. Here, we observe, the density acquires an isobaric profile around  $z^{max}_t$.
In general, for $z^{max}$ $> z^{max}_t$ %
the density or pressure profile does not change from its own modelled profile at $z^{max}_t$, or the pressure and density scale height of the disk remain constant. 

\begin{figure*}
\centering
\begin{tabular}{lr}
  \includegraphics[width=0.46\textwidth]{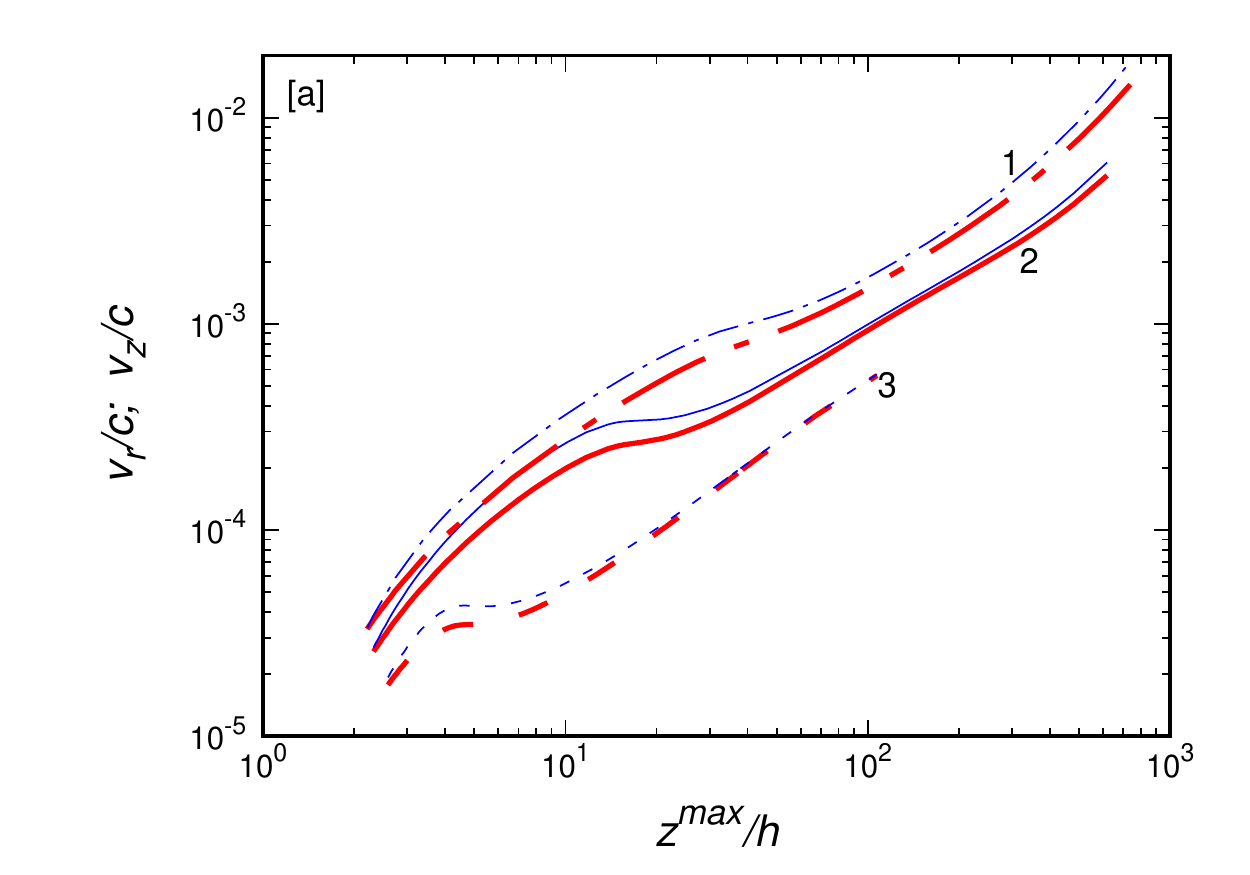}&\hspace{-0.90cm}
\includegraphics[width=0.46\textwidth]{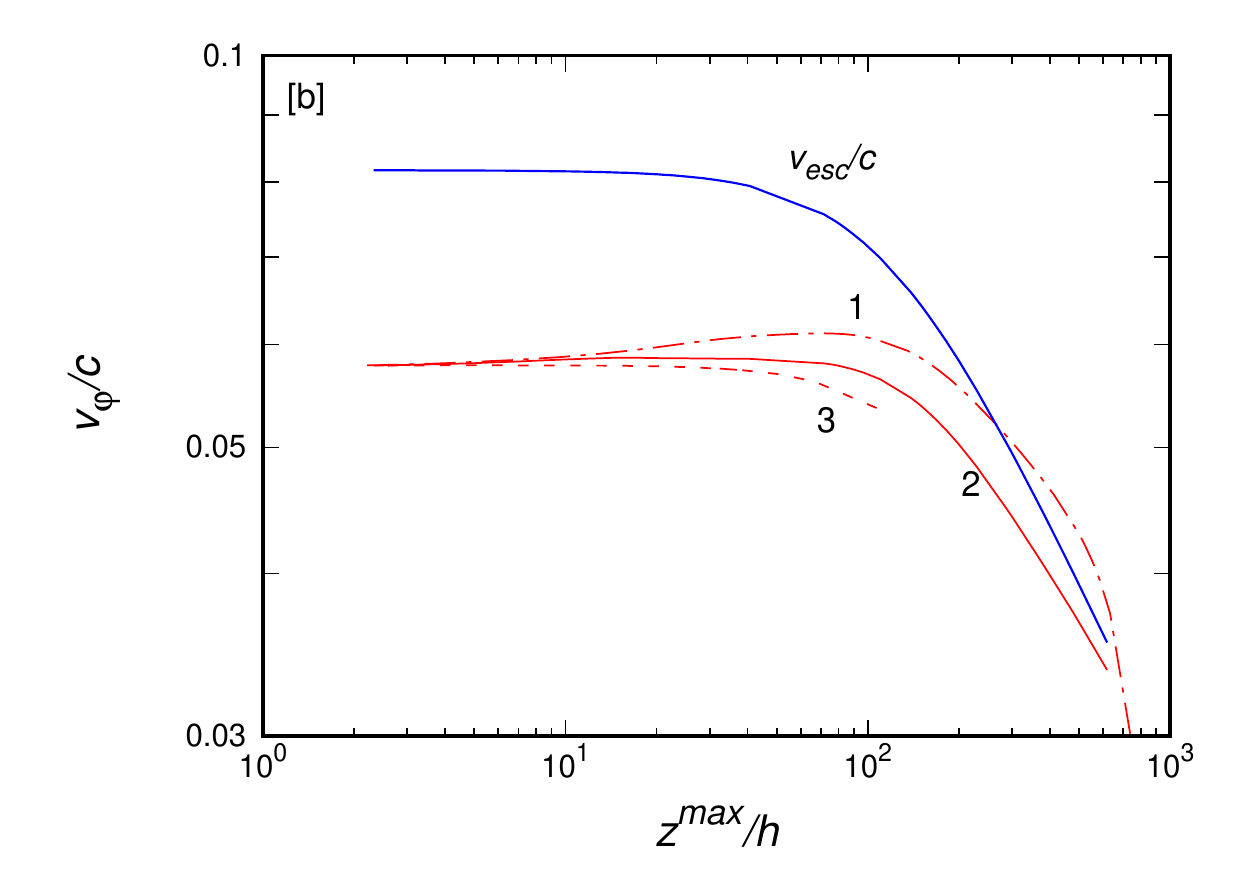}\\ 
  \includegraphics[width=0.46\textwidth]{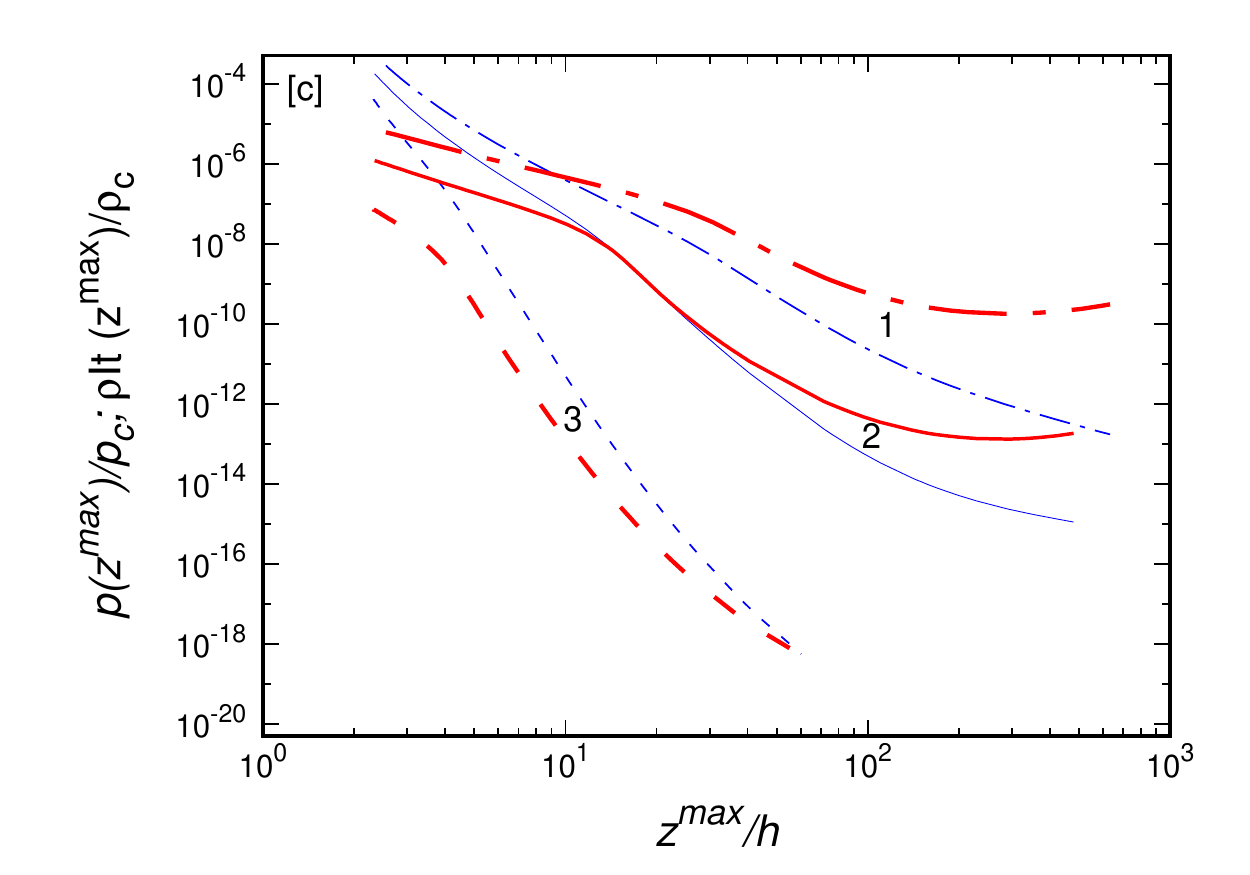}&\hspace{-0.90cm}
  \includegraphics[width=0.46\textwidth]{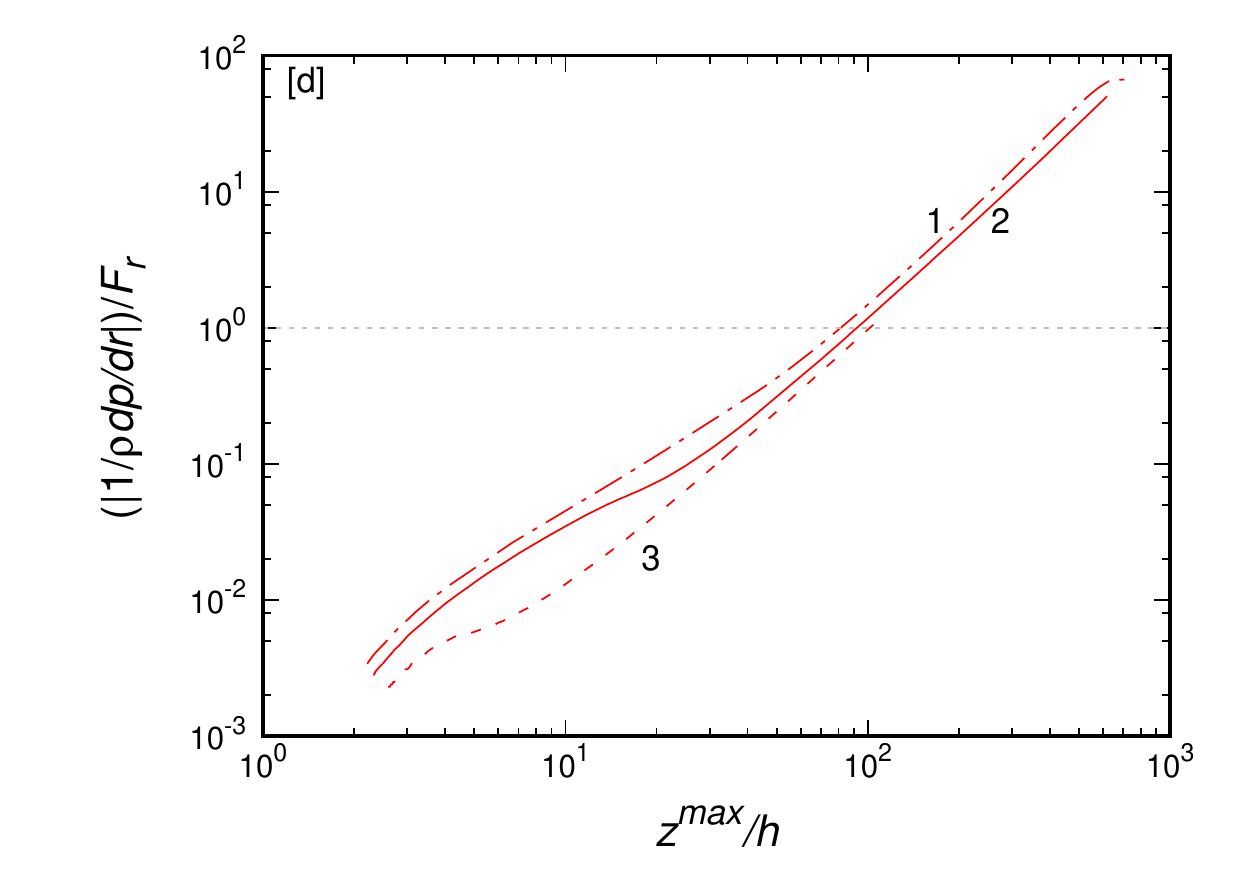}
\end{tabular}\vspace{-0.3cm}
\caption{The solutions of our model equations at $z$ = $z^{max}$ (or given $x$), $r$ = 300$R_g$ for three different $f_v$. In all panels, the curves 3, 2 and 1 are for $f_v$ = 0.1, 1 and 3, which are shown by dashed, solid and dot-dashed curves respectively. In upper panels, the different velocities have been shown, where $v_z$ (thick curve) and $v_r$ (thin curve) are shown in panel (a) and $v_\phi$ along with $v_{esc}$ in panel (b). In panel (c), the pressure (thick
  curves) and density (thin curves) profile has been shown. In panel [d] the ratio of radial pressure gradient to radial gravitational force ($\frac{1}{\rho}\frac{\partial p}{\partial r}$\big/$F_r$) has been shown.    
}
\label{fig:rr300}
\end{figure*}

In  Figure \ref{fig:fv_x} we have shown one-one mapping between $x$ and $z^{max}$ for different $f_v$, next we study the variation of flow variables at 
$z^{max}$ (or $x$) for different choice of $f_v$. We consider three values of $f_v$ = 0.1, 1 and 3; the results 
are shown by curves 3, 2 and 1 respectively in all panels of  Figure \ref{fig:rr300}. 
In  Figure \ref{fig:rr300}a,  $v_z$ (thick curves) and $v_r$ (thin curves) have
been studied; $p$ (thick curves) and $\rho$ (thin curves) have been shown in
 Figure \ref{fig:rr300}c; and $v_\phi$ and $\left|\frac{1}{\rho}\frac{\partial p}{\partial r}\right|\Big/F_r$ have been presented in   Figure \ref{fig:rr300}b and \ref{fig:rr300}d respectively.
The density or pressure at a given $z^{max}$ increases with increasing $f_v$  
and they decrease rapidly with $z^{max}$ for smaller $f_v$, while for a higher $z^{max}$ the pressure varies slightly. 

The radial component of pressure gradient becomes greater than the radial
gravitational force,  $\frac{1}{\rho}\frac{\partial p}{\partial r}$ $\gtrsim$ $F_r$, around $z^{max}$ $\gtrsim$ 100$h$, for all three values of $f_v$ (here,
$r=$  $118h$ = 300$R_g$). The pressure scale heights are $\sqrt{1.2}$, $\sqrt{3.8}$ and $\sqrt{4.2} h$  for $f_v$ = 0.1, 1 and 3 respectively at $x = x^{max}$
(or  $z^{max}$ $\gtrsim$ 100 $h$). 
Like earlier, just above $z^{max}$, where the radial pressure gradient is comparable to the radial gravitational force, 
the fluid will be 
ejected from the systems either tangentially or tilted upward direction depending on the fluid speed $ \sqrt{v_\phi^2+v_r^2+v_z^2} = v_{wind} $.
The height $z^{max}$, where fluid is rotationally unbound, is termed as a wind outflow ejection height.
For example, for $f_v$ = 3 at $z^{max}$ $\sim$ 270$h$ the wind will escape to infinity almost tangentially (as $v_{wind}$ $\sim v_\phi$ $> v_{esc}$), 
but for $z^{max} >$ 270$h$, $v_{wind} > v_\phi$, the wind direction will make less than $90\degr$ from the vertical $z$-axis or $\theta_w < 90\degr$, here $\theta_w$ is the angle between the wind direction and  $z$-axis. 
The wind outflow direction changes from parallel (to the disk plane) direction to the upward direction, when the height of releasing site of wind increases. Recently, \citet{Kumar2017} modeled the observed high energy power-law spectra in HS state in bulk Comptonization for relativistic conical wind, where the 
change of wind direction is similar to what is found 
here for $\theta_w < 90\degr$ 
\cite[see also,][]{Kumar2018}.

At a given $z^{max}$,  $v_z$, $v_r$ and $v_\phi$ increase with $f_v$. 
The increment in velocities is not linear, e.g., at $z^{max}$ = 100$h$, $v_z$ 
increases almost by 1.8 and 1.5 times when $f_v$ increases from 0.1 to 1 and from 1 to 3 respectively. 
In general, $v_z$ and $v_r$ are increasing with $z^{max}$ while $v_\phi$ is decreasing for a given $f_v$. 
$v_z$ or $v_r$ becomes comparable to $v_\phi$ at a smaller $z^{max}$ when $f_v$ increases. 
 $v_\phi$ becomes larger than $v_{esc}$ at smaller $z^{max}$ for a bigger $f_v$, e.g., at $z^{max}$  $\sim$ 270, 700$h$ %
for $f_v$ = 10, 3 respectively.  Hence, the increment of initial vertical speed (with  restriction $f_v <$ 10) boosts the wind outflow.

\section{Wind solutions}

Wind outflow model is usually characterized with density, speed and launching radius,
and these parameters should be consistent with the parameters derived from the photo-ionization model for given ion species, like, ionization parameters $\xi$, wind column density $N_{h}$.
In the previous section, we have found that at maximum attainable height for acceleration, $z^{max}$, if the radial pressure gradient is comparable to the radial gravitational force, then the disk material would be blown off with speed $\sqrt{v_\phi^2+v_z^2+v_r^2}$ and termed as wind outflow. We have studied the generic properties of wind solutions 
considering $x$ and $f_v$ as parameters, for fixed launching radius $r$, mass accretion rate and viscosity parameter.
Now in the present section, we  explore the wind characteristic with $r$, $\dot{M}$, $\alpha$. Finally we compare the modeled wind characteristic with observations.      
Wind is observed usually in HS state or left side of the hardness-intensity diagram (q-diagram), where the luminosity varies more than two orders of magnitude
\cite[e.g.,][]{Dunn-etal2010, Ponti-etal2012}.
We consider a wide range of mass accretion rate $\dot{M}$= 0.2 - 0.005 $\dot{M}_{Edd}$, where $\dot{M}_{Edd}$ = $L_{Edd}/(c^2 \eta)$, is the Eddington accretion rate,  $L_{Edd}$ is the Eddington luminosity and $\eta$ is the efficiency. A typical range of viscosity parameter $\alpha$ in a thin accretion disk is $\sim$ 0.1 - 0.4 \cite[][]{King-etal2007}, we take the full range of $\alpha$ in our calculations.  
To explore the wind parameters, without loss of generality, we take a 10 $M_{\odot}$ compact object, which gives the Eddington accretion rate  $\dot{M}_{Edd}$ $\sim$  2 $\times 10 ^{19}$ g/s for $\eta$ = 0.1.

\subsection{Wind launching radius}
\begin{figure*}
\centering
\begin{tabular}{lr}
 \includegraphics[width=0.46\textwidth]{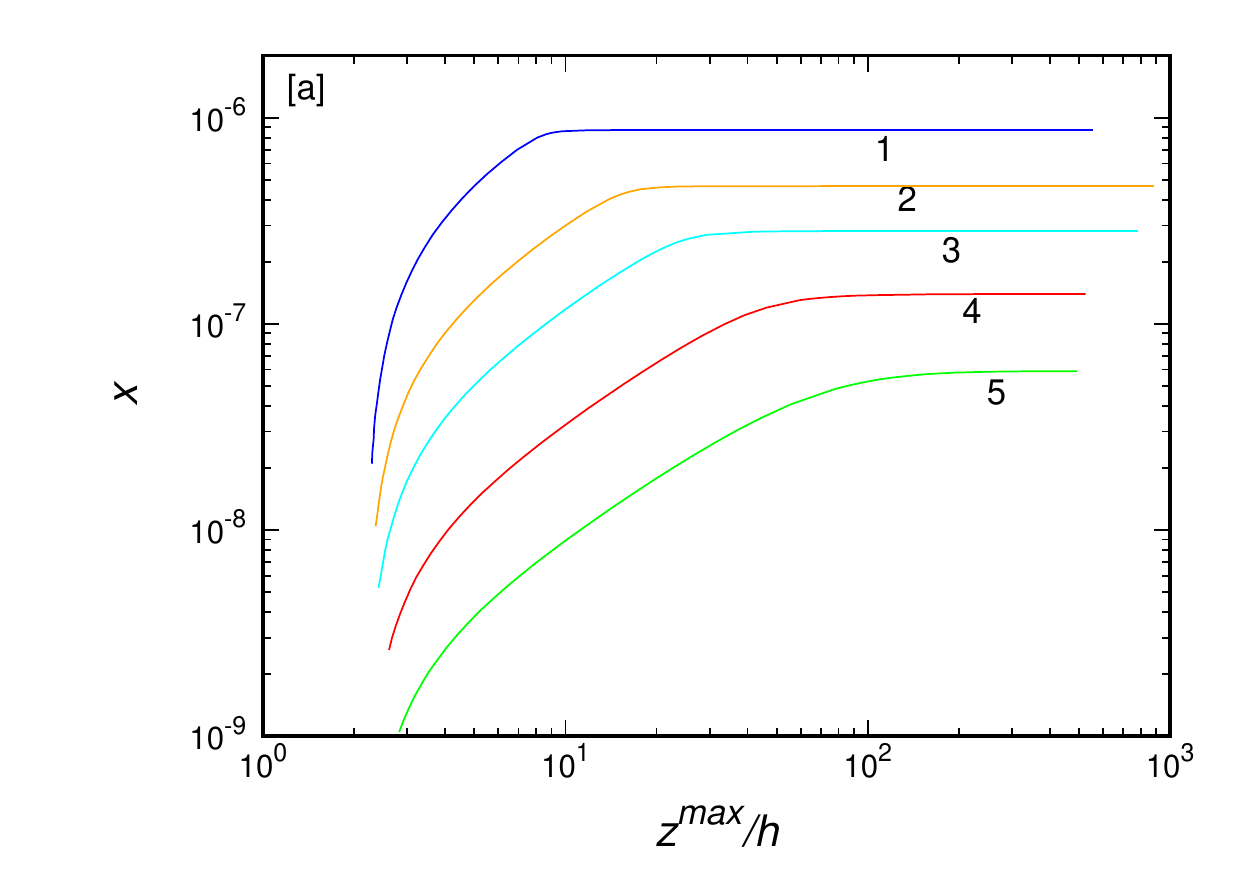}&\hspace{-0.90cm}
  \includegraphics[width=0.46\textwidth]{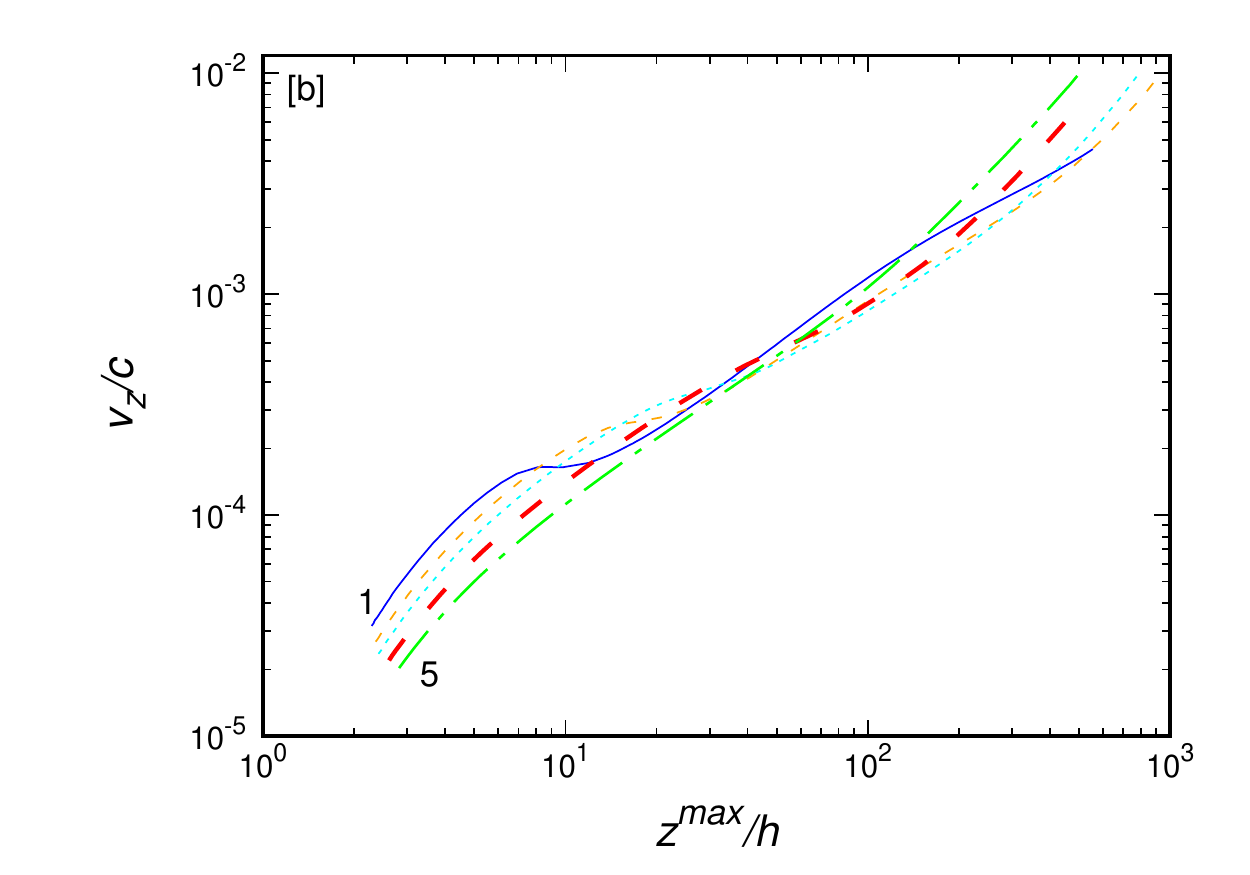}\\
  \includegraphics[width=0.46\textwidth]{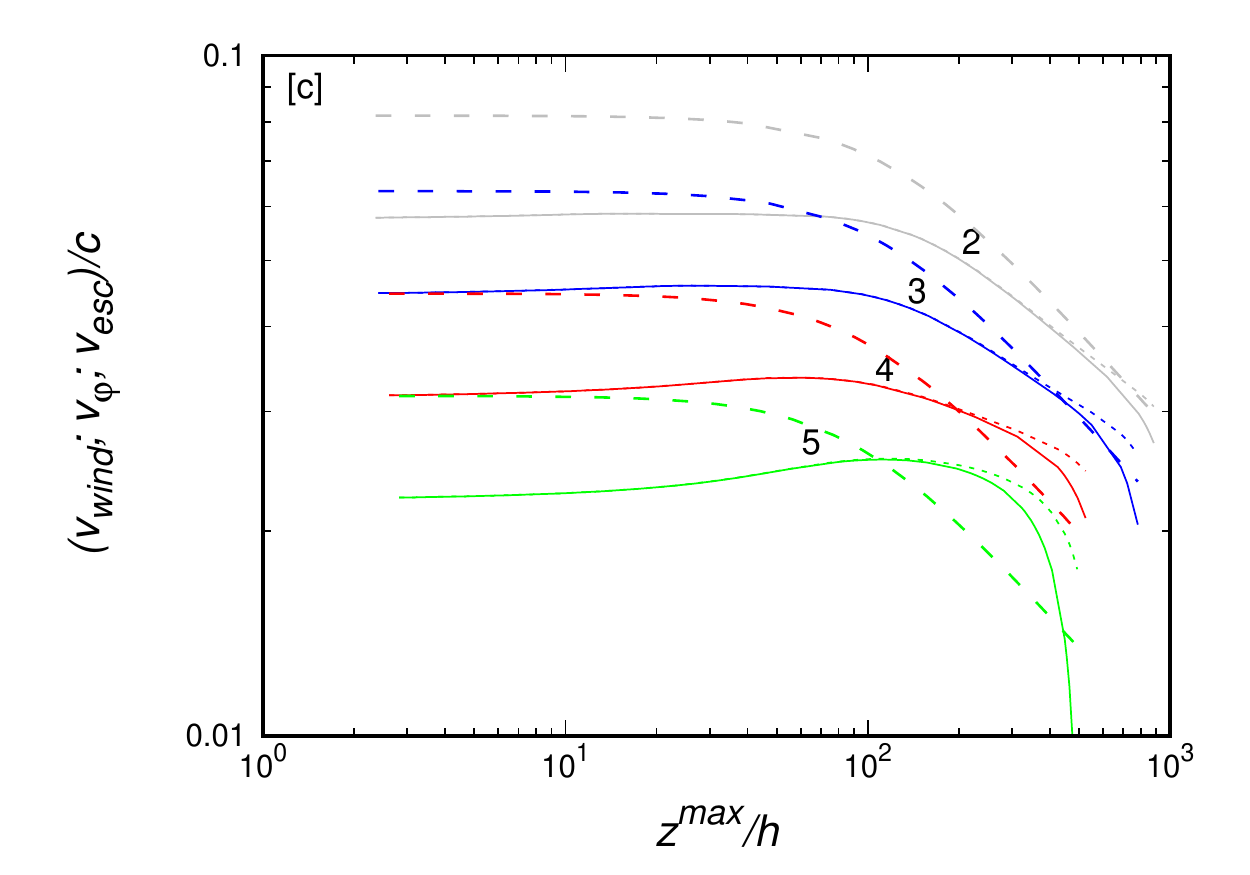}&\hspace{-0.90cm} 
 \includegraphics[width=0.46\textwidth]{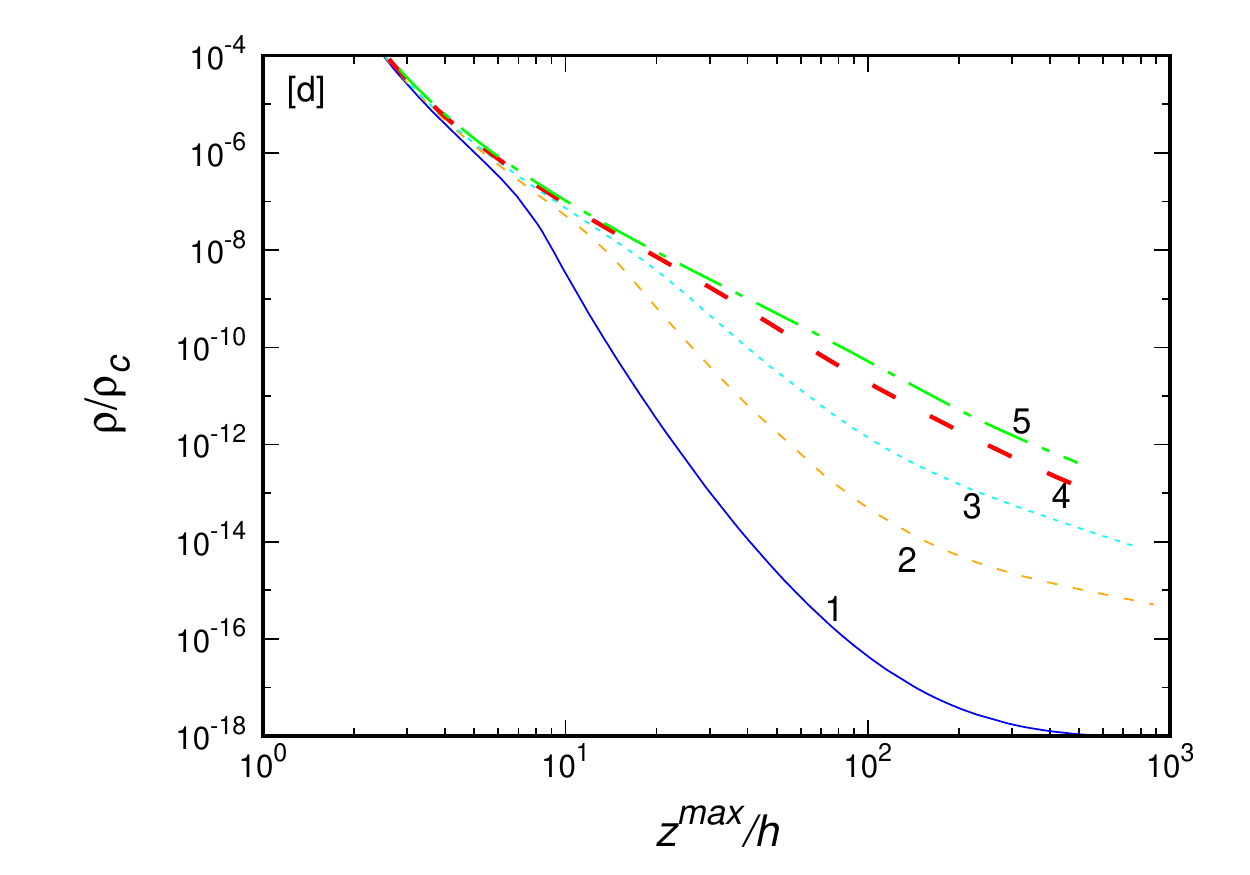}\\ 
\end{tabular}
\caption{The solutions of our model equations at $z$ = $z^{max}$ (or a given $x$), $f_v$ $\sim$ 1 for five different $r$. In all panels the curves 1, 2, 3, 4 and 5 are for $r$ = 150, 300, 500, 1000 and 2000$R_g$ respectively. The panel (a) shows $x$ vs. $z^{max}$. In panel (b) the variation of $v_z$ with $z^{max}$ has been shown. In panel [c] the variations of $v_\phi$ (solid curves), $v_{wind}$ ($\sqrt{v_z^2+v_r^2+v_\phi^2}$, dotted curves) and $v_{esc}$ (dashed curves)
  have been shown. The density profile has been shown in panel [d]. 
  }
\label{fig:wd-r}
\end{figure*}

To explore the favorable wind launching site, 
 we take a large range of launching radius $r$ = 150 $-$ 2000$R_g$.
Since in the Keplerian disk, the radial velocity $v_r$ increases  with decreasing $r$, 
even for the same $f_v$ the initial guess value of $v_z$ increases with decreasing $r$. Like previous section (see  Figure \ref{fig:fv_x}a), we attain a large $x$ range for a smaller launching radius $r$.
The results are shown in Figure \ref{fig:wd-r} for mass accretion rate $\dot{M}$ = 0.005$\dot{M}_{Edd}$, $\alpha$ = 0.1 and $f_v$ $\sim$ 1. In all panels, the curves 1, 2, 3, 4 and 5 are for launching radius $r$ = 150, 300, 500, 1000 and 2000$R_g$ respectively. In panel [a], $x$ and corresponding $z^{max}$ have been shown. The quantity $z^{max}_t$, the minimum $z^{max}$ where $x$ tends to acquire a maximum value, is  $\sim$10, 25, 45, 80 and 150$h$ for curves 1, 2, 3, 4 and 5 respectively.       
Herewith,
 we only show the $v_z$ variation with $z^{max}$ at a given launching radius (shown in panel [b]), as we have noted earlier that at $z^{max}$,  $v_z$ and $v_r$ are comparable to the sound speed with $v_z < |v_r|$, (e.g., in Figure \ref{fig:rr300}[a]). 
 In panel [c], the variations of $v_\phi$ and wind speed $v_{wind} = \sqrt{v_z^2+v_r^2+v_\phi^2}$ are presented along with the escape velocity $v_{esc}$. Here, $v_{esc}$ is different for different $r$, just because of that $z$ is measured in
 the unit of scale height $h$, and $h/r$ decreases with increasing $r$.
In panel [d], the density as a function of $z^{max}$ is shown, here for a given $z^{max}$ the density decreases with decreasing $r$.

For clarity, we specify three different values of $z^{max}$ (like $z^{max}_t$) as, $z^{max}_b :$ a minimum $z^{max}$ where $\frac{1}{\rho}\frac{\partial p}{\partial r}$ $>$ $F_r$; $z^{max}_p :$  a minimum $z^{max}$ where $v_{wind} > v_\phi$; $z^{max}_e :$ a minimum $z^{max}$ where $v_{wind} > v_{esc}$.
For $z^{max}$ $>$ $z^{max}_b$ the gas is rotationally unbound 
and a wind outflow launches (however, wind can also launch from the lower
  height $z^{max} <$ $z^{max}_b$ where the radial component of pressure gradient significantly contributes in balancing the rotation along with the radial gravitational force);
for $z^{max}_b <$ $z^{max}$ $<$ $z^{max}_p$ the wind outflow is mainly ejected
tangentially, or parallel to the disk plane in all directions;  and above $z^{max}_p$ the wind 
 launches with $\theta_w < 90 \degr$. 
In panel [c], for launching radius $r$ = 300, 500, 1000 and 2000$R_g$,  $z^{max}_p$ are
$\sim$340, 230, 170 and 120$h$ and $z^{max}_e$ (corresponding $v_{wind}$) are 710 (0.033c), 405 (0.032c), 190 (0.03c) and 105$h$ (0.025c) respectively. 
 $z^{max}_b$ is  100, 80, 70 and 65$h$ for $r$ = 300, 500, 1000 and 2000$R_g$ respectively. Hence $z^{max}_b$ is smaller than $z^{max}_t$ for $r$ $\gtrsim$ 800$R_g$ at $\dot{M} = 0.005\dot{M}_{Edd}$ and $\alpha$ = 0.1, it meant that the wind outflow can occur for smaller value of $x$ ($< x^{max}$).
In short, 
$z^{max}_b$, $z^{max}_p$  and $z^{max}_e$ decrease with increasing launching radius $r$ and particularly, after some large $r$ ($> 800 R_g$), $z^{max}_b$ becomes smaller than $z^{max}_t$. Thus the wind can launch easily from the outer radius of the disk.

\subsection{ $\dot{M}$ and $\alpha$ for wind}

\begin{figure*}
\centering
\begin{tabular}{lr}
 \includegraphics[width=0.46\textwidth]{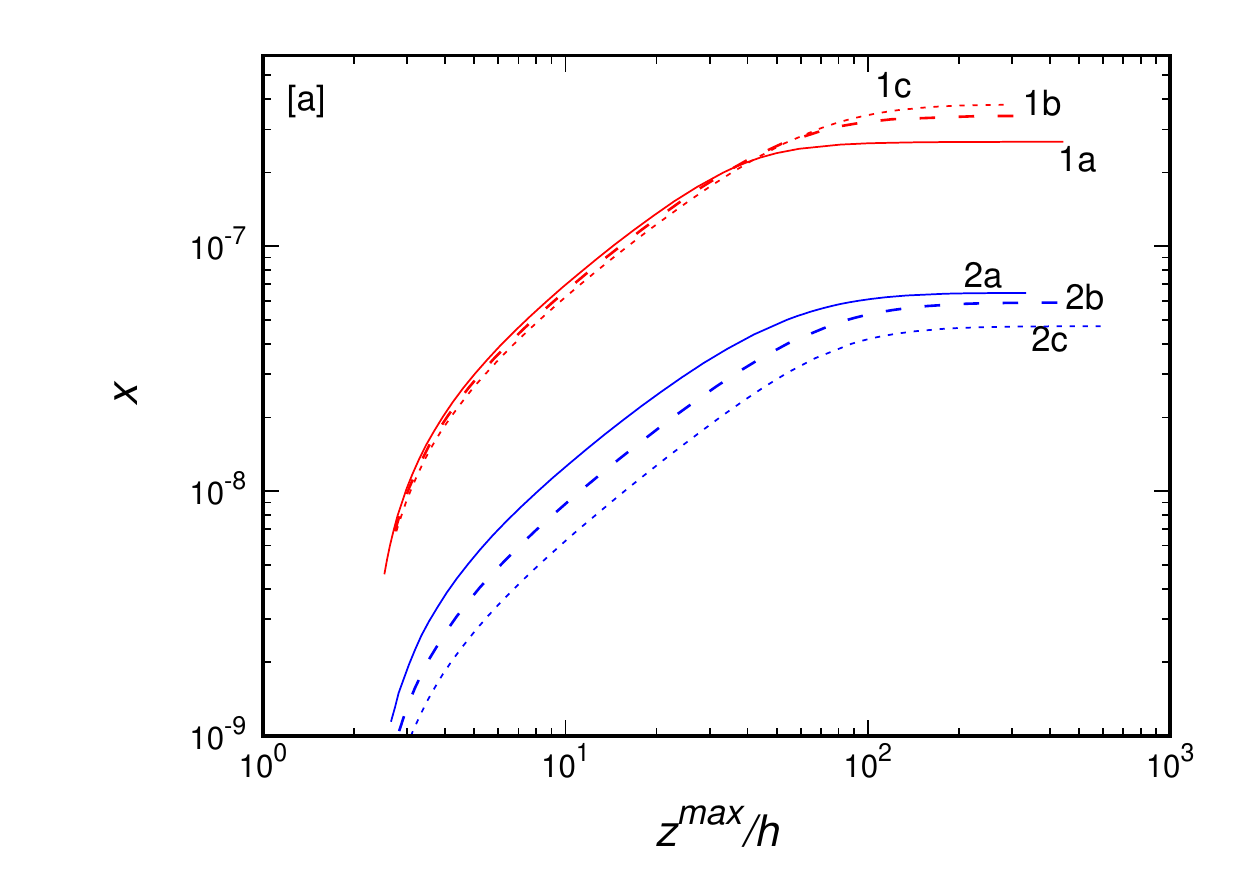}&\hspace{-0.90cm} 
  \includegraphics[width=0.46\textwidth]{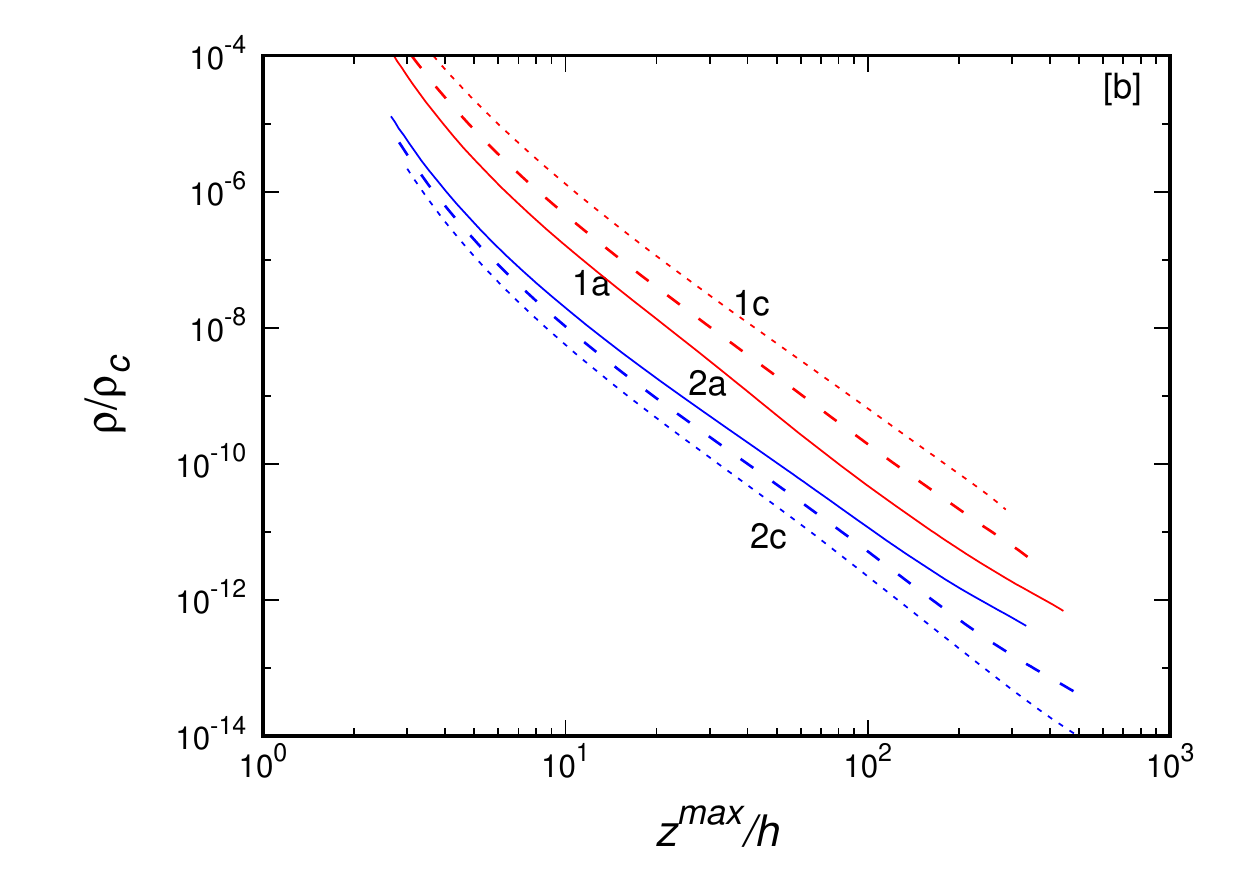}\\ 
  \includegraphics[width=0.46\textwidth]{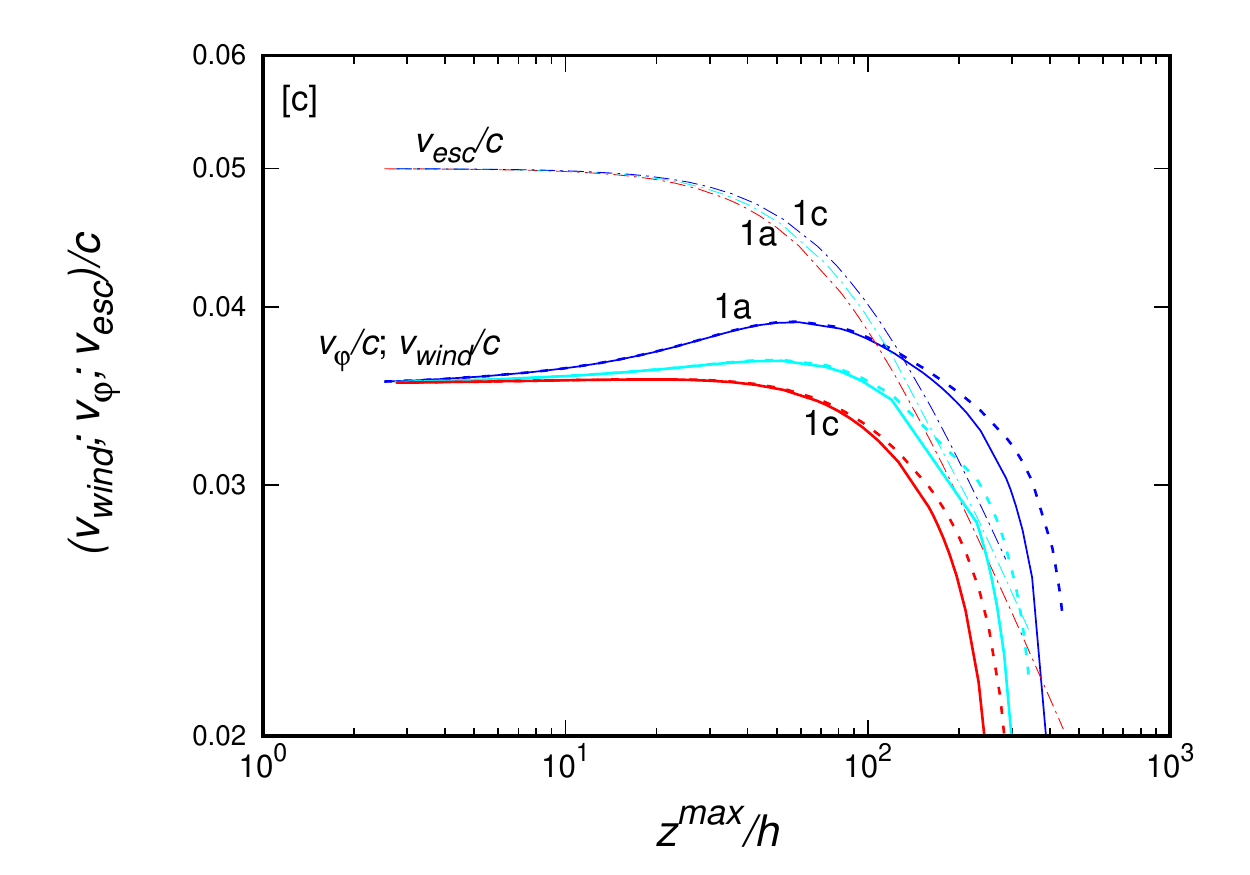}&\hspace{-0.90cm} 
 \includegraphics[width=0.46\textwidth]{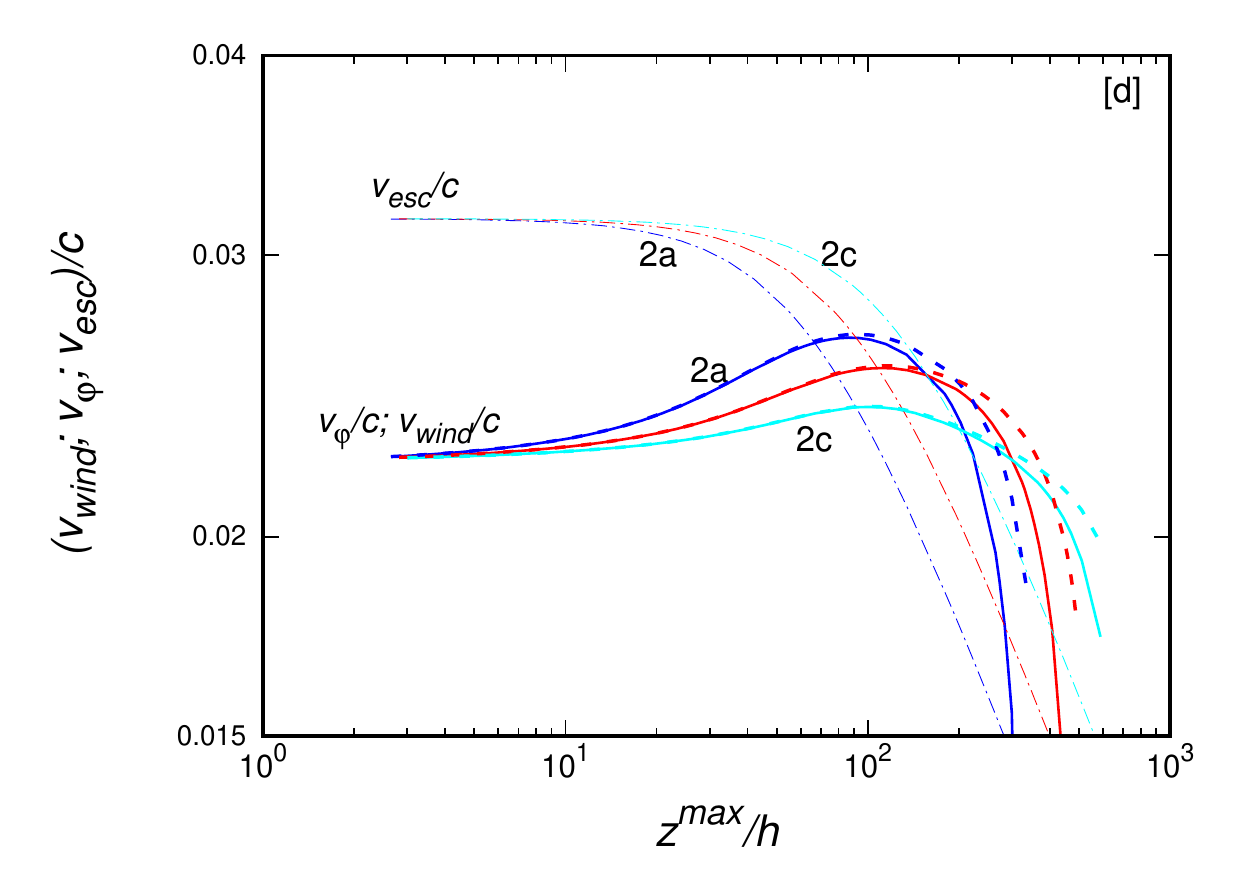}\\ 
\end{tabular}\vspace{-0.3cm}
\caption{The solutions of our model equations at $z$ = $z^{max}$ (or a given $x$), $f_v$ $\sim$ 1 for three different $\alpha$ (shown by curves 1a, 1b and 1c for $\alpha$ = 0.1, 0.2 and 0.4 respectively) and three different $\dot{M}$ (shown by curves 2a, 2b and 2c for $\dot{M}$ = 
0.05, 0.005 and 0.0005 $\dot{M}_{Edd}$ respectively). The curves marked by 1 and 2 are for $r$ = 800 and 2000$R_g$ respectively.
The $x$ vs. $z^{max}$ curves are shown in panel [a] and density profile in panel [b]. The density curves 2a, 2b and 2c are shifted down by factor 10 for clarity. In bottom panels the variations of $v_\phi$ (solid curve), $v_{wind}$ (dotted curve) and $v_{esc}$ (dot-dashed curve) are shown. 
}
\label{fig:wd-md-a}
\end{figure*}

Next, we explore the behavior of wind outflow characteristic over the mass accretion rate and viscosity.
For this we take two launching  radii 800 and 2000$R_g$, where the wind launching is easier. %
Without loss of generality, the  dependence of wind characteristics on viscosity is examined for $r$ = 800$R_g$, with taking three different values of
$\alpha$, while the dependence related to mass accretion rate done at
$r$ = 2000$R_g$ 
with three different value of $\dot{M}$.
The results are shown in Figure \ref{fig:wd-md-a}. In all panels of  Figure
\ref{fig:wd-md-a} the curves marked 1 and 2 are for $r$ = 800 and 2000$R_g$ respectively.
The curves 1a, 1b and 1c are for $\alpha$ = 0.1, 0.2 and 0.4 respectively (fixed $\dot{M}$ = 0.05 $\dot{M}_{Edd}$), and the curves 2a, 2b and 2c are for
$\dot{M}$ = 0.05, 0.005 and 0.0005 $\dot{M}_{Edd}$ respectively (fixed $\alpha$ = 0.1).
In panel [a], the $x$ versus $z^{max}$ curve is shown and in panel [b], the density variation with $z^{max}$ has been shown. The  densities for $r$ = 800 and 2000 $R_g$ are comparable (which is also shown earlier for $r \gtrsim 1000R_g$  in
Figure \ref{fig:wd-r}[d] by curves 4 and 5), for clarity the curves 2a $-$ 2c are lowered by factor 10.  
At a given $z^{max}$, the density increases with increasing either $\alpha$ or $\dot{M}$. For $r$ = 2000$R_g$, at $z^{max}$ = 100$h$, the density
increases by factor 5 
by increasing the accretion rate from 0.0005 to 0.05$\dot{M}_{Edd}$. For $r$ = 800$R_g$ at $z^{max}$ = 100$h$, the density increases almost 12 times with increasing $\alpha$ from 0.1 to 0.4.

The wind speed, $v_{esc}$ and $v_\phi$ are shown in  panels [c] and [d], where the panel [c] is for different  $\alpha$ (i.e., $r$ = 800$R_g$) and panel [d] for different $\dot{M}$ (i.e.,  $r$ = 2000$R_g$). 
Since, the scale height $h$ at a given $r$ varies differently with $\dot{M}$
and $\alpha$, mainly $h$ changes small by varying  $\alpha$ in the Keplerian disk.
$v_{esc}$ as a function of $z^{max}/h$
is different for different $\alpha$ and $\dot{M}$ which is shown by the upper
 curves 1a - 1c and 2a - 2c of panels [c] and [d] respectively.     
The quantity $z^{max}_p$ %
 decreases either by increasing $\alpha$ or by increasing $\dot{M}$. 
The quantity $z^{max}_e$ %
 increases with increasing $\alpha$, while decreases with increasing $\dot{M}$. 
 For $r$ = 2000$R_g$, $\alpha$ = 0.1, the $z^{max}_e$ = 66, 106 and 197$h$
 for $\dot{M}$ = 0.05, 0.005 and 0.0005$M_{Edd}$ (where, $r/h$ $\sim$65, 92 and 130) respectively.
 The wind escapes the system from lower height $z^{max}$ for higher accretion rate at a given launching radius, while wind may not escape the system for higher $\alpha$ (lower curve 1c in panel [c]).
 This complex behaviour of wind solution with respect to $\alpha$ may be due to
   the  turbulent windy medium (see \S3.3). The condition for turbulent windy medium  changes by changing $\alpha$, while it is fixed
   for different $\dot{M}$  (because of fixed $\alpha$).
Hence, the increment of accretion rate helps the wind launching 
by lowering the height
$z^{max}_p$ and $z^{max}_e$, while the increment of $\alpha$ may oppose the wind launching by elevating  $z^{max}_e$.

\subsection{Comparison with observations}
The absorption line features of ion species in X-ray spectrum of LMXBs reveal the presence of wind outflow. 
The primarily source for photoionization of wind matter is the inner region of the disk. From the observed absorption line features, one can determine the ion species, wind hydrogen column density $N_{h}$, wind velocity $v_{wind}$ and also ionization parameter $\xi$ for ion species. The ionization parameter is defined
as $\xi = \frac{L}{n_h r_*^2}$, where $L$ is the ionizing luminosity of the
source, $r_*$ is the distance between the ionizing source (where from
irradiation comes, here inner accretion disk) and wind matters,
$n_{h}$ = $\frac{\rho}{\mu\ m_p}$ is the hydrogen number density of wind
matter. The wind column density is defined
as $N_{h} = n_h r_*$.  The luminosity of the source, usually, can be deduced
from the observed spectrum. For known $\xi$ and $N_h$ (also $L$), one can estimate
$r_*$, or specifically one can guess the rough estimate of the launching
radius $r$ \cite[e.g.,][]{Gatuzz-etal2019, Miller-etal2015, Kaastra-etal2014}.
In the present model, we know the wind launching radius $r$, the wind ejection height $z^{max}$, wind density $\rho (r, z=z^{max})$ and wind speed $v_{wind}$. By comparison to the wind parameters (extracted from wind absorption features, like $N_{h}$, $\xi$), we can constrain the range of disk free parameters like $f_v$, $\dot{M}$, $\alpha$.

 X-ray spectra of LMXBs, mainly, exhibit strong absorption lines of Fe {\sc xxv} (He-like) and Fe {\sc xxvi} (H-like).    
The typical range of $N_{h}$  and $\log{\xi}$ for Fe {\sc xxv} and Fe {\sc xxvi} are  $\sim10^{21} - 10^{23} cm^{-2}$ and 3 - 6 erg cm s$^{-1}$ respectively 
\cite[e.g.,][]{Kubota-etal2007, Miller-etal2015, Chakravorty-etal2016, Gatuzz-etal2019, Trigo-Boirin2016}.
With the advantage of known wind outflow location, we define the ionization parameter $\xi$, following \citet[][]{Ross-Fabian1993}, as

\beqn\label{eq:ionpar}
\xi =\left(\frac{r_{in}}{r_1}\right)^2 \frac{F_x}{n_h},
\eeqn
where $F_x$ is the ionizing flux which is emitted from the inner region of the disk at radius $r_{in}$, $r_1$ = $\sqrt{r^2+(z^{max})^2}$ is 
the distance between ionizing source and wind matters with $r \gg r_{in}$.

The estimated blackbody temperature $T_{bb}$ (by spectral modeling) for HS state in LMXBs  varies in $\sim$1 - 1.5 keV 
whereas the bolometric luminosity is in range  $\sim 10^{37} - 10^{39}$ erg/s
\cite[e.g.,][]{King-etal2013, Miller-etal2006, Gatuzz-etal2019}.
In the Keplerian disk the above range of $T_{bb}$ can be generated at radius $r_{in}$
$\sim$ 20 - 50$R_g$ either having a radiation pressure dominated regime or having a gas pressure dominated regime with Thomson scattering as discussed by \citet{Shakura-Sunyaev1973} \cite[see also,][]{Novikov-Thorne1973}.
For above range of $T_{bb}$, and $r_{in}$, the flux $F_x$ can vary in
$10^{20} - 10^{22}$ erg $cm^{-2} s^{-1}$ with having mass accretion rate
$\dot{M}$ = 0.005-0.2 $\dot{M}_{Edd}$.
Using equation (\ref{eq:ionpar}) we estimate $n_h$  $\sim$ $10^{11} - 10^{14} cm^{-3}$ by fixing the average value of $r_{in}$ = 30$R_g$, $r_1$ = 3000$R_g$ and $F_x$ $\approx$ $10^{21}$ erg $cm^{-2} s^{-1}$ for a mentioned range of $\log{\xi}$ for Fe {\sc xxv} {\sc xxvi}.  
However, the above range for $n_h$ is  maximal in this sense that (a) few percent  ($< 10 \%$) of ionizing flux will illuminate the wind matter, (b) 
the photon can ionize Fe {\sc xxiv} and Fe {\sc xxv}, which has energy larger or equal to their ionization energy, 
where
the ionization energies for Fe {\sc xxiv} and Fe {\sc xxv} are 2.04 and 8.8 keV respectively.
For further calculations, we fix the reasonable limits of the observed $n_h$
for wind outflow to  $\sim$ $10^{9} - 10^{15} cm^{-3}$. This wide range of $n_h$ is consistent with the thermal stability curve, as the wind is thermodynamically stable (i.e, the slope of the temperature versus the pressure ($\xi/T$) curve is positive) in HS state for a wide range of wind density \cite[see][]{Chakravorty-etal2013}.

\begin{figure*}
\centering
\begin{tabular}{lr}
  \includegraphics[width=0.46\textwidth]{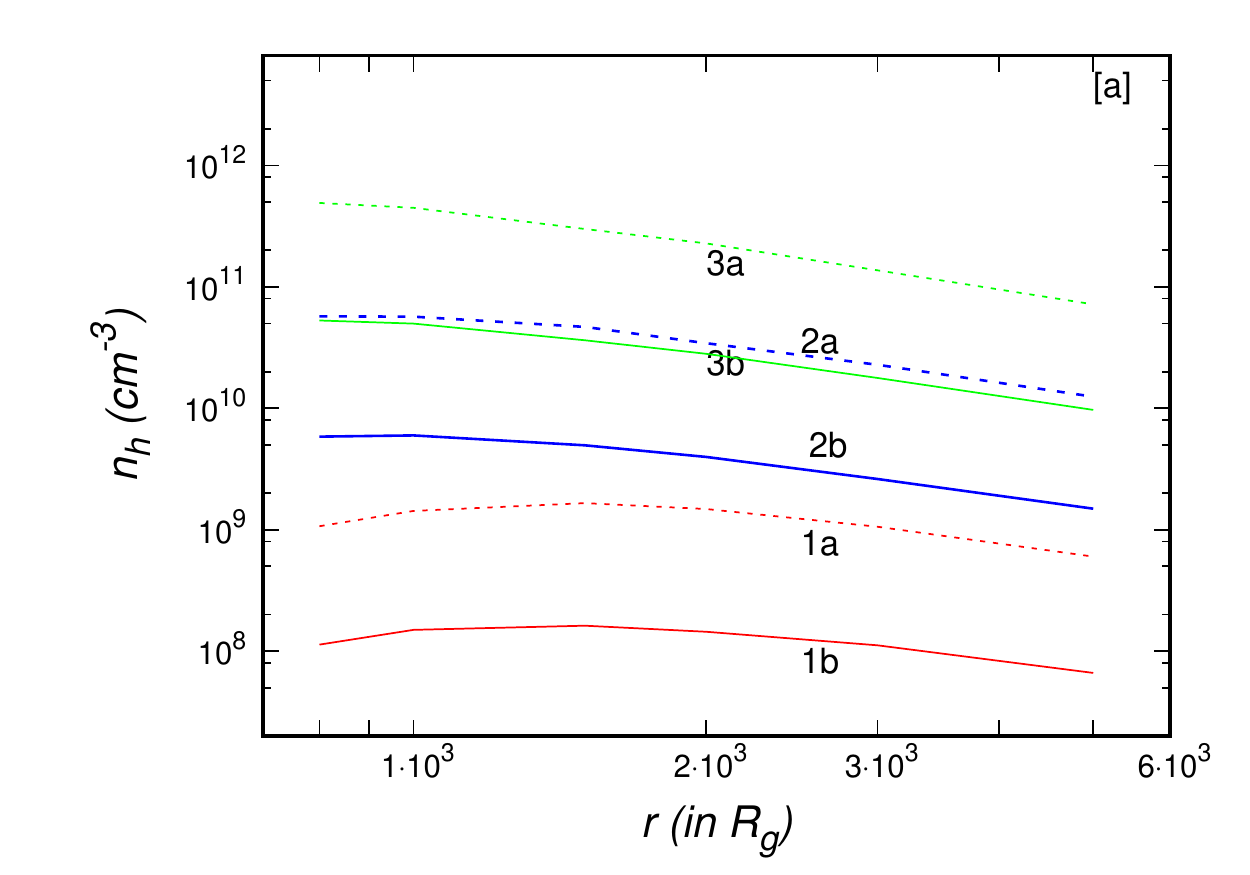}&\hspace{-0.90cm} 
  \includegraphics[width=0.46\textwidth]{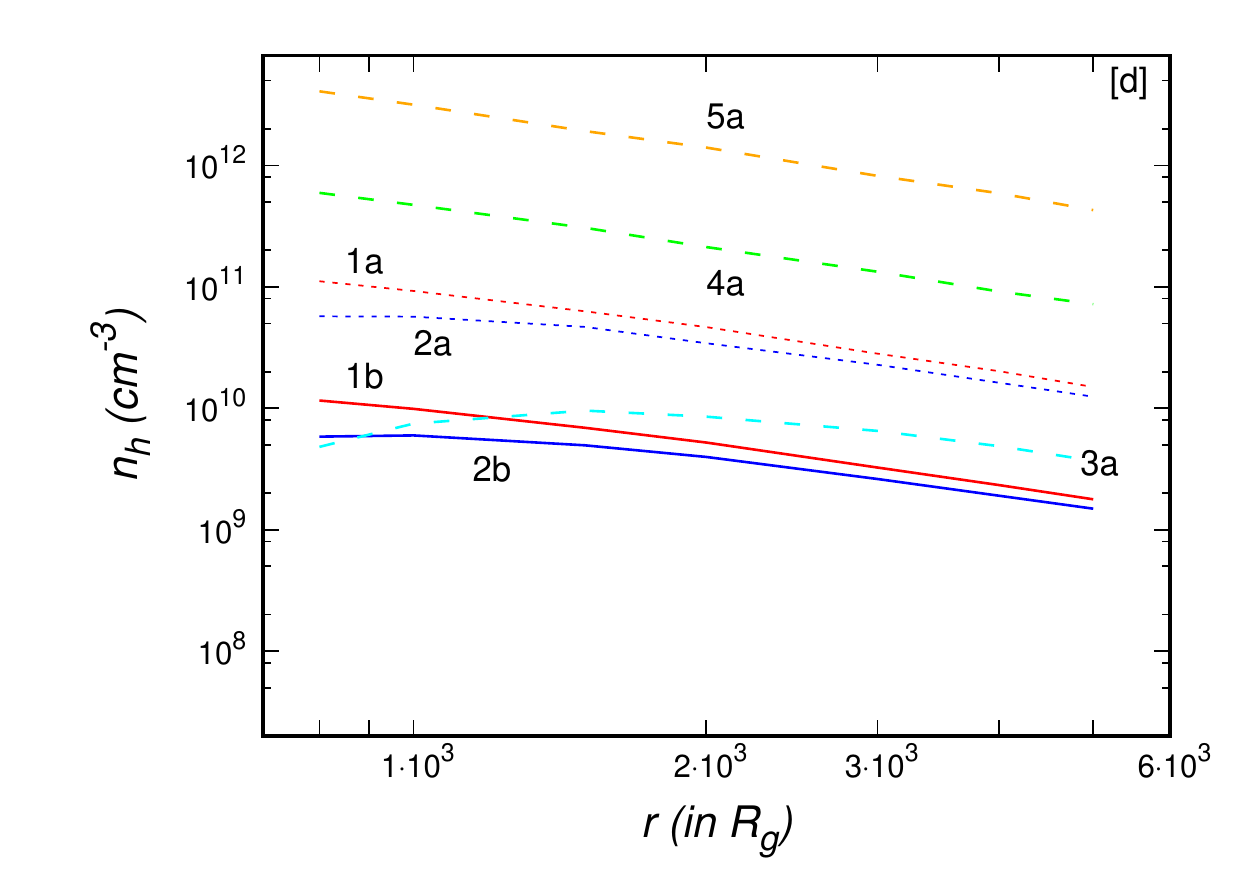}\\ 
  \includegraphics[width=0.46\textwidth]{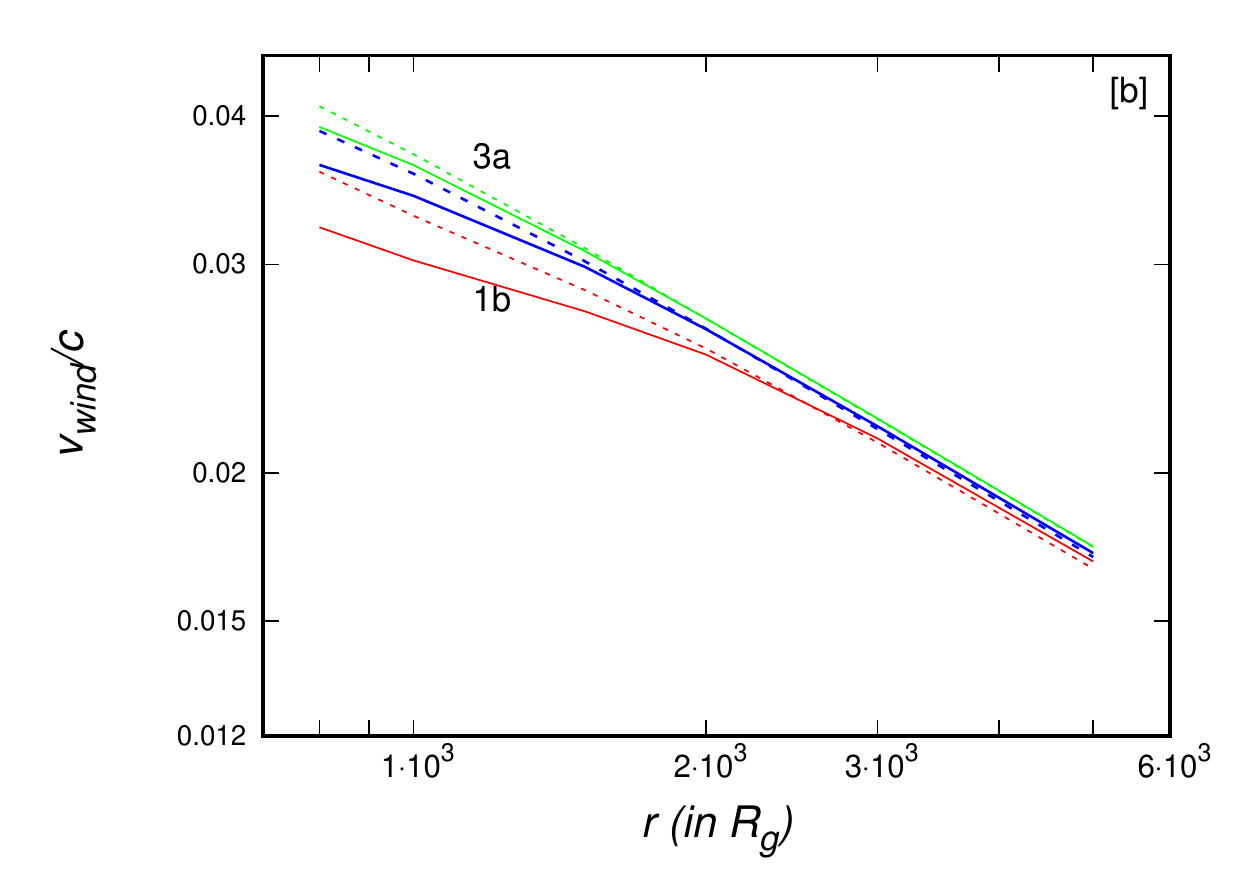}&\hspace{-0.90cm} 
  \includegraphics[width=0.46\textwidth]{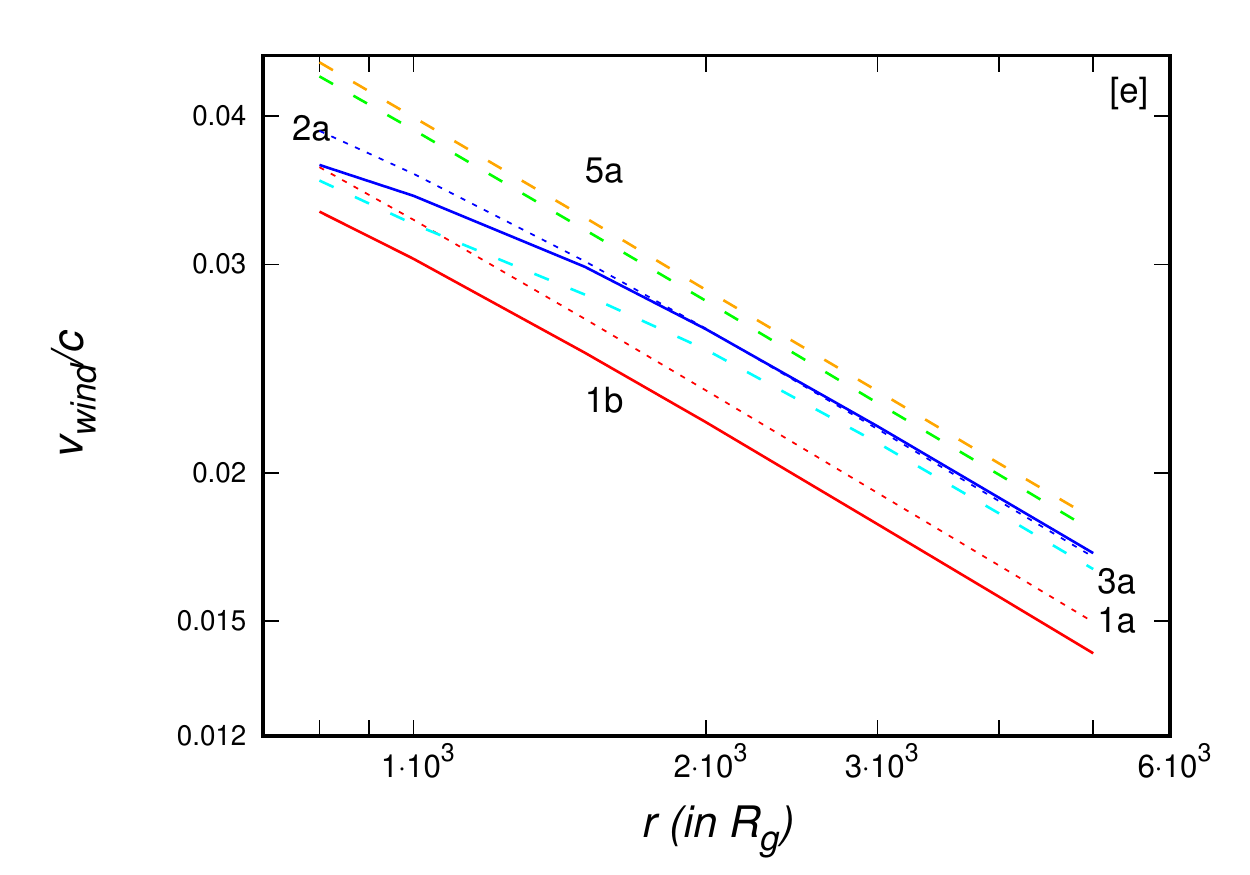}\\ 
  \includegraphics[width=0.46\textwidth]{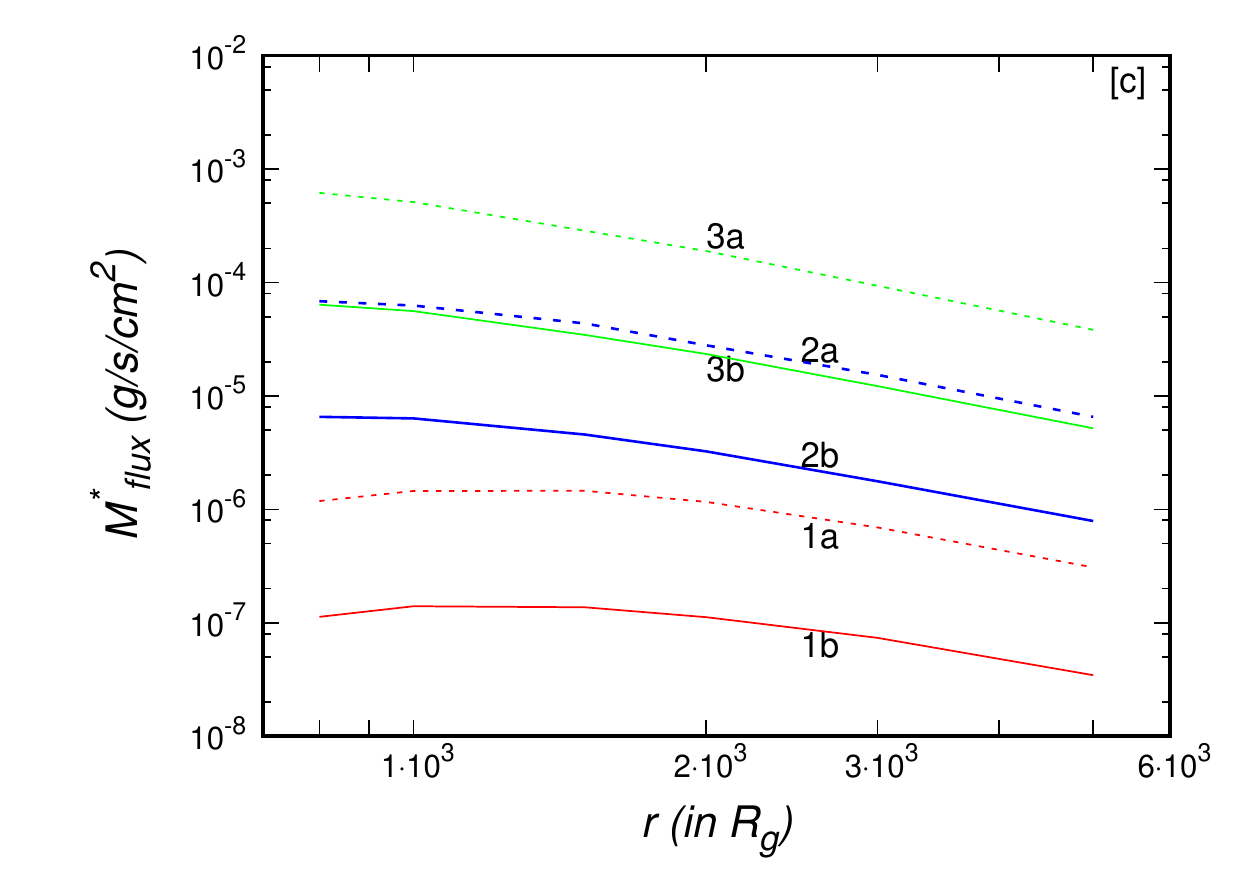}&\hspace{-0.90cm} 
  \includegraphics[width=0.46\textwidth]{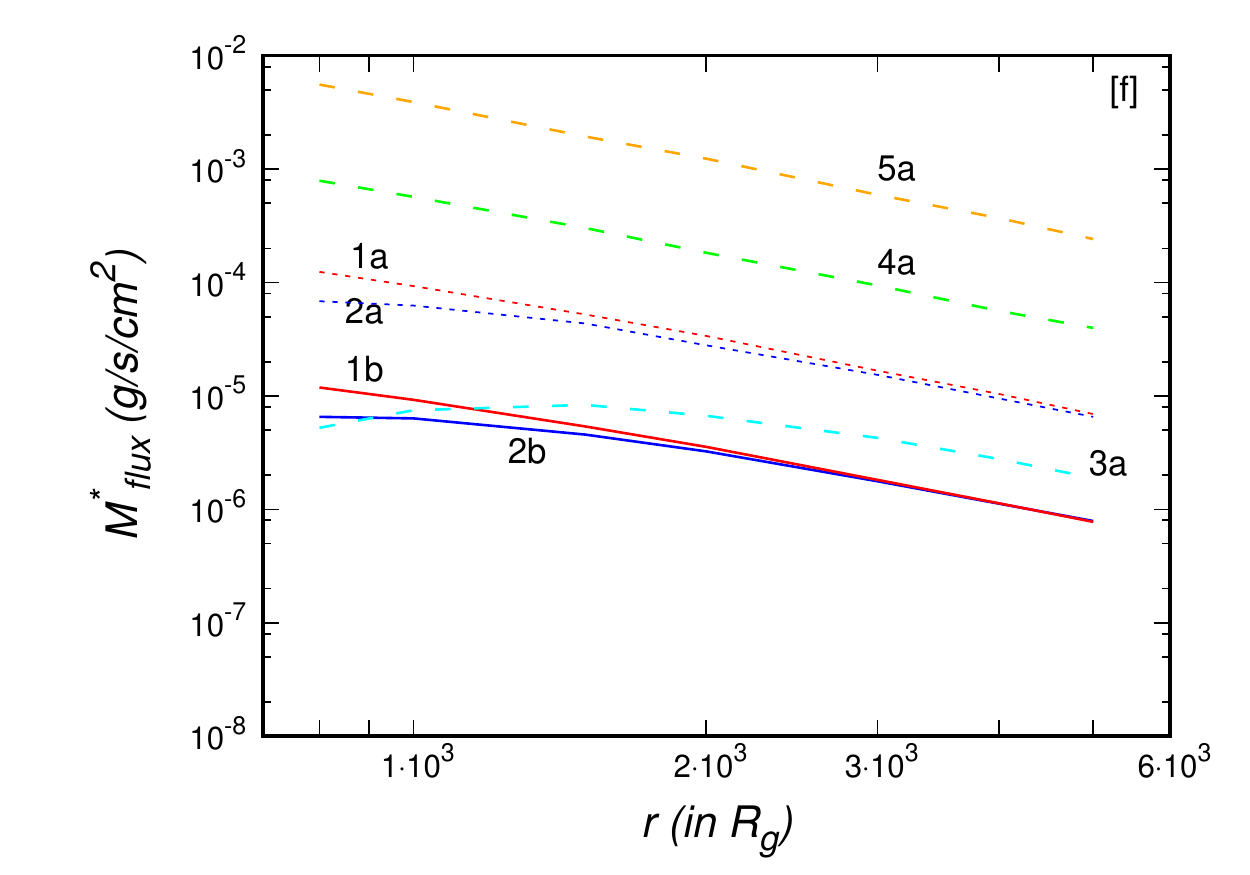}\\ 
\end{tabular}\vspace{-0.3cm}
\caption{Wind solutions for launching radius $r$ = 800 - 5000$R_g$ at two wind ejection heights $z^{max}$ = $r$ (curves marked by a) and 2$r$ (curves marked by b).
  The upper, middle and lower rows are for hydrogen number density, wind speed and mass outflow rate per unit area respectively.
  The left column is for different mass accretion rate with $\alpha$ = 0.1, $f_v$ $\sim$1, where curves marked by 1, 2 and 3 are for $\dot{M}$ = 0.005, 0.05 and
  0.2 $\dot{M}_{Edd}$ respectively.
  The right column is for different $\alpha$ and  $f_v$, in which the curve
  marked by 1 is for $\alpha$ = 0.2 and $f_v$ = 1 and the curves marked by
  2, 3 and 4 are for $f_v$ = 1, 0.4 and 3 respectively and $\alpha$ = 0.1. In 
the right column,  curve 5a is for $\dot{M}$ = 0.2$\dot{M}_{Edd}$, $f_v$ = 3, $\alpha$ = 0.1, while for other curves  $\dot{M}$ = 0.05$\dot{M}_{Edd}$.
}
\label{fig:wd-par}
\end{figure*}

The mass outflow rate for wind $\dot{M}_{out}$ is, usually, defined as \cite[][]{King-etal2013} $\dot{M}_{out}$ = $\Omega \rho r_1^2 v_{wind} \ C_v$; where $\Omega$ is the covering factor (0 $<$ $\Omega$ $<$ 4$\pi$), and $C_v$ is the line-of sight global filling factor with the assumption of non-spherical wind outflow. 
To avoid the uncertainty over numerical values of $\Omega$ and $C_v$, we calculate the mass outflow rate per unit area, mass flux rate $\dot{M}_{flux}$, which is defined as 
\beqn \label{eq:m_out}
\dot{M}_{flux} = \rho v_{wind}   = \frac{\dot{M}_{out}}{\Omega r_1^2 C_v} \eeqn

\subsubsection{Wind characteristics}
We compute the wind characteristics for a wide range of launching radius
  $r$ = 800 $-$ 5000 $R_g$ with 7 different radii $r$ = 800, 1000, 1500, 2000, 3000, 4000 and 5000$R_g$. For simplicity, we do not take an account for the 
possible decrease of mass accretion rate with decreasing $r$ due to a mass loss by  wind outflow, which we 
intend to study in future. The  wind characteristics are computed for a same mass accretion rate for the above mentioned range of $r$. 
In Figure \ref{fig:wd-par}, the upper, middle and lower rows are for the hydrogen column density $n_h$, wind speed $v_{wind}$ and wind outflow rate per unit area $\dot{M}_{flux}$ respectively. We explore the wind properties at two values of
wind ejection height, $z^{max}$ = $r$ and 2$r$, the results are shown by
curves marked with a and b respectively in all panels. 
In the left column, the curves marked as 1, 2 and 3 are for three different
mass accretion rates $\dot{M}$ = 0.005, 0.05 and 0.2 $\dot{M}_{Edd}$
respectively with $\alpha$ = 0.1, $f_v$ $\sim$ 1.
The estimated hydrogen column density $n_h$ varies between $10^8$ to $10^{12} cm^{-3}$.  
For low accretion rate, $\dot{M} < 0.005$$\dot{M}_{Edd}$,  $n_h$ is less than $10^{9} cm^{-3}$ for $z^{max} > r$ (even $z^{max} \sim r/2$), which is well below 
the observation limit.  
It hence seems that the lower accretion rate $\dot{M} < 0.005 \dot{M}_{Edd}$ is not viable to launch the observed dense wind outflow in LMXBs.
Further, to elevate $n_h$ magnitude, we increase $\alpha$ and $f_v$, the results are presented in the right column.

In the right column 
the curves marked as 1, 2, 3 and 4 are for $\dot{M}$ = 0.05$\dot{M}_{Edd}$
and the curve 5 is for $\dot{M}$ = 0.2$\dot{M}_{Edd}$. The curves 2a and 2b in
the right column are the same as the curves 2a and 2b in the left column. 
The curves marked as 2 and 1 are for $\alpha$ = 0.1 and 0.2 respectively
with fixed $f_v$ $\sim$ 1. $n_h$  enhances almost by a factor 2 by increasing
$\alpha$ from 0.1 to 0.2.  The curves 3a, 2a and 4a are for $f_v$ = 0.4, 1
and 3 respectively with fixed $\alpha$ = 0.1. The hydrogen column density
enhances almost by one order by increasing $f_v$ by a factor 3, which
can also be noticed with curve 3a of left column and curve 5a of right
column.  
In short, for a fixed lower limit of the hydrogen number density $n_h = 10^9 cm^{-3}$, the accretion rate $\dot{M} > 0.05 \dot{M}_{Edd}$ well describes
the wind properties for any $\alpha$ and $f_v$ (even with $f_v < 1$) while we
need the higher $\alpha$ and $f_v$ for $0.005\dot{M}_{Edd} < \dot{M} < 0.05 \dot{M}_{Edd}$. 

For all curves of Figure \ref{fig:wd-par}, we find $z^{max} > z^{max}_b$, 
i.e., all are representing a wind solutions,  and $z^{max} < z^{max}_p$,
i.e., the wind matter is ejected tangentially in all directions  with speed
$v_{wind}$ $\sim v_\phi$ (= 0.01 $-$ 0.04 c). 
Here, $z^{max}$ = $r$ and $2r$, 
are referring that essentially we are calculating the wind characteristics along two lines of sight $\theta_l = 45 \degr$ and $29 \degr$ respectively.    
In addition, for $z^{max} = r/2$ (or, $\theta_l \sim$ $74 \degr$) we find that,
  wind is ejected 
  with speed $v_{wind} \sim v_\phi < v_{esc} $ for $\dot{M} = 0.05 M_{Edd}$, $f_v =$  1 and $\alpha$ = 0.1. And the hydrogen number density of wind outflow
  ejected from $z^{max} = r/2$ is around 5 times greater than that ejected
  from $z^{max} = r$. 
For $\theta_l >$ 29$^\circ$, the winds are moving mainly parallel to the disk in all directions, which will lead to generate both blue and red shifted absorption
lines. However, the winds which are moving towards us are more visible than
those in the other directions.
We observe a blue shifted line with the velocity component towards our line of
sight, though in some sources a double dip absorption line profiles are also
observed in high resolved spectra \cite[e.g.,][]{Miller-etal2015}. Hence our estimated wind speed is maximal and within the observable range.     
It is also the reason (mainly due to the wind directions almost parallel to the equatorial plane of the disk, $\theta_w = 90 \degr$) that  wind is observed preferentially in high-inclination sources \cite[e.g.,][]{Ponti-etal2012}.

For $z^{max} < 2r$ or $\theta_l > 29\degr$, the wind-outflow is mainly an equatorial wind with small opening angle, which is inferred also from many observations \cite[e.g.,][]{Trigo-Boirin2016, Allen-etal2018}. We find that the wind opening angle increases from $\sim 1\degr$ to $\sim 5.5\degr$, when the wind ejection height $z^{max}$ increases from $r/2$ to $2r$ at the fixed launching radius $r = 2000R_g$. For the curve 2b ($z^{max}=2r$) of the left panel of Figure \ref{fig:wd-par}, it increases from $\sim 4\degr$ to $\sim 6\degr$ by increasing $r$ from 1000 to 5000$R_g$. Here, we define the wind opening angle $\theta_{op}$ as, $\theta_{op} = \tan^{-1}\left(v_z \left/  \sqrt{v_\phi^2+v_r^2}\right.  \right)$. In general, the wind opening angle increases with increasing either the wind ejection height, or launching radius, or mass accretion rate, or viscosity.

For non-parallel wind outflow 
($\theta_w <  90 \degr$)
we have to 
consider higher $z^{max}$ ($> 2r$) such that $z^{max} >$  $z^{max}_p$ or $v_{wind} > v_\phi$. Since the wind density decreases with increasing $z^{max}$, to achieve the observable lower limit of $n_h$ we have to go for sufficiently large 
$\dot{M}$ and $f_v$. For example, we obtain $v_{wind}$ $\approx$ 1.1 $v_\phi$ ($\theta_w \sim 77 \degr$, or $\theta_{op} \sim 13 \degr$) and 1.35 $v_\phi$ ($\theta_w \sim 66 \degr$, or $\theta_{op} \sim 24 \degr$) at $z^{max}$ = $3r$ and $4r$ respectively for  $r$ = 2000 $R_g$, $\dot{M}$ = 0.2$\dot{M}_{Edd}$, $f_v$ = 3 and $\alpha$ = 0.1. 
Hence for extreme cases, wind can be observed at larger height $z^{max} >z^{max}_p> 2r$ with wind speed greater than  $v_\phi$, thence the wind can be observed in low- inclination sources \cite[e.g.,][]{Degenaar-etal2016}. In general, in
the present model the wind outflow can be observed in high- inclination sources for
rich parameters sets (i.e., $z^{max}_b < z^{max} < 2r < z^{max}_p$)
while extreme parameter sets (i.e., $z^{max} > z^{max}_p > 2r$) are
needed to observe a wind outflow in low-inclination sources.      

The column density $N_{h}$ is measured along the line of sight. 
Here for any line of sight ($\theta_l$) the hydrogen number density $n_h$ increases with decreasing launching radius $r$, and for a given $r$,   $n_h$ decreases with decreasing $\theta_l$. 
The column density can be defined as $ N_{h}$ = $\langle n_h \rangle \langle r \rangle$, here $\langle n_h \rangle$ is an average hydrogen number density, 
$\langle r \rangle$ is an average thickness of the wind for the given line of sight. For $\langle r \rangle$ = 3000$R_g$ and $n_h \equiv$[$10^9, 10^{14} cm^{-3}$],  the column density ranges from  $10^{19}$ to $10^{24} cm^{-2}$. Hence, the estimated range for column density is within the observed range. 
Like the hydrogen column density $n_h$ (the right panel of  Figure \ref{fig:wd-par}), the acceptable range for the mass outflow rate per unit area $\dot{M}_{flux}$ is $10^{-6} - 10^{-1}$ g $s^{-1} cm^{-2}$ for $r/2 < z^{max} < 2r$. The mass outflow rate $\dot{M}_{out}$
has been computed by using equation (\ref{eq:m_out}) for above range of $\dot{M}_{flux}$. The estimated range of $\dot{M}_{out}/C_v$ for average launching radius $r$ = 3000$R_g$ is $\approx 10^{13} - 10^{18}$ g/s. %
Here, the upper limit of mass outflow rate corresponds to the lower 
limit of $z^{max} \sim r/2$. In order to obtain this upper limit 
$\dot{M}_{out}/C_v$ = $10^{18}$ g/s, the chosen other parameters are  $\dot{M}$ $\sim 4 \times 10^{18}$ g/s, $r \sim 1000R_g$, $f_v \sim 3$ and $\alpha =0.1$. In general, the mass outflow rate is comparatively less than the mass inflow rate for $z^{max} >r$, while they are comparable for $z^{max} \lesssim r/2$.
However, for comparable mass outflow and inflow rates, the wind matter density is many orders of magnitude lesser than the midplane density.
For example,  for curve 2a of Figure \ref{fig:wd-md-a} ($r = 2000 R_g$, $\dot{M} = 10^{18}$ g/s, $f_v =1$; $r/h$ = 65.2) the mass outflow rate is $\dot{M}_{out}/C_v$ $\sim 10^{16}$ and $10^{17}$ g/s and the wind matter density  $\rho/\rho_c$ $\sim 3 \times 10^{-9}$ and $3 \times 10^{-8}$ for $z^{max}$ = $r/2$ and $r/4$ respectively. 



\subsubsection{Wind power and discussions} 

The power or kinetic luminosity of the wind $L_{wind}$ is defined as
\beqn \label{eq:w_power}
L_{wind} = \frac{1}{2} \dot{M}_{out}  v_{wind}^2
\eeqn
In the present model, in the range of launching radius $r$ = 800 $-$ 5000$R_g$,
we obtain a wide range of wind speed 0.01 $< v_{wind}/c <$ 0.04. On average, the
mass outflow rate is in the range of $\approx 10^{13} - 10^{18}$ g/s.
The wind power for average launching radius $r$ $\sim$ 3000$R_g$ and average wind speed $\sim$ 0.02c is in the range of $10^{31} - 10^{37}$ erg/s. The wind power is a few orders of magnitude less than the observed luminosity
for $z^{max} > r$, however the maximum wind power is comparable to the luminosity for $z^{max} \sim r/2$. These are consistent with the observed wind power
when one considers a non-spherical wind outflow, as reported by
\citet{King-etal2013} \cite[see also,][]{Miller-etal2015, Ponti-etal2016}.

In the present model, the kinetic luminosity of wind outflow can be less than,
 greater than or comparable to the observed source luminosity depending upon the wind ejection height, e.g., for $z^{max} < r/2$, $L_{wind} > L$, on the contrary to the interpretation of \citet{Allen-etal2018}\cite[see also,][]{Ponti-etal2016}.
Since for $z^{max} < 2r$ the wind speed is mainly an azimuthal speed, $v_{wind} \sim v_\phi$; and the wind density increases with decreasing $z^{max}$ for a given $r$. 
Note, the wind ejection height will be determined by how much irradiated energy impinges upon the particular launching radius. Basically, the required energy for ejecting the wind at a height $z^{max}$ (or for a particular $x$), or the 
enhancement in the internal energy due to $x$, must be supplied by an irradiation or an external heating. 
We first compute the vertically averaged enhancement in the internal energy per unit volume per unit time due to $x$ at a given launching radius $r$ as follows:
  \begin{align} \label{eq:int_enhance}
    \epsilon_{exess}^{x} &= \left. \frac{2\pi}{z^{max} \ t_w} \int \frac{3}{2}c_s(z)^2\rho(z) dz \right|_{{\rm arbitrary}~x} \nonumber   \\  &\ \qquad - \left. \frac{2\pi}{z^{max} \ t_w}\int \frac{3}{2}c_s(z)^2\rho(z) dz \right|_{x=0},
  \end{align}  
where the second term in RHS is an internal energy per unit volume 
	without irradiation ($x=0$),  $t_w = z^{max}/\frac{1}{z^{max}}\int v_z dz$ is the time scale for the wind ejection, and $c_s(z)^2\rho(z)=3kT\rho/2\mu m_p$ is the internal energy per unit volume. Next we compute the rate of irradiated energy per unit 
volume at a launching radius $r$ by an inner region of temperature $T_{in}$ 
at a radius $r_{in}$, given by
\beqn \label{eq:irr_en}
\epsilon^{irr} =\frac{r_{in}^2}{r^2 \ h} \sigma T_{in}^4(1-\beta)C_{sph},
\eeqn
where 
$\beta$
is the albedo, 
$C_{sph}$ is a constant ($\ll$ 1) which determines how much fraction of irradiated energy (from inner region at $r_{in}$) falls normally on the surface area $2\pi r dr$ at height $h$. The radiation pressure at $r$ due to the irradiation  can be expressed as, $p_{rad}^{irr} = \frac{C_{sph} \ r_{in}^2}{r^2 \ c} \sigma T_{in}^4 $; clearly in outer region, $ p_{rad}^{irr} \ll p$.
We find that within uncertainty, like over wind-geometry,  $\epsilon_{exess}^{x}$ and $\epsilon^{irr}$ are comparable, e.g., for $r_{in}$ = 30$R_g$, $kT_{in}$ = 1keV, $r=3000 R_g$, $x \sim x^{max}$, $f_v$ =1, $\dot{M} = 0.05\dot{M}_{Edd}$. In addition, for a fixed line of sight, we find $x$ $\propto$ $1/r^2$, similar to the  flux variation over distance. Thus, the inner disk irradiation is capable for launching a wind outflow from an outer region of the Keplerian disk.


\begin{figure*}
  \centering
\begin{tabular}{l}\hspace{-0.9cm}
  \includegraphics[width=0.75\textwidth]{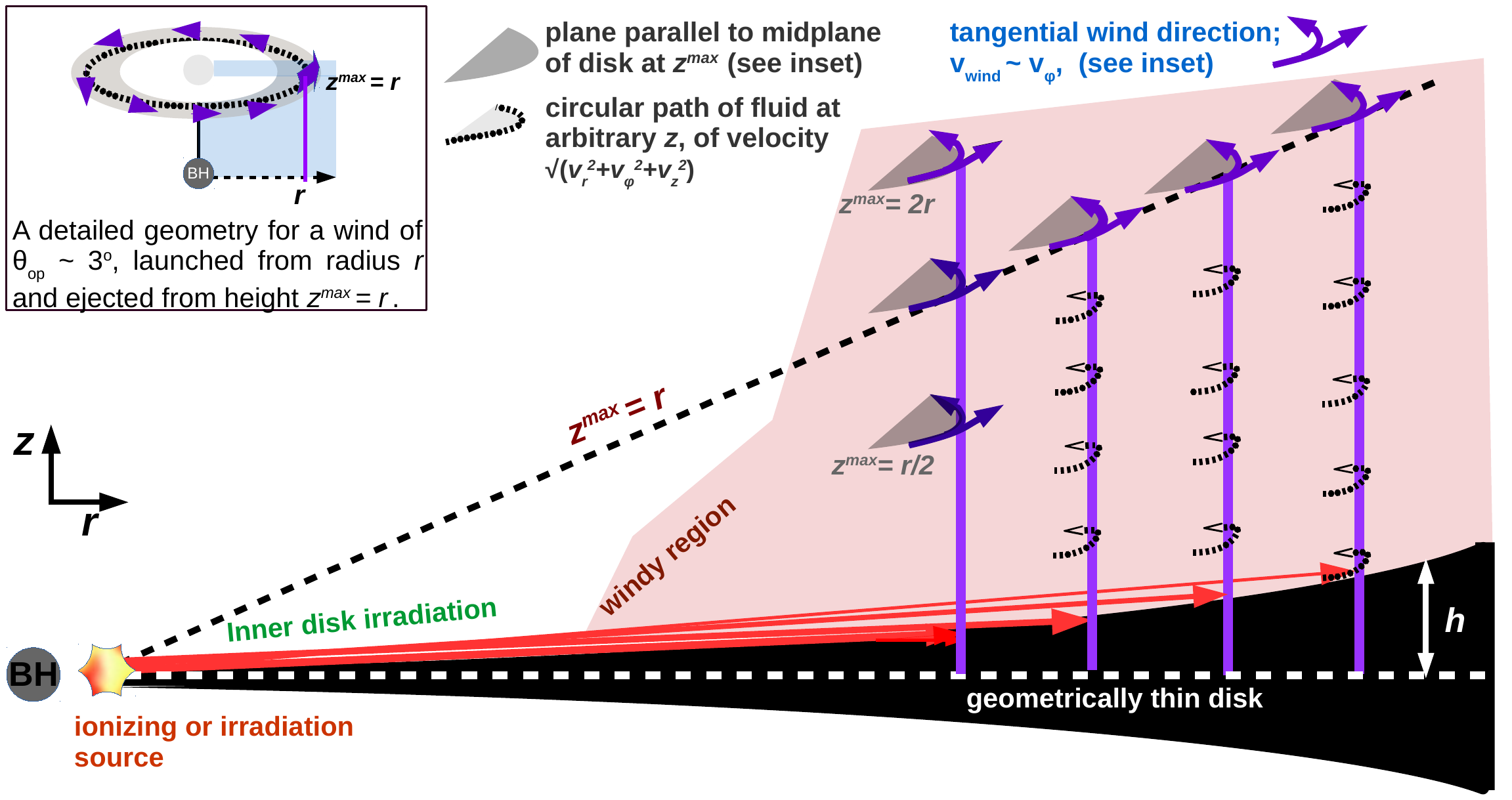}
\end{tabular}\vspace{-0.3cm}
\caption{A cross-sectional schematic view of the presented wind-outflow (driven by inner disk irradiation) model in a geometrically thin disk, i.e., in ($r,z$)-plane around a black hole (BH).
	We solve the governing equations at a fixed $r$ along the $z-$axis (which is not a streamline), and the fluids are moving along the circular path at a height $z$, shown by dotted curves, with speed 
	$\sqrt{v_r^2+v_\phi^2+v_z^2}$ dominated by $v_\phi$. However this approach reproduces solution approximately similar to that of grid-points approach (see Figure \ref{fig:cons_num_schm}).
    For a given magnitude of an external heating, the fluid reaches upto a maximum height $z^{max}$ where an equipartition of energy between internal and kinetic energies of fluid is attained.
  Above $z_f$, the radial pressure gradient acts radially inward (see e.g., Figure \ref{fig:cons_num_schm}), and if it opposes the rotation 
 significantly along with the radial gravitational force, then the fluid is ejected from $z^{max}$ with fluid velocity along the perimeter, i.e., ejected in all direction (see inset). The wind is an equatorial wind with small opening angle $\theta_{op}$ ($=tan^{-1}\big[v_z/\sqrt{v_\phi^2+v_r^2}\big]$), 
for $z^{max}<2r$; $v_{wind} \sim v_\phi$. However $\theta_{op}$ increases with increasing $z^{max}$ (see text for details).
Here, we show the wind-outflow for a fixed ejection height, $z^{max} = r$, for different $r$. Also, we show an example, where the wind is ejected from different height (i.e., for different $x$ or different magnitude of the external
heating) for a fixed $r$. 
}
\label{fig:wind_geo}
\end{figure*}

In the present work, we have explored the thermal irradiation induced wind outflow model in a geometrically thin disk.  
We find an equatorial wind with a small
opening angle. A schematic diagram of the present wind-outflow model
has been shown in Figure \ref{fig:wind_geo}.
Apart from the magnetically driven wind, the disk emission line can potentially
  launch the wind. In LMXBs, however, the line driven wind is not possible \cite[][]{Proga-Kallman2002}. Recently, \cite{Giustini-Proga2019} have shown that the line driven
  wind is also not possible in low-luminous AGNs (LLAGNs), and in general those AGNs with black hole mass $M_{BH} < 10^8 M_\odot$ and mass accretion rate $\dot{M} < 10^{-2}M_{Edd}$. The present model is more applicable for LMXBs  and LLAGNs.
We are in the process of extending this model for LLAGNs, mainly to emphasize that the fraction of decrement in mass accretion rate occurs in the outer region of the disk of LLAGNs (or, thin disk) (Kumar $\&$ Mukhopadhyay 2020, in preparation). 
The wind-outflow launches
close to the black hole, with a lower bound $r = 800 R_g$ (almost two orders of magnitude
less than the
Compton radius $R_{IC}$ of thermal-wind model for $10^8$K Compton temperature),
which is favorable for a dense
outflow \cite[e.g.,][and references therein]{Reynolds2012, Neilsen-2013}.
However, \cite{Done-etal2018} have modified the thermal-wind model with
the inclusion of radiation pressure and argued that Compton radius will 
decrease when the source luminosity becomes comparable to the Eddington 
luminosity.
Moreover, there is a thermal wind model where wind starts to launch very far away from the black hole almost around the Bondi radius (where the gravitational pull of the black hole is comparable to the internal thermal energy of the gas) (e.g., \citealt{Dyda-etal2017}, see also \citealt{Clarke-Alexander2016, Ballabio-etal2020}).

The present wind-outflow solutions are always subsonic in contrast to the thermal-wind model \cite[see, for a general discussion on sonic points for
  disk winds (thermal-wind),][]{Waters-Proga2012}. Although, like thermal-wind model, we have a critical
point for
$v_r^2 \rightarrow \Gamma_1 c_s^2$ with $v_r \sim v_z$, as discussed in point (d) of \S \ref{sub:sol}, at this critical point the fluid arrives at an isobaric regime and further there is no acceleration.
However in our model, due to the irradiation the internal energy or the sound speed of the medium increases with height. 
In addition, the radiation pressure due to irradiation in the outer region is negligible in comparison with the gas pressure, hence we do not include the 
radiation pressure term in the governing equations \cite[see for the radiation pressure term, e.g.,][]{Dannen-etal2020}. Note that
in the present model, the wind outflow medium is turbulent. However, \cite{Woods-etal1996} assumed that the base of wind outflow is above the disk midplane, mainly to avoid the uncertainties over the viscosity \cite[see also,][]{Proga-Kallman2002}. Importantly, for a viscous flow, the Bernoulli parameter along the particle trajectory (or streamline) is not constant \cite[e.g.,][]{Yuan-etal2015}. 
In the existing literature, the common approach for a wind solution is to assume a fixed streamline \cite[see, e.g.,][]{Begelman-etal1983, Waters-Proga2012}. 
However, we solve the governing equations along the $z$-axis for a fixed launching radius of wind
and naturally obtain an observed equatorial wind with small opening angle (see Figure \ref{fig:wind_geo}).

\section{Summary}%

We have formulated a steady, axisymmetric disk in cylindrical coordinates
and solved for wind outflow solutions along the vertical axis at a given launching
radius from the midplane. We have assumed a tiny vertical speed $v_z$, which
is some small factor $f_v$ of the radial speed $v_r$ and very less compared
to the sound speed $c_s$; $v_z = f_v v_r \ll c_s$, at the launching radius. We
have included the viscous effects by considering both tangential shearing
stresses $W_{\phi r}$ and $W_{\phi z}$ and assumed the other shearing stress
negligible compared to the tangential shearing stress, i.e., $W_{rz} \sim 0$. 
We have incorporated the external heating
in vertical hydrostatic 
equation, as an effect that the flows are not in vertical mechanical
equilibrium and it is parameterized by a number $x$, where $x$ = 0 stands for
a hydrostatic equilibrium. The primary source of external heating is the
irradiation by the inner
disk. 
We have also taken an account, the effect of radial pressure gradient (in
addition to the radial component of gravitational force $F_r$) on rotations of
the fluids.  Like the Keplerian disks, we have assumed that viscous
generated heat immediately radiates out vertically by blackbody emission
(i.e., the medium is optically thick). 
With having $v_z \ll c_s$ at the launching radius, the present framework
reduces to the Keplerian disk, at least, near to the midplane of the disk.
Hence, we initialize the flow variables with their respective  Keplerian
values at a given $r$, at which the pressure is gas dominated and the opacity
comes mainly from the free-free absorptions. 
We have compared  the model predicted vertical structure for $x$ = 0 to the
Keplerian disk, and found 
that like Keplerian disk the pressure and density profiles follow an
isothermal profile but with different scale heights. %

We have obtained 
an acceleration solutions 
for a finite range of $x$ $\equiv$[0, $x^{max}$] for a given $f_v$, and it
accelerates upto
a maximum height (termed as $z^{max}$) for a given $x$. At $z^{max}$, $v_z$
and $v_r$ are comparable to the sound speed of the medium $c_s$, which
signifies that we reach at an isobaric phase, i.e.,  
above $z^{max}$ there is no pressure gradient. As well as it assures that the
wind outflow is thermally driven.
The quantity
$z^{max}$ increases with $x$.
The accessible range of $x$ (for an acceleration) increases 
with increasing $f_v$, thus both $x$ and initial vertical speed are intimately
related with the external heating.
We have observed that by increasing $x$, the pressure profile in the vertical
direction remains to be isothermal profile, only pressure scale height
increases with $x$, while
the density profile transits from the isothermal profile to the isobaric
profile and its scale height decreases to the lowest value
at $x^{max}$.
We have found that the radial pressure gradient flips the sign to positive
(acts inwardly or opposes the rotations) around pressure scale height. 
For sufficiently larger $z^{max}$, it becomes comparable to the radial
gravitational force, and above
$z^{max}$, 
$F_r$ cannot balance the rotational effect alone, eventually the fluid matters are
blown off
with speed $v_{wind} (=\sqrt{v_r^2+v_z^2+v_\phi^2})$.
In general for $\frac{\partial p}{\partial r} \ll F_r$ at
$z^{max}$,  the matter is rotationally bound otherwise unbound. 

We have found that the wind outflow can be launched easily from the outer
region of the disk ($> 800 R_g$). The quantities $z^{max}_b$
(a minimum $z^{max}$ where radial pressure gradient is comparable to the
radial gravitational force), $z^{max}_p$
(a minimum $z^{max}$ where $v_{wind} >$ $v_\phi$), and  $z^{max}_e$ (a minimum
$z^{max}$ where $v_{wind} >$ $v_{esc}$) 
decrease with increasing launching radius $r$. Moreover, the heights
$z^{max}_p$ and  $z^{max}_e$ decrease with increasing $\dot{m}$, while
$z^{max}_e$ increases with increasing viscosity parameter $\alpha$. Hence the
increment of accretion rate helps the wind launching while the increment of
$\alpha$ is not. The density at a given $z^{max}$ increases
with increasing $\dot{m}$, $\alpha$ and $f_v$.

We have explored the wind characteristics for two ejection
heights $z^{max}$ = $r$ and $2r$ (or for two line of sights $\theta_l = 45
\degr$ and $29 \degr$ respectively) for launching radius range $r \equiv$
800 $-$ 5000$R_g$. We have found that both the ejection heights are far
below to $z^{max}_p$ ($v_{wind} \sim v_\phi$), while for some cases it is 
higher than  $z^{max}_e$ ($v_{wind} > v_{esc}$).
Hence, For $z^{max}$ $< 2r$ or $\theta_l > 29 \degr$, 
  the winds are
  ejected tangentially or parallel to the equatorial plane of the disk in all directions with speed $v_\phi$ ($\sim$0.01c $-$ 0.04c), which explains mainly two
  things: (a) the winds are preferentially observed in
high-inclination sources, (b) formation of red and blue shifted absorption line profiles; the double dipped absorption lines of Fe {\sc xxv}, {\sc xxvi} have been observed in high resolved spectra of a few LMXBs. 
However, in the present model the wind can  also be observed in low-inclination 
sources if it is ejected from the larger height $z^{max} > 2r$, in this case
$v_{wind} > v_\phi$.
The wind hydrogen density decreases with increasing $r$ for a given line of
sight and it decreases with decreasing line of sight for a given $r$.

We have estimated a range for wind hydrogen density, for known ionizing flux
from the observation, wind location from the present model, 
ionization parameter of Fe {\sc xxv}, {\sc xxvi}, which is $10^9
< n_h/cm^{-3} < 10^{15}$.  
For $n_h$ $> 10^9 cm^{-3}$, 
the accretion rate $\dot{M} >$ 0.05$\dot{M}_{Edd}$ well  describes
the wind properties for any $\alpha$ and $f_v$, but for accretion
rate  0.005$\dot{M}_{Edd} < \dot{M} < 0.05\dot{M}_{Edd}$, one needs
a larger $\alpha$
and $f_v$, and with $\dot{M} < 0.005\dot{M}_{Edd}$ one cannot produce high dense wind.
The estimated maximum possible mass outflow rate is a few
factors less than the mass inflow rate for $z^{max} > r/2$. 
The maximum wind power is a
few orders of magnitude less than the observed luminosity of the source
when wind is ejected from a higher height ($z^{max} > r$), while they are
comparable for $z^{max} = r/2$.


\section*{Acknowledgements}
NK is supported by University Grant Commission (UGC), New Delhi, India through Dr. D.S. Kothari Post-Doctoral Fellowship (201718-PH/17-18/0013). 
The work is partly supported by 
a project of Department of Science and Technology (DST), India, with Grant No. DSTO/PPH/BMP/1946 (EMR/2017/001226).

\subsection*{Data availability} No datasets are analysed.

\def\aap{A\&A}%
\def\aapr{A\&A~Rev.}%
\def\aaps{A\&AS}%
\def\aj{AJ}%
\def\actaa{Acta Astron.}%
\def\araa{ARA\&A}%
\def\apj{ApJ}%
\def\apjl{ApJ}%
\def\apjs{ApJS}%
\def\apspr{Astrophys.~Space~Phys.~Res.}%
\def\ao{Appl.~Opt.}%
\def\aplett{Astrophys.~Lett.}%
\def\apss{Ap\&SS}%
\def\azh{AZh}%
\def\bain{Bull.~Astron.~Inst.~Netherlands}%
\def\baas{BAAS}%
\def\bac{Bull. astr. Inst. Czechosl.}%
\def\caa{Chinese Astron. Astrophys.}%
\def\cjaa{Chinese J. Astron. Astrophys.}%
\def\fcp{Fund.~Cosmic~Phys.}%
\def\gafd{Geophys.\ Astrophys.\ Fluid Dyn.}
\def\gca{Geochim.~Cosmochim.~Acta}%
\def\grl{Geophys.~Res.~Lett.}%
\def\iaucirc{IAU~Circ.}%
\def\icarus{Icarus}%
\def\jcap{J. Cosmology Astropart. Phys.}%
\def\jcp{J.~Chem.~Phys.}%
\def\jfm{JFM}
\def\jgr{J.~Geophys.~Res.}%
\def\jqsrt{J.~Quant.~Spec.~Radiat.~Transf.}%
\def\jrasc{JRASC}%
\def\mnras{MNRAS}%
\def\memras{MmRAS}%
\def\memsai{Mem.~Soc.~Astron.~Italiana}%
\def\na{New A}%
\def\nar{New A Rev.}%
\def\nat{Nature}%
\def\natas{Nature Astronomy}%
\def\nphysa{Nucl.~Phys.~A}%
\def\pasa{PASA}%
\def\pasj{PASJ}%
\def\pasp{PASP}%
\def\physrep{Phys.~Rep.}%
\def\physscr{Phys.~Scr}%
\def\planss{Planet.~Space~Sci.}%
\def\pra{Phys.~Rev.~A}%
\def\prb{Phys.~Rev.~B}%
\def\prc{Phys.~Rev.~C}%
\def\prd{Phys.~Rev.~D}%
\def\pre{Phys.~Rev.~E}%
\def\prl{Phys.~Rev.~Lett.}%
\def\procspie{Proc.~SPIE}%
\def\qjras{QJRAS}%
\def\rmxaa{Rev. Mexicana Astron. Astrofis.}%
\def\sgg{Stud.\ Geoph.\ et\ Geod.}
\def\skytel{S\&T}%
\def\solphys{Sol.~Phys.}%
\def\sovast{Soviet~Ast.}%
\def\ssr{Space~Sci.~Rev.}%
\def\zap{ZAp}%
\def\memsai{Memorie della Societa Astronomica Italiana}

\bibliographystyle{mnras}  

\bibliography{wind_1}

\begin{thebibliography}{}
\makeatletter
\relax
\def\mn@urlcharsother{\let\do\@makeother \do\$\do\&\do\#\do\^\do\_\do\%\do\~}
\def\mn@doi{\begingroup\mn@urlcharsother \@ifnextchar [ {\mn@doi@}
  {\mn@doi@[]}}
\def\mn@doi@[#1]#2{\def\@tempa{#1}\ifx\@tempa\@empty \href
  {http://dx.doi.org/#2} {doi:#2}\else \href {http://dx.doi.org/#2} {#1}\fi
  \endgroup}
\def\mn@eprint#1#2{\mn@eprint@#1:#2::\@nil}
\def\mn@eprint@arXiv#1{\href {http://arxiv.org/abs/#1} {{\tt arXiv:#1}}}
\def\mn@eprint@dblp#1{\href {http://dblp.uni-trier.de/rec/bibtex/#1.xml}
  {dblp:#1}}
\def\mn@eprint@#1:#2:#3:#4\@nil{\def\@tempa {#1}\def\@tempb {#2}\def\@tempc
  {#3}\ifx \@tempc \@empty \let \@tempc \@tempb \let \@tempb \@tempa \fi \ifx
  \@tempb \@empty \def\@tempb {arXiv}\fi \@ifundefined
  {mn@eprint@\@tempb}{\@tempb:\@tempc}{\expandafter \expandafter \csname
  mn@eprint@\@tempb\endcsname \expandafter{\@tempc}}}

\bibitem[\protect\citeauthoryear{{Alexander}, {Clarke}  \&
  {Pringle}}{{Alexander} et~al.}{2006}]{Alexander-etal2006}
{Alexander} R.~D.,  {Clarke} C.~J.,   {Pringle} J.~E.,  2006, \mn@doi [\mnras]
  {10.1111/j.1365-2966.2006.10293.x}, \href
  {https://ui.adsabs.harvard.edu/abs/2006MNRAS.369..216A} {369, 216}

\bibitem[\protect\citeauthoryear{{Allen}, {Schulz}, {Homan}, {Neilsen}, {Nowak}
   \& {Chakrabarty}}{{Allen} et~al.}{2018}]{Allen-etal2018}
{Allen} J.~L.,  {Schulz} N.~S.,  {Homan} J.,  {Neilsen} J.,  {Nowak} M.~A.,
  {Chakrabarty} D.,  2018, \mn@doi [\apj] {10.3847/1538-4357/aac2d1}, \href
  {https://ui.adsabs.harvard.edu/abs/2018ApJ...861...26A} {861, 26}

\bibitem[\protect\citeauthoryear{{Ballabio}, {Alexander}  \&
  {Clarke}}{{Ballabio} et~al.}{2020}]{Ballabio-etal2020}
{Ballabio} G.,  {Alexander} R.~D.,   {Clarke} C.~J.,  2020, \mn@doi [\mnras]
  {10.1093/mnras/staa1767}, \href
  {https://ui.adsabs.harvard.edu/abs/2020MNRAS.496.2932B} {496, 2932}

\bibitem[\protect\citeauthoryear{{Begelman}, {McKee}  \& {Shields}}{{Begelman}
  et~al.}{1983}]{Begelman-etal1983}
{Begelman} M.~C.,  {McKee} C.~F.,   {Shields} G.~A.,  1983, \mn@doi [\apj]
  {10.1086/161178}, \href
  {https://ui.adsabs.harvard.edu/abs/1983ApJ...271...70B} {271, 70}

\bibitem[\protect\citeauthoryear{{Bhattacharya}, {Ghosh}  \&
  {Mukhopadhyay}}{{Bhattacharya} et~al.}{2010}]{Bhattacharya-etal2010}
{Bhattacharya} D.,  {Ghosh} S.,   {Mukhopadhyay} B.,  2010, \mn@doi [\apj]
  {10.1088/0004-637X/713/1/105}, \href
  {https://ui.adsabs.harvard.edu/abs/2010ApJ...713..105B} {713, 105}

\bibitem[\protect\citeauthoryear{{Bisnovatyi-Kogan} \&
  {Lovelace}}{{Bisnovatyi-Kogan} \& {Lovelace}}{2001}]{Bisnovatyi-Lovelace2001}
{Bisnovatyi-Kogan} G.~S.,  {Lovelace} R.~V.~E.,  2001, \mn@doi [\nar]
  {10.1016/S1387-6473(01)00146-4}, \href
  {https://ui.adsabs.harvard.edu/abs/2001NewAR..45..663B} {45, 663}

\bibitem[\protect\citeauthoryear{{Chakrabarti} \& {Titarchuk}}{{Chakrabarti} \&
  {Titarchuk}}{1995}]{Chakrabarti-Titarchuk1995}
{Chakrabarti} S.,  {Titarchuk} L.~G.,  1995, \mn@doi [\apj] {10.1086/176610},
  \href {https://ui.adsabs.harvard.edu/abs/1995ApJ...455..623C} {455, 623}

\bibitem[\protect\citeauthoryear{{Chakravorty}, {Lee}  \&
  {Neilsen}}{{Chakravorty} et~al.}{2013}]{Chakravorty-etal2013}
{Chakravorty} S.,  {Lee} J.~C.,   {Neilsen} J.,  2013, \mn@doi [\mnras]
  {10.1093/mnras/stt1593}, \href
  {https://ui.adsabs.harvard.edu/abs/2013MNRAS.436..560C} {436, 560}

\bibitem[\protect\citeauthoryear{{Chakravorty} et~al.,}{{Chakravorty}
  et~al.}{2016}]{Chakravorty-etal2016}
{Chakravorty} S.,  et~al., 2016, \mn@doi [\aap] {10.1051/0004-6361/201527163},
  \href {https://ui.adsabs.harvard.edu/abs/2016A&A...589A.119C} {589, A119}

\bibitem[\protect\citeauthoryear{{Clarke} \& {Alexander}}{{Clarke} \&
  {Alexander}}{2016}]{Clarke-Alexander2016}
{Clarke} C.~J.,  {Alexander} R.~D.,  2016, \mn@doi [\mnras]
  {10.1093/mnras/stw1178}, \href
  {https://ui.adsabs.harvard.edu/abs/2016MNRAS.460.3044C} {460, 3044}

\bibitem[\protect\citeauthoryear{{Dannen}, {Proga}, {Waters}  \&
  {Dyda}}{{Dannen} et~al.}{2020}]{Dannen-etal2020}
{Dannen} R.~C.,  {Proga} D.,  {Waters} T.,   {Dyda} S.,  2020, \mn@doi [\apjl]
  {10.3847/2041-8213/ab87a5}, \href
  {https://ui.adsabs.harvard.edu/abs/2020ApJ...893L..34D} {893, L34}

\bibitem[\protect\citeauthoryear{{Degenaar} et~al.,}{{Degenaar}
  et~al.}{2016}]{Degenaar-etal2016}
{Degenaar} N.,  et~al., 2016, \mn@doi [\mnras] {10.1093/mnras/stw1593}, \href
  {https://ui.adsabs.harvard.edu/abs/2016MNRAS.461.4049D} {461, 4049}

\bibitem[\protect\citeauthoryear{{D{\'\i}az Trigo} \& {Boirin}}{{D{\'\i}az
  Trigo} \& {Boirin}}{2016}]{Trigo-Boirin2016}
{D{\'\i}az Trigo} M.,  {Boirin} L.,  2016, \mn@doi [Astronomische Nachrichten]
  {10.1002/asna.201612315}, \href
  {https://ui.adsabs.harvard.edu/abs/2016AN....337..368D} {337, 368}

\bibitem[\protect\citeauthoryear{{D{\'\i}az Trigo}, {Migliari}, {Miller-Jones}
  \& {Guainazzi}}{{D{\'\i}az Trigo} et~al.}{2014}]{trigo-etal2014}
{D{\'\i}az Trigo} M.,  {Migliari} S.,  {Miller-Jones} J.~C.~A.,   {Guainazzi}
  M.,  2014, \mn@doi [\aap] {10.1051/0004-6361/201424554}, \href
  {https://ui.adsabs.harvard.edu/abs/2014A&A...571A..76D} {571, A76}

\bibitem[\protect\citeauthoryear{{Done}, {Gierli{\'n}ski}  \& {Kubota}}{{Done}
  et~al.}{2007}]{Done-etal2007}
{Done} C.,  {Gierli{\'n}ski} M.,   {Kubota} A.,  2007, \mn@doi [\aapr]
  {10.1007/s00159-007-0006-1}, \href
  {http://adsabs.harvard.edu/abs/2007A%26ARv..15....1D} {15, 1}

\bibitem[\protect\citeauthoryear{{Done}, {Tomaru}  \& {Takahashi}}{{Done}
  et~al.}{2018}]{Done-etal2018}
{Done} C.,  {Tomaru} R.,   {Takahashi} T.,  2018, \mn@doi [\mnras]
  {10.1093/mnras/stx2400}, \href
  {https://ui.adsabs.harvard.edu/abs/2018MNRAS.473..838D} {473, 838}

\bibitem[\protect\citeauthoryear{{Dunn}, {Fender}, {K{\"o}rding}, {Belloni}  \&
  {Cabanac}}{{Dunn} et~al.}{2010}]{Dunn-etal2010}
{Dunn} R.~J.~H.,  {Fender} R.~P.,  {K{\"o}rding} E.~G.,  {Belloni} T.,
  {Cabanac} C.,  2010, \mn@doi [\mnras] {10.1111/j.1365-2966.2010.16114.x},
  \href {https://ui.adsabs.harvard.edu/abs/2010MNRAS.403...61D} {403, 61}

\bibitem[\protect\citeauthoryear{{Dyda}, {Dannen}, {Waters}  \& {Proga}}{{Dyda}
  et~al.}{2017}]{Dyda-etal2017}
{Dyda} S.,  {Dannen} R.,  {Waters} T.,   {Proga} D.,  2017, \mn@doi [\mnras]
  {10.1093/mnras/stx406}, \href
  {https://ui.adsabs.harvard.edu/abs/2017MNRAS.467.4161D} {467, 4161}

\bibitem[\protect\citeauthoryear{{Frank}, {King}  \& {Raine}}{{Frank}
  et~al.}{2002}]{Frank-etal2002}
{Frank} J.,  {King} A.,   {Raine} D.~J.,  2002, {Accretion Power in
  Astrophysics: Third Edition}

\bibitem[\protect\citeauthoryear{{Gatuzz}, {D{\'\i}az Trigo}, {Miller-Jones}
  \& {Migliari}}{{Gatuzz} et~al.}{2019}]{Gatuzz-etal2019}
{Gatuzz} E.,  {D{\'\i}az Trigo} M.,  {Miller-Jones} J.~C.~A.,   {Migliari} S.,
  2019, \mn@doi [\mnras] {10.1093/mnras/sty2850}, \href
  {https://ui.adsabs.harvard.edu/abs/2019MNRAS.482.2597G} {482, 2597}

\bibitem[\protect\citeauthoryear{{Ghosh} \& {Mukhopadhyay}}{{Ghosh} \&
  {Mukhopadhyay}}{2009}]{Ghosh-Mukhopadhyay2009}
{Ghosh} S.,  {Mukhopadhyay} B.,  2009, \mn@doi [Research in Astronomy and
  Astrophysics] {10.1088/1674-4527/9/2/005}, \href
  {https://ui.adsabs.harvard.edu/abs/2009RAA.....9..157G} {9, 157}

\bibitem[\protect\citeauthoryear{{Giustini} \& {Proga}}{{Giustini} \&
  {Proga}}{2019}]{Giustini-Proga2019}
{Giustini} M.,  {Proga} D.,  2019, \mn@doi [\aap]
  {10.1051/0004-6361/201833810}, \href
  {https://ui.adsabs.harvard.edu/abs/2019A&A...630A..94G} {630, A94}

\bibitem[\protect\citeauthoryear{{Higginbottom}, {Knigge}, {Long}, {Matthews},
  {Sim}  \& {Hewitt}}{{Higginbottom} et~al.}{2018}]{Higginbottom-etal2018}
{Higginbottom} N.,  {Knigge} C.,  {Long} K.~S.,  {Matthews} J.~H.,  {Sim}
  S.~A.,   {Hewitt} H.~A.,  2018, \mn@doi [\mnras] {10.1093/mnras/sty1599},
  \href {https://ui.adsabs.harvard.edu/abs/2018MNRAS.479.3651H} {479, 3651}

\bibitem[\protect\citeauthoryear{{Homan}, {Neilsen}, {Allen}, {Chakrabarty},
  {Fender}, {Fridriksson}, {Remillard}  \& {Schulz}}{{Homan}
  et~al.}{2016}]{Homan-etal2016}
{Homan} J.,  {Neilsen} J.,  {Allen} J.~L.,  {Chakrabarty} D.,  {Fender} R.,
  {Fridriksson} J.~K.,  {Remillard} R.~A.,   {Schulz} N.,  2016, \mn@doi
  [\apjl] {10.3847/2041-8205/830/1/L5}, \href
  {https://ui.adsabs.harvard.edu/abs/2016ApJ...830L...5H} {830, L5}

\bibitem[\protect\citeauthoryear{{Jiao} \& {Wu}}{{Jiao} \&
  {Wu}}{2011}]{Jiao-Wu2011}
{Jiao} C.-L.,  {Wu} X.-B.,  2011, \mn@doi [\apj] {10.1088/0004-637X/733/2/112},
  \href {https://ui.adsabs.harvard.edu/abs/2011ApJ...733..112J} {733, 112}

\bibitem[\protect\citeauthoryear{{Kaastra} et~al.,}{{Kaastra}
  et~al.}{2014}]{Kaastra-etal2014}
{Kaastra} J.~S.,  et~al., 2014, arXiv e-prints, \href
  {https://ui.adsabs.harvard.edu/abs/2014arXiv1412.1171K} {p. arXiv:1412.1171}

\bibitem[\protect\citeauthoryear{{King}, {Pringle}  \& {Livio}}{{King}
  et~al.}{2007}]{King-etal2007}
{King} A.~R.,  {Pringle} J.~E.,   {Livio} M.,  2007, \mn@doi [\mnras]
  {10.1111/j.1365-2966.2007.11556.x}, \href
  {https://ui.adsabs.harvard.edu/abs/2007MNRAS.376.1740K} {376, 1740}

\bibitem[\protect\citeauthoryear{{King} et~al.,}{{King}
  et~al.}{2013}]{King-etal2013}
{King} A.~L.,  et~al., 2013, \mn@doi [\apj] {10.1088/0004-637X/762/2/103},
  \href {https://ui.adsabs.harvard.edu/abs/2013ApJ...762..103K} {762, 103}

\bibitem[\protect\citeauthoryear{{Knigge}, {Woods}  \& {Drew}}{{Knigge}
  et~al.}{1995}]{Knigge-etal1995}
{Knigge} C.,  {Woods} J.~A.,   {Drew} J.~E.,  1995, \mn@doi [\mnras]
  {10.1093/mnras/273.2.225}, \href
  {https://ui.adsabs.harvard.edu/abs/1995MNRAS.273..225K} {273, 225}

\bibitem[\protect\citeauthoryear{{Kubota} et~al.,}{{Kubota}
  et~al.}{2007}]{Kubota-etal2007}
{Kubota} A.,  et~al., 2007, \mn@doi [\pasj] {10.1093/pasj/59.sp1.S185}, \href
  {https://ui.adsabs.harvard.edu/abs/2007PASJ...59S.185K} {59, 185}

\bibitem[\protect\citeauthoryear{{Kumar}}{{Kumar}}{2017}]{Kumar2017}
{Kumar} N.,  2017, arXiv e-prints, \href
  {https://ui.adsabs.harvard.edu/abs/2017arXiv170804427K} {p. arXiv:1708.04427}

\bibitem[\protect\citeauthoryear{{Kumar}}{{Kumar}}{2018}]{Kumar2018}
{Kumar} N.,  2018, \mn@doi [Journal of Astrophysics and Astronomy]
  {10.1007/s12036-017-9508-z}, \href
  {https://ui.adsabs.harvard.edu/abs/2018JApA...39...13K} {39, 13}

\bibitem[\protect\citeauthoryear{{Kumar} \& {Gu}}{{Kumar} \&
  {Gu}}{2018}]{Kumar-Gu2018}
{Kumar} R.,  {Gu} W.-M.,  2018, \mn@doi [\apj] {10.3847/1538-4357/aac328},
  \href {https://ui.adsabs.harvard.edu/abs/2018ApJ...860..114K} {860, 114}

\bibitem[\protect\citeauthoryear{{Lee}, {Reynolds}, {Remillard}, {Schulz},
  {Blackman}  \& {Fabian}}{{Lee} et~al.}{2002}]{Lee-etal2002}
{Lee} J.~C.,  {Reynolds} C.~S.,  {Remillard} R.,  {Schulz} N.~S.,  {Blackman}
  E.~G.,   {Fabian} A.~C.,  2002, \mn@doi [\apj] {10.1086/338588}, \href
  {https://ui.adsabs.harvard.edu/abs/2002ApJ...567.1102L} {567, 1102}

\bibitem[\protect\citeauthoryear{{Miller} et~al.,}{{Miller}
  et~al.}{2006}]{Miller-etal2006}
{Miller} J.~M.,  et~al., 2006, \mn@doi [\apj] {10.1086/504673}, \href
  {https://ui.adsabs.harvard.edu/abs/2006ApJ...646..394M} {646, 394}

\bibitem[\protect\citeauthoryear{{Miller}, {Fabian}, {Kaastra}, {Kallman},
  {King}, {Proga}, {Raymond}  \& {Reynolds}}{{Miller}
  et~al.}{2015}]{Miller-etal2015}
{Miller} J.~M.,  {Fabian} A.~C.,  {Kaastra} J.,  {Kallman} T.,  {King} A.~L.,
  {Proga} D.,  {Raymond} J.,   {Reynolds} C.~S.,  2015, \mn@doi [\apj]
  {10.1088/0004-637X/814/2/87}, \href
  {https://ui.adsabs.harvard.edu/abs/2015ApJ...814...87M} {814, 87}

\bibitem[\protect\citeauthoryear{{Misra} \& {Taam}}{{Misra} \&
  {Taam}}{2001}]{Misra-Taam2001}
{Misra} R.,  {Taam} R.~E.,  2001, \mn@doi [\apj] {10.1086/320978}, \href
  {https://ui.adsabs.harvard.edu/abs/2001ApJ...553..978M} {553, 978}

\bibitem[\protect\citeauthoryear{{Mondal} \& {Mukhopadhyay}}{{Mondal} \&
  {Mukhopadhyay}}{2019}]{Mondal-Mukhopadhyay2018}
{Mondal} T.,  {Mukhopadhyay} B.,  2019, \mn@doi [\mnras]
  {10.1093/mnrasl/sly165}, \href
  {https://ui.adsabs.harvard.edu/abs/2019MNRAS.482L..24M} {482, L24}

\bibitem[\protect\citeauthoryear{{Mondal} \& {Mukhopadhyay}}{{Mondal} \&
  {Mukhopadhyay}}{2020}]{Mondal-Mukhopadhyay2020}
{Mondal} T.,  {Mukhopadhyay} B.,  2020, \mn@doi [\mnras]
  {10.1093/mnras/staa1161}, \href
  {https://ui.adsabs.harvard.edu/abs/2020MNRAS.495..350M} {495, 350}

\bibitem[\protect\citeauthoryear{{Narayan} \& {Yi}}{{Narayan} \&
  {Yi}}{1995}]{Narayan-Yi1995}
{Narayan} R.,  {Yi} I.,  1995, \mn@doi [\apj] {10.1086/175599}, \href
  {https://ui.adsabs.harvard.edu/abs/1995ApJ...444..231N} {444, 231}

\bibitem[\protect\citeauthoryear{{Neilsen}}{{Neilsen}}{2013}]{Neilsen-2013}
{Neilsen} J.,  2013, \mn@doi [Advances in Space Research]
  {10.1016/j.asr.2013.04.021}, \href
  {https://ui.adsabs.harvard.edu/abs/2013AdSpR..52..732N} {52, 732}

\bibitem[\protect\citeauthoryear{{Novikov} \& {Thorne}}{{Novikov} \&
  {Thorne}}{1973}]{Novikov-Thorne1973}
{Novikov} I.~D.,  {Thorne} K.~S.,  1973, in Black Holes (Les Astres Occlus). pp
  343--450

\bibitem[\protect\citeauthoryear{{Pereyra}, {Kallman}  \& {Blondin}}{{Pereyra}
  et~al.}{1997}]{Pereyra-etal1997}
{Pereyra} N.~A.,  {Kallman} T.~R.,   {Blondin} J.~M.,  1997, \mn@doi [\apj]
  {10.1086/303671}, \href
  {https://ui.adsabs.harvard.edu/abs/1997ApJ...477..368P} {477, 368}

\bibitem[\protect\citeauthoryear{{Pinto}, {Middleton}  \& {Fabian}}{{Pinto}
  et~al.}{2016}]{Pinto-etal2016}
{Pinto} C.,  {Middleton} M.~J.,   {Fabian} A.~C.,  2016, \mn@doi [\nat]
  {10.1038/nature17417}, \href
  {https://ui.adsabs.harvard.edu/abs/2016Natur.533...64P} {533, 64}

\bibitem[\protect\citeauthoryear{{Ponti}, {Fender}, {Begelman}, {Dunn},
  {Neilsen}  \& {Coriat}}{{Ponti} et~al.}{2012}]{Ponti-etal2012}
{Ponti} G.,  {Fender} R.~P.,  {Begelman} M.~C.,  {Dunn} R.~J.~H.,  {Neilsen}
  J.,   {Coriat} M.,  2012, \mn@doi [\mnras]
  {10.1111/j.1745-3933.2012.01224.x}, \href
  {https://ui.adsabs.harvard.edu/abs/2012MNRAS.422L..11P} {422, L11}

\bibitem[\protect\citeauthoryear{{Ponti}, {Bianchi}, {Mu{\~n}oz-Darias}, {De},
  {Fender}  \& {Merloni}}{{Ponti} et~al.}{2016}]{Ponti-etal2016}
{Ponti} G.,  {Bianchi} S.,  {Mu{\~n}oz-Darias} T.,  {De} K.,  {Fender} R.,
  {Merloni} A.,  2016, \mn@doi [Astronomische Nachrichten]
  {10.1002/asna.201612339}, \href
  {https://ui.adsabs.harvard.edu/abs/2016AN....337..512P} {337, 512}

\bibitem[\protect\citeauthoryear{{Pringle}}{{Pringle}}{1981}]{Pringle1981}
{Pringle} J.~E.,  1981, \mn@doi [\araa] {10.1146/annurev.aa.19.090181.001033},
  \href {https://ui.adsabs.harvard.edu/abs/1981ARA&A..19..137P} {19, 137}

\bibitem[\protect\citeauthoryear{{Proga} \& {Kallman}}{{Proga} \&
  {Kallman}}{2002}]{Proga-Kallman2002}
{Proga} D.,  {Kallman} T.~R.,  2002, \mn@doi [\apj] {10.1086/324534}, \href
  {https://ui.adsabs.harvard.edu/abs/2002ApJ...565..455P} {565, 455}

\bibitem[\protect\citeauthoryear{{Proga}, {Stone}  \& {Drew}}{{Proga}
  et~al.}{1998}]{Proga-etal1998}
{Proga} D.,  {Stone} J.~M.,   {Drew} J.~E.,  1998, \mn@doi [\mnras]
  {10.1046/j.1365-8711.1998.01337.x}, \href
  {https://ui.adsabs.harvard.edu/abs/1998MNRAS.295..595P} {295, 595}

\bibitem[\protect\citeauthoryear{{Rajesh} \& {Mukhopadhyay}}{{Rajesh} \&
  {Mukhopadhyay}}{2010}]{Rajesh-Mukhopadhyay2010}
{Rajesh} S.~R.,  {Mukhopadhyay} B.,  2010, \mn@doi [\mnras]
  {10.1111/j.1365-2966.2009.15925.x}, \href
  {https://ui.adsabs.harvard.edu/abs/2010MNRAS.402..961R} {402, 961}

\bibitem[\protect\citeauthoryear{{Remillard} \& {McClintock}}{{Remillard} \&
  {McClintock}}{2006}]{Remillard-McClintock2006}
{Remillard} R.~A.,  {McClintock} J.~E.,  2006, \mn@doi [\araa]
  {10.1146/annurev.astro.44.051905.092532}, \href
  {https://ui.adsabs.harvard.edu/abs/2006ARA&A..44...49R} {44, 49}

\bibitem[\protect\citeauthoryear{{Reynolds}}{{Reynolds}}{2012}]{Reynolds2012}
{Reynolds} C.~S.,  2012, \mn@doi [\apjl] {10.1088/2041-8205/759/1/L15}, \href
  {https://ui.adsabs.harvard.edu/abs/2012ApJ...759L..15R} {759, L15}

\bibitem[\protect\citeauthoryear{{Romanova}, {Ustyugova}, {Koldoba}  \&
  {Lovelace}}{{Romanova} et~al.}{2009}]{Romanova-etal2009}
{Romanova} M.~M.,  {Ustyugova} G.~V.,  {Koldoba} A.~V.,   {Lovelace} R.~V.~E.,
  2009, \mn@doi [\mnras] {10.1111/j.1365-2966.2009.15413.x}, \href
  {https://ui.adsabs.harvard.edu/abs/2009MNRAS.399.1802R} {399, 1802}

\bibitem[\protect\citeauthoryear{{Ross} \& {Fabian}}{{Ross} \&
  {Fabian}}{1993}]{Ross-Fabian1993}
{Ross} R.~R.,  {Fabian} A.~C.,  1993, \mn@doi [\mnras]
  {10.1093/mnras/261.1.74}, \href
  {https://ui.adsabs.harvard.edu/abs/1993MNRAS.261...74R} {261, 74}

\bibitem[\protect\citeauthoryear{{Shakura} \& {Sunyaev}}{{Shakura} \&
  {Sunyaev}}{1973}]{Shakura-Sunyaev1973}
{Shakura} N.~I.,  {Sunyaev} R.~A.,  1973, \aap, \href
  {https://ui.adsabs.harvard.edu/abs/1973A&A....24..337S} {500, 33}

\bibitem[\protect\citeauthoryear{{Tatum}, {Turner}, {Sim}, {Miller}, {Reeves},
  {Patrick}  \& {Long}}{{Tatum} et~al.}{2012}]{Tatum-etal2012}
{Tatum} M.~M.,  {Turner} T.~J.,  {Sim} S.~A.,  {Miller} L.,  {Reeves} J.~N.,
  {Patrick} A.~R.,   {Long} K.~S.,  2012, \mn@doi [\apj]
  {10.1088/0004-637X/752/2/94}, \href
  {https://ui.adsabs.harvard.edu/abs/2012ApJ...752...94T} {752, 94}

\bibitem[\protect\citeauthoryear{{Tombesi}, {Tazaki}, {Mushotzky}, {Ueda},
  {Cappi}, {Gofford}, {Reeves}  \& {Guainazzi}}{{Tombesi}
  et~al.}{2014}]{Tombesi-etal2014}
{Tombesi} F.,  {Tazaki} F.,  {Mushotzky} R.~F.,  {Ueda} Y.,  {Cappi} M.,
  {Gofford} J.,  {Reeves} J.~N.,   {Guainazzi} M.,  2014, \mn@doi [\mnras]
  {10.1093/mnras/stu1297}, \href
  {https://ui.adsabs.harvard.edu/abs/2014MNRAS.443.2154T} {443, 2154}

\bibitem[\protect\citeauthoryear{{Tombesi}, {Mel{\'e}ndez}, {Veilleux},
  {Reeves}, {Gonz{\'a}lez-Alfonso}  \& {Reynolds}}{{Tombesi}
  et~al.}{2015}]{Tombesi-etal2015}
{Tombesi} F.,  {Mel{\'e}ndez} M.,  {Veilleux} S.,  {Reeves} J.~N.,
  {Gonz{\'a}lez-Alfonso} E.,   {Reynolds} C.~S.,  2015, \mn@doi [\nat]
  {10.1038/nature14261}, \href
  {https://ui.adsabs.harvard.edu/abs/2015Natur.519..436T} {519, 436}

\bibitem[\protect\citeauthoryear{{Waters} \& {Proga}}{{Waters} \&
  {Proga}}{2012}]{Waters-Proga2012}
{Waters} T.~R.,  {Proga} D.,  2012, \mn@doi [\mnras]
  {10.1111/j.1365-2966.2012.21823.x}, \href
  {https://ui.adsabs.harvard.edu/abs/2012MNRAS.426.2239W} {426, 2239}

\bibitem[\protect\citeauthoryear{{Woods}, {Klein}, {Castor}, {McKee}  \&
  {Bell}}{{Woods} et~al.}{1996}]{Woods-etal1996}
{Woods} D.~T.,  {Klein} R.~I.,  {Castor} J.~I.,  {McKee} C.~F.,   {Bell} J.~B.,
   1996, \mn@doi [\apj] {10.1086/177101}, \href
  {https://ui.adsabs.harvard.edu/abs/1996ApJ...461..767W} {461, 767}

\bibitem[\protect\citeauthoryear{{Yuan} \& {Narayan}}{{Yuan} \&
  {Narayan}}{2014}]{Yuan-Narayan2014}
{Yuan} F.,  {Narayan} R.,  2014, \mn@doi [\araa]
  {10.1146/annurev-astro-082812-141003}, \href
  {https://ui.adsabs.harvard.edu/abs/2014ARA&A..52..529Y} {52, 529}

\bibitem[\protect\citeauthoryear{{Yuan}, {Gan}, {Narayan}, {Sadowski}, {Bu}  \&
  {Bai}}{{Yuan} et~al.}{2015}]{Yuan-etal2015}
{Yuan} F.,  {Gan} Z.,  {Narayan} R.,  {Sadowski} A.,  {Bu} D.,   {Bai} X.-N.,
  2015, \mn@doi [\apj] {10.1088/0004-637X/804/2/101}, \href
  {https://ui.adsabs.harvard.edu/abs/2015ApJ...804..101Y} {804, 101}

\makeatother
\end{thebibliography}

\appendix


\section{Consistency of numerical scheme} 

\begin{figure}
  \centering
\begin{tabular}{l}\hspace{-0.9cm}
  \includegraphics[width=0.45\textwidth]{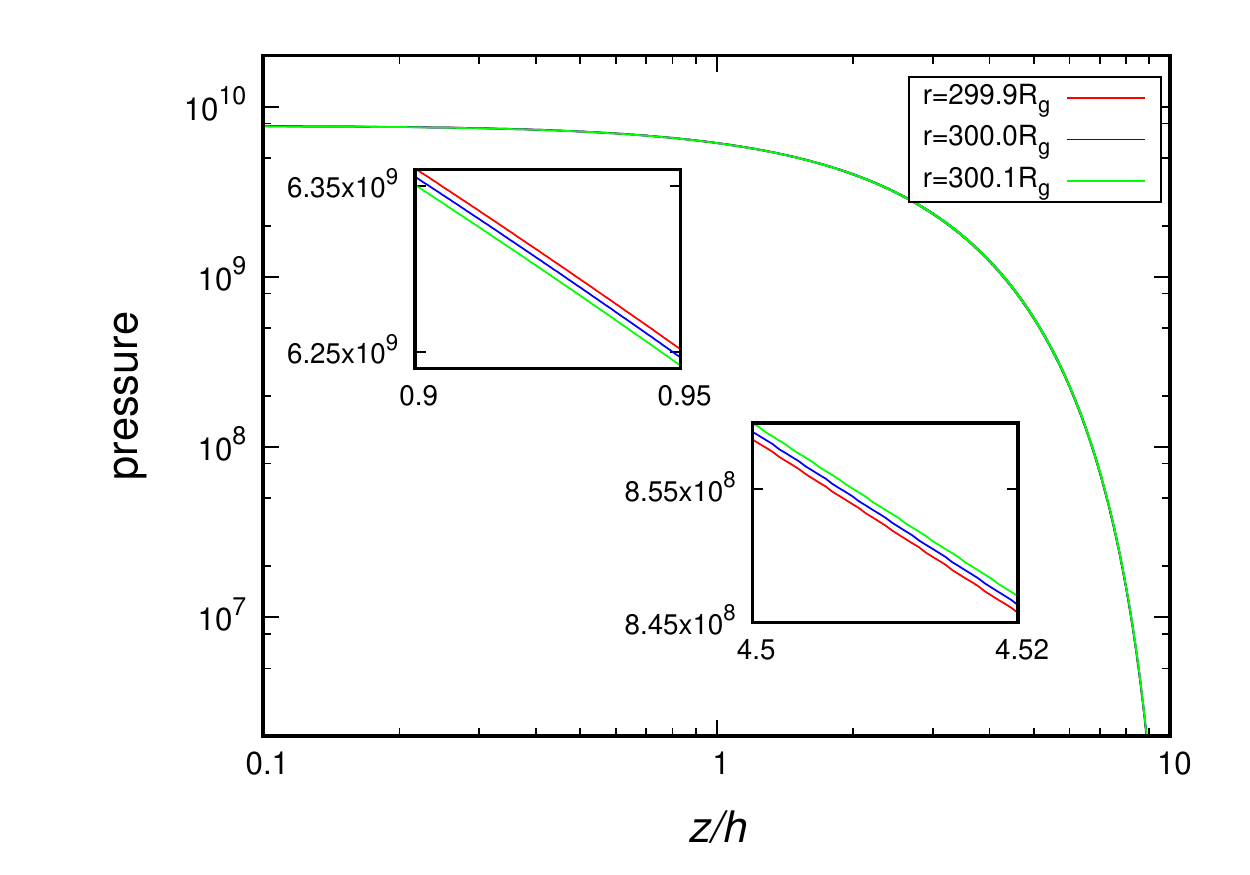}
\end{tabular}\vspace{-0.3cm}
\caption{ The pressure as a function of height for three adjacent $r$, 299.9, 300.0 and 300.1$R_g$. Here the middle curve (or $r=300R_g$) is same as the curve 2 of Figure \ref{fig:p_x}, and other two curves are solved for the same set 
	of parameters of the middle curve. Two insets show that for $z<z_f$ the inner region pressure is greater than the outer region while the opposite is true for $z>z_f$.
That is, $\frac{\partial p}{\partial r}$ flips the sign above $z_f$. 
	Thus the present solution for a fixed $r$ is consistent approximately 
	with the solutions would have obtained with varying the radial
	grid points.
    Here, $z_f$ is $\sim 3.1h$, while for the curve 2 of Figure \ref{fig:p_x} $z_f \sim 2.2h$.  }  
\label{fig:cons_num_schm}
\end{figure}

With the aim of understanding outflow at a fixed launching radius $r$, 
we have solved the  governing equations along the $z$-axis (by adopting a 2.5-dimensional accretion disk formalism, \citealp[e.g.,][]{Mondal-Mukhopadhyay2018, Mondal-Mukhopadhyay2020}). Here, 
we check the consistency of these solutions based for a fixed $r$ by considering two additional grid points around 
$r$ as $r-\Delta r$ and $r+\Delta r$ with $\frac{\Delta r}{r}\ll 1$. For this, we consider a fact that $\frac{\partial p}{\partial r}$ flips its sign from negative to positive at a height $z_f$, and $z_f$ is related to the pressure scale height $h_p$ as $z_f = h_p^2/h$ (see the discussion point (b) of subsection \S2.1). We take the set of free parameters the same as that corresponding to the curve 2 of Figure \ref{fig:p_x}, i.e., $r = 300R_g$, $f_v\sim 1$, $x = 3.05414 \times 10^{-7}$, $M_c = 10M_\odot$, $\alpha = 0.1$, $\dot{M} = 0.005\dot{M}_{Edd}$.
In Figure \ref{fig:cons_num_schm}, we show the pressure as a function of height $z$ for three adjacent launching radii $r$ = 299.9, 300.0 and 300.1$R_g$ (or for three nearby grid points in $r$). 
In insets of Figure \ref{fig:cons_num_schm}, we show that at a height $0.9h<z<0.95h$ the inner region pressure is larger than the outer region pressure, while for $4.5h<z<4.52h$ it is opposite.
In general, for $z<z_f$ the inner region pressure is larger than the outer 
region pressure and for $z>z_f$ it is opposite. In another way, the radial 
pressure gradient flips the sign at a height $z_f$. However,
the magnitude of $z_f$ estimated based on three radial grid points is slightly 
larger than that obtained for curve 2 of Figure \ref{fig:p_x}.
Thus, in the present method of solution at a fixed $r$, we are also effectively 
taking an account of the variation of flow variables in the radial direction, as in any case we solve them by treating as partial differentials.
In short, even if we are solving the governing equations for a fixed $r$, this 
solution effectively represents approximately a similar picture when one solves the governing equations with taking the grid in both the directions $r$ and $z$.
However, commonly the wind solution is obtained by first defining a streamline
for the wind. In the contrary to first define a streamline, we solve the governing equation along the $z$-axis (which is not a streamline) for a fixed launching radius and obtain an equatorial wind of a small opening angle.
A schematic diagram for the present approach and solution is shown in Figure \ref{fig:wind_geo}.
 \\

\bsp
\label{lastpage}
\end{document}